\documentclass[longauth]{aa}  

\usepackage{graphicx}
\usepackage{txfonts}
\usepackage{orcidlink}
\usepackage{siunitx}
\usepackage{booktabs}
\usepackage{multirow}
\usepackage{arydshln}
\usepackage[flushleft]{threeparttable}
\usepackage{threeparttable}
\usepackage{placeins}
\usepackage{float}

\newcommand{\rjup}{R$_\mathrm{J}$}
\newcommand{\mjup}{M$_\mathrm{J}$}

\newcommand{\teff}{$T_{\rm eff}$}
\newcommand{\logg}{log\,$g_\star$}

\newcommand{\feh}{[Fe/H]}

\newcommand{\ars}{$a/R_\star$}

\newcommand{\mpl}{$M_{\rm p}$}
\newcommand{\rpl}{$R_{\rm p}$}

\newcommand{\fiveoneBmag}{$11.6 \pm 0.2$}
\newcommand{\fiveoneVmag}{$10.792 \pm 0.011$}

\newcommand{\fiveonepmRA}{$6.13 \pm 0.07$}
\newcommand{\fiveonepmDEC}{$-65.60 \pm 0.06$}
\newcommand{\fiveoneTmag}{$10.233 \pm 0.006$}

\newcommand{\fiveoned}{$230 \pm 2$}
\newcommand{\fivesixBmag}{$11.55 \pm 0.10$}
\newcommand{\fivesixVmag}{$10.860 \pm 0.007$}

\newcommand{\fivesixpmRA}{$-8.69 \pm 0.04$}
\newcommand{\fivesixpmDEC}{$-35.20 \pm 0.05$}
\newcommand{\fivesixTmag}{$10.449 \pm 0.006$}

\newcommand{\fivesixd}{$305 \pm 2$}
\newcommand{\twooneBmag}{$11.66 \pm 0.09$}
\newcommand{\twooneVmag}{$10.887 \pm 0.006$}

\newcommand{\twoonepmRA}{$-44.00 \pm 0.04$}
\newcommand{\twoonepmDEC}{$7.89 \pm 0.07$}
\newcommand{\twooneTmag}{$10.167 \pm 0.006$}

\newcommand{\twooned}{$198.5 \pm 1.4$}
\newcommand{\fiveoneGAIAmag}{$10.667 \pm 0.003$}
\newcommand{\fiveonegaiabp}{$10.992 \pm 0.003$}
\newcommand{\fiveonegaiarp}{$10.175 \pm 0.004$}
\newcommand{\fiveoneplx}{$4.496 \pm 0.016$}
\newcommand{\fivesixGAIAmag}{$10.857 \pm 0.003$}
\newcommand{\fivesixgaiabp}{$11.156 \pm 0.003$}
\newcommand{\fivesixgaiarp}{$10.391 \pm 0.004$}
\newcommand{\fivesixplx}{$3.303 \pm 0.011$}
\newcommand{\twooneGAIAmag}{$10.673 \pm 0.003$}
\newcommand{\twoonegaiabp}{$11.071 \pm 0.003$}
\newcommand{\twoonegaiarp}{$10.108 \pm 0.004$}
\newcommand{\twooneplx}{$5.116 \pm 0.013$}
\newcommand{\twooneRV}{$-108.55 \pm 0.17$}
\newcommand{\fiveoneRV}{$16.33 \pm 0.53$}
\newcommand{\fivesixRV}{$6.06 \pm 0.26$}
\newcommand{\twooneRUWE}{$1.003$}
\newcommand{\fiveoneRUWE}{$1.034$}
\newcommand{\fivesixRUWE}{$0.871$}

\newcommand{\fiveonelogg}{$4.18 \pm 0.07$}

\newcommand{\fiveonevmac}{$3.7$}
\newcommand{\fiveonevsini}{$4.8 \pm 0.9$}
\newcommand{\fiveoneTeff}{$5763 \pm 90$}
\newcommand{\fiveoneFeH}{$0.2 \pm 0.07$}
\newcommand{\fiveonevmic}{$1.04$}

\newcommand{\fivesixlogg}{$4.05 \pm 0.07$}

\newcommand{\fivesixvmac}{$4.34$}
\newcommand{\fivesixvsini}{$4.3 \pm 1.2$}
\newcommand{\fivesixTeff}{$5905 \pm 100$}
\newcommand{\fivesixFeH}{$0.11 \pm 0.05$}
\newcommand{\fivesixvmic}{$1.12$}

\newcommand{\twoonelogg}{$4.19 \pm 0.05$}

\newcommand{\twoonevmac}{$3.49$}
\newcommand{\twoonevsini}{$3.7 \pm 0.5$}
\newcommand{\twooneTeff}{$5673 \pm 50$}
\newcommand{\twooneFeH}{$0.47 \pm 0.08$}
\newcommand{\twoonevmic}{$1$}

\newcommand{\fiveoneradbasta}{$1.42 \pm 0.05$}
\newcommand{\fiveonemassbasta}{$1.12 \pm 0.05$}
\newcommand{\fiveonerhobasta}{$0.55 \pm 0.07$}
\newcommand{\fiveoneagebasta}{$7.5 \pm 1.5$}
\newcommand{\fiveonelumbasta}{$1.98 \pm 0.09$}
\newcommand{\fivesixradbasta}{$1.64 \pm 0.06$}
\newcommand{\fivesixmassbasta}{$1.13 \pm 0.04$}
\newcommand{\fivesixrhobasta}{$0.36 \pm 0.04$}
\newcommand{\fivesixagebasta}{$6.6 \pm 0.7$}
\newcommand{\fivesixlumbasta}{$2.87 \pm 0.14$}
\newcommand{\twooneradbasta}{$1.32 \pm 0.04$}
\newcommand{\twoonemassbasta}{$1.12 \pm 0.05$}
\newcommand{\twoonerhobasta}{$0.69 \pm 0.06$}
\newcommand{\twooneagebasta}{$7.7 \pm 1.6$}
\newcommand{\twoonelumbasta}{$1.57 \pm 0.06$}
\newcommand{\midfiveone}[1][]{$2460267.9303 \pm 0.0003$}
\newcommand{\midfivesix}[1][]{$2459704.1558 \pm 0.0009$}
\newcommand{\midtwoone}[1][]{$2459018.9195 \pm 0.0008$}
\newcommand{\perfiveone}[1][]{$24.91645 \pm 0.00002$}
\newcommand{\perfivesix}[1][]{$22.01850 \pm 0.00004$}
\newcommand{\pertwoone}[1][]{$8.60079 ^{+0.00002} _{-0.00001}$}
\newcommand{\arfiveone}[1][]{$26.1 \pm 1.0$}
\newcommand{\arfivesix}[1][]{$21 ^{+2} _{-3}$}
\newcommand{\artwoone}[1][]{$11.5 \pm 0.4$}
\newcommand{\rpfiveone}[1][]{$0.0703 \pm 0.0005$}
\newcommand{\rpfivesix}[1][]{$0.0624 \pm 0.0012$}
\newcommand{\rptwoone}[1][]{$0.0702 \pm 0.0007$}
\newcommand{\cosifiveone}[1][]{$0.014 ^{+0.004} _{-0.003}$}
\newcommand{\cosifivesix}[1][]{$0.055 ^{+0.012} _{-0.015}$}
\newcommand{\cositwoone}[1][]{$0.074 \pm 0.004$}
\newcommand{\ecosfiveone}[1][]{$0.21 ^{+0.04} _{-0.03}$}
\newcommand{\ecosfivesix}[1][]{$-0.02 \pm 0.16$}
\newcommand{\ecostwoone}[1][]{$0.12 ^{+0.05} _{-0.04}$}
\newcommand{\esinfiveone}[1][]{$0.24 ^{+0.06} _{-0.05}$}
\newcommand{\esinfivesix}[1][]{$0.53 ^{+0.09} _{-0.08}$}
\newcommand{\esintwoone}[1][]{$0.06 ^{+0.12} _{-0.09}$}
\newcommand{\kampfiveone}[1][]{$79.6 \pm 2.0$}
\newcommand{\kampfivesix}[1][]{$16.9 ^{+1.8} _{-1.7}$}
\newcommand{\kamptwoone}[1][]{$79.8 \pm 1.8$}
\newcommand{\kampcfiveone}[1][]{$\cdots$}
\newcommand{\kampcfivesix}[1][]{$\cdots$}
\newcommand{\kampctwoone}[1][]{$41 \pm 3$}
\newcommand{\midcfiveone}[1][]{$\cdots$}
\newcommand{\midcfivesix}[1][]{$\cdots$}
\newcommand{\midctwoone}[1][]{$2459620 ^{+18} _{-14}$}
\newcommand{\percfiveone}[1][]{$\cdots$}
\newcommand{\percfivesix}[1][]{$\cdots$}
\newcommand{\perctwoone}[1][]{$414.5 ^{+1.5} _{-1.6}$}
\newcommand{\ecoscfiveone}[1][]{$\cdots$}
\newcommand{\ecoscfivesix}[1][]{$\cdots$}
\newcommand{\ecosctwoone}[1][]{$-0.37 ^{+0.09} _{-0.10}$}
\newcommand{\esincfiveone}[1][]{$\cdots$}
\newcommand{\esincfivesix}[1][]{$\cdots$}
\newcommand{\esinctwoone}[1][]{$-0.64 ^{+0.07} _{-0.08}$}
\newcommand{\gammafiveone}[1][]{$47781 \pm 2$}
\newcommand{\gammafivesix}[1][]{$65122.9 ^{+1.5} _{-1.2}$}
\newcommand{\gammatwoone}[1][]{$-64817 ^{+18} _{-12}$}
\newcommand{\gammatfiveone}[1][]{$38.7 ^{+1.8} _{-1.7}$}
\newcommand{\gammatfivesix}[1][]{$-45 ^{+4} _{-3}$}
\newcommand{\gammattwoone}[1][]{$-35 ^{+21} _{-16}$}
\newcommand{\gammattfiveone}[1][]{$\cdots$}
\newcommand{\gammattfivesix}[1][]{$\cdots$}
\newcommand{\gammatttwoone}[1][]{$-40 ^{+17} _{-12}$}
\newcommand{\jitterfiveone}[1][]{$13 \pm 3$}
\newcommand{\jitterfivesix}[1][]{$0.006 ^{+0.027} _{-0.006}$}
\newcommand{\jittertwoone}[1][]{$20 \pm 3$}
\newcommand{\jittertfiveone}[1][]{$0.02 ^{+0.24} _{-0.02}$}
\newcommand{\jittertfivesix}[1][]{$0.007 ^{+0.027} _{-0.007}$}
\newcommand{\jitterttwoone}[1][]{$0.3 ^{+3.0} _{-0.3}$}
\newcommand{\jitterttfiveone}[1][]{$\cdots$}
\newcommand{\jitterttfivesix}[1][]{$\cdots$}
\newcommand{\jittertttwoone}[1][]{$6.7 ^{+1.5} _{-1.7}$}
\newcommand{\kampdfiveone}[1][]{$\cdots$}
\newcommand{\kampdfivesix}[1][]{$\cdots$}
\newcommand{\kampdtwoone}[1][]{$91 ^{+11} _{-16}$}
\newcommand{\middfiveone}[1][]{$\cdots$}
\newcommand{\middfivesix}[1][]{$\cdots$}
\newcommand{\middtwoone}[1][]{$2460505 ^{+81} _{-115}$}
\newcommand{\perdfiveone}[1][]{$\cdots$}
\newcommand{\perdfivesix}[1][]{$\cdots$}
\newcommand{\perdtwoone}[1][]{$3336 ^{+378} _{-538}$}
\newcommand{\circmidfiveone}[1][]{$2460267.9302 \pm 0.0003$}
\newcommand{\circmidfivesix}[1][]{$2459704.1558 \pm 0.0008$}
\newcommand{\circperfiveone}[1][]{$24.91645 \pm 0.00002$}
\newcommand{\circperfivesix}[1][]{$22.01850 \pm 0.00004$}
\newcommand{\circarfiveone}[1][]{$28.2 ^{+0.8} _{-0.7}$}
\newcommand{\circarfivesix}[1][]{$28.5 ^{+1.9} _{-2.4}$}
\newcommand{\circrpfiveone}[1][]{$0.0703 \pm 0.0005$}
\newcommand{\circrpfivesix}[1][]{$0.0625 ^{+0.0012} _{-0.0011}$}
\newcommand{\circcosifiveone}[1][]{$0.012 \pm 0.003$}
\newcommand{\circcosifivesix}[1][]{$0.029 \pm 0.003$}
\newcommand{\circkampfiveone}[1][]{$77 \pm 2$}
\newcommand{\circkampfivesix}[1][]{$15.5 ^{+1.6} _{-1.7}$}
\newcommand{\circgammafiveone}[1][]{$47781 \pm 3$}
\newcommand{\circgammafivesix}[1][]{$65122.9 \pm 1.3$}
\newcommand{\circgammatfiveone}[1][]{$38.7 ^{+2.0} _{-1.8}$}
\newcommand{\circgammatfivesix}[1][]{$-45 \pm 3$}
\newcommand{\circjitterfiveone}[1][]{$17 ^{+2} _{-3}$}
\newcommand{\circjitterfivesix}[1][]{$0.009 ^{+0.059} _{-0.009}$}
\newcommand{\circjittertfiveone}[1][]{$0.19 ^{+3.11} _{-0.19}$}
\newcommand{\circjittertfivesix}[1][]{$0.006 ^{+0.027} _{-0.006}$}
\newcommand{\eccfiveone}[1][]{$0.10 \pm 0.02$}
\newcommand{\eccfivesix}[1][]{$0.30 ^{+0.09} _{-0.08}$}
\newcommand{\ecctwoone}[1][]{$0.028 ^{+0.013} _{-0.014}$}
\newcommand{\wwfiveone}[1][]{$49 ^{+12} _{-10}$}
\newcommand{\wwfivesix}[1][]{$92 ^{+18} _{-17}$}
\newcommand{\wwtwoone}[1][]{$26 ^{+46} _{-35}$}
\newcommand{\incfiveone}[1][]{$89.20 ^{+0.18} _{-0.21}$}
\newcommand{\incfivesix}[1][]{$86.9 ^{+0.9} _{-0.7}$}
\newcommand{\inctwoone}[1][]{$85.8 ^{+0.3} _{-0.2}$}
\newcommand{\impfiveone}[1][]{$0.34 ^{+0.08} _{-0.06}$}
\newcommand{\impfivesix}[1][]{$0.82 ^{+0.04} _{-0.03}$}
\newcommand{\imptwoone}[1][]{$0.840 ^{+0.012} _{-0.011}$}
\newcommand{\impoccfiveone}[1][]{$0.39 ^{+0.10} _{-0.07}$}
\newcommand{\impoccfivesix}[1][]{$1.5 \pm 0.3$}
\newcommand{\impocctwoone}[1][]{$0.85 \pm 0.03$}
\newcommand{\semifiveone}[1][]{$0.172 \pm 0.007$}
\newcommand{\semifivesix}[1][]{$0.160 \pm 0.019$}
\newcommand{\semitwoone}[1][]{$0.070 \pm 0.002$}
\newcommand{\teqfiveone}[1][]{$796 \pm 18$}
\newcommand{\teqfivesix}[1][]{$906 \pm 55$}
\newcommand{\teqtwoone}[1][]{$1173 \pm 23$}
\newcommand{\totalfiveone}[1][]{$6.87 \pm 0.03$}
\newcommand{\totalfivesix}[1][]{$3.95 ^{+0.10} _{-0.09}$}
\newcommand{\totaltwoone}[1][]{$3.77 \pm 0.04$}
\newcommand{\gressfiveone}[1][]{$0.51 \pm 0.03$}
\newcommand{\gressfivesix}[1][]{$0.65 ^{+0.12} _{-0.13}$}
\newcommand{\gresstwoone}[1][]{$0.75 \pm 0.06$}
\newcommand{\radpfiveone}[1][]{$0.969 \pm 0.017$}
\newcommand{\radpfivesix}[1][]{$0.996 \pm 0.015$}
\newcommand{\radptwoone}[1][]{$0.899 \pm 0.015$}
\newcommand{\mpfiveone}[1][]{$1.23 \pm 0.05$}
\newcommand{\mpfivesix}[1][]{$0.25 \pm 0.03$}
\newcommand{\mptwoone}[1][]{$0.87 \pm 0.04$}
\newcommand{\finsofiveone}[1][]{$66 \pm 6$}
\newcommand{\finsofivesix}[1][]{$112 \pm 10$}
\newcommand{\finsotwoone}[1][]{$215 \pm 15$}
\newcommand{\ecccfiveone}[1][]{$\cdots$}
\newcommand{\ecccfivesix}[1][]{$\cdots$}
\newcommand{\eccctwoone}[1][]{$0.54 \pm 0.04$}
\newcommand{\wwcfiveone}[1][]{$\cdots$}
\newcommand{\wwcfivesix}[1][]{$\cdots$}
\newcommand{\wwctwoone}[1][]{$-120 \pm 9$}
\newcommand{\mpcfiveone}[1][]{$\cdots$}
\newcommand{\mpcfivesix}[1][]{$\cdots$}
\newcommand{\mpctwoone}[1][]{$1.62 \pm 0.13$}
\newcommand{\semicfiveone}[1][]{$\cdots$}
\newcommand{\semicfivesix}[1][]{$\cdots$}
\newcommand{\semictwoone}[1][]{$0.95 \pm 0.08$}
\newcommand{\eccdfiveone}[1][]{$\cdots$}
\newcommand{\eccdfivesix}[1][]{$\cdots$}
\newcommand{\eccdtwoone}[1][]{$\cdots$}
\newcommand{\wwdfiveone}[1][]{$\cdots$}
\newcommand{\wwdfivesix}[1][]{$\cdots$}
\newcommand{\wwdtwoone}[1][]{$\cdots$}
\newcommand{\mpdfiveone}[1][]{$\cdots$}
\newcommand{\mpdfivesix}[1][]{$\cdots$}
\newcommand{\mpdtwoone}[1][]{$7.2 \pm 1.1$}
\newcommand{\semidfiveone}[1][]{$\cdots$}
\newcommand{\semidfivesix}[1][]{$\cdots$}
\newcommand{\semidtwoone}[1][]{$4.5 \pm 0.7$}
\newcommand{\LConeqsumfiveone}[1][]{$0.62 \pm 0.09$}
\newcommand{\LConeqonefiveone}[1][]{$0.39 \pm 0.04$}
\newcommand{\LConeqtwofiveone}[1][]{$0.23 \pm 0.04$}
\newcommand{\LConeqdifffiveone}[1][]{$0.16$}
\newcommand{\LCthreeqsumfiveone}[1][]{$0.83 ^{+0.08} _{-0.09}$}
\newcommand{\LCthreeqonefiveone}[1][]{$0.71 ^{+0.04} _{-0.05}$}
\newcommand{\LCthreeqtwofiveone}[1][]{$0.12 ^{+0.04} _{-0.05}$}
\newcommand{\LCthreeqdifffiveone}[1][]{$0.59$}
\newcommand{\LCfiveqsumfiveone}[1][]{$0.52 \pm 0.09$}
\newcommand{\LCfiveqonefiveone}[1][]{$0.32 ^{+0.05} _{-0.04}$}
\newcommand{\LCfiveqtwofiveone}[1][]{$0.20 ^{+0.05} _{-0.04}$}
\newcommand{\LCfiveqdifffiveone}[1][]{$0.12$}
\newcommand{\LCsevenqsumfiveone}[1][]{$0.57 \pm 0.09$}
\newcommand{\LCsevenqonefiveone}[1][]{$0.38 \pm 0.04$}
\newcommand{\LCsevenqtwofiveone}[1][]{$0.19 \pm 0.04$}
\newcommand{\LCsevenqdifffiveone}[1][]{$0.19$}
\newcommand{\LCeightqsumfiveone}[1][]{$0.67 \pm 0.09$}
\newcommand{\LCeightqonefiveone}[1][]{$0.48 ^{+0.05} _{-0.04}$}
\newcommand{\LCeightqtwofiveone}[1][]{$0.19 ^{+0.05} _{-0.04}$}
\newcommand{\LCeightqdifffiveone}[1][]{$0.29$}
\newcommand{\LCnineqsumfiveone}[1][]{$0.68 \pm 0.09$}
\newcommand{\LCnineqonefiveone}[1][]{$0.53 ^{+0.05} _{-0.04}$}
\newcommand{\LCnineqtwofiveone}[1][]{$0.15 ^{+0.05} _{-0.04}$}
\newcommand{\LCnineqdifffiveone}[1][]{$0.38$}
\newcommand{\LConeqsumfivesix}[1][]{$0.59 \pm 0.07$}
\newcommand{\LConeqonefivesix}[1][]{$0.28 ^{+0.03} _{-0.04}$}
\newcommand{\LConeqtwofivesix}[1][]{$0.32 ^{+0.03} _{-0.04}$}
\newcommand{\LConeqdifffivesix}[1][]{$-0.04$}
\newcommand{\LCfourqsumfivesix}[1][]{$0.43 \pm 0.07$}
\newcommand{\LCfourqonefivesix}[1][]{$0.26 \pm 0.04$}
\newcommand{\LCfourqtwofivesix}[1][]{$0.16 \pm 0.04$}
\newcommand{\LCfourqdifffivesix}[1][]{$0.10$}
\newcommand{\LConeqsumtwoone}[1][]{$0.60 ^{+0.06} _{-0.05}$}
\newcommand{\LConeqonetwoone}[1][]{$0.32 \pm 0.03$}
\newcommand{\LConeqtwotwoone}[1][]{$0.28 \pm 0.03$}
\newcommand{\LConeqdifftwoone}[1][]{$0.05$}
\newcommand{\LCfiveqsumtwoone}[1][]{$0.98 \pm 0.08$}
\newcommand{\LCfiveqonetwoone}[1][]{$0.51 \pm 0.04$}
\newcommand{\LCfiveqtwotwoone}[1][]{$0.46 \pm 0.04$}
\newcommand{\LCfiveqdifftwoone}[1][]{$0.05$}
\newcommand{\LCthreeqsumtwoone}[1][]{$0.62 ^{+0.09} _{-0.08}$}
\newcommand{\LCthreeqonetwoone}[1][]{$0.68 \pm 0.04$}
\newcommand{\LCthreeqtwotwoone}[1][]{$-0.06 \pm 0.04$}
\newcommand{\LCthreeqdifftwoone}[1][]{$0.74$}
\newcommand{\LCfourqsumtwoone}[1][]{$0.45 \pm 0.09$}
\newcommand{\LCfourqonetwoone}[1][]{$0.30 \pm 0.05$}
\newcommand{\LCfourqtwotwoone}[1][]{$0.15 \pm 0.05$}
\newcommand{\LCfourqdifftwoone}[1][]{$0.15$}
\newcommand{\tsmfiveone}[1][]{$ 17.4 \pm 1.7 $}
\newcommand{\esmfiveone}[1][]{$ 14.9 \pm 0.9 $}
\newcommand{\tsmfivesix}[1][]{$ 77 \pm 12 $}
\newcommand{\esmfivesix}[1][]{$ 14.5 \pm 1.8 $}
\newcommand{\tsmtwoone}[1][]{$ 36 \pm 3 $}
\newcommand{\esmtwoone}[1][]{$ 38.1 \pm 1.7 $}
\newcommand{\hbmalpha}[1][]{$0.77 \pm 0.06$}
\newcommand{\hbmlambda}[1][]{$3.1 \pm 0.5$}
\newcommand{\hbmsigma}[1][]{$0.029 \pm 0.006$}
\newcommand{\Nhbm}[1][]{$N=188$}
\newcommand{\hbmalphasingle}[1][]{$0.87 \pm 0.07$}
\newcommand{\hbmalphacompanion}[1][]{$0.66 \pm 0.09$}
\newcommand{\hbmalphasub}[1][]{$0.65 \pm 0.15$}
\newcommand{\hbmalphasuper}[1][]{$0.83 \pm 0.07$}
\newcommand{\hbmlambdalogit}[1][]{$13 \pm 3$}
\newcommand{\hbmsigmalogit}[1][]{$0.33 \pm 0.04$}
\newcommand{\hbmbetazerologit}[1][]{$-0.5 \pm 0.4$}
\newcommand{\hbmbetaonelogit}[1][]{$2.2 \pm 0.9$}
\newcommand{\hbmcross}[1][]{$0.22^{+0.22}_{-0.13}$}

\usepackage[]{hyperref}
\usepackage{xcolor}
\hypersetup{
    colorlinks,
    linkcolor={red!50!black},
    citecolor={blue!50!black},
    urlcolor={blue!80!black}
}

\newcommand{\fref}[1]{Figure~\ref{#1}}
\newcommand{\tref}[1]{Table~\ref{#1}}

\DeclareMathOperator{\logit}{logit}
\DeclareSIUnit{\hertz}{Hz}
\DeclareSIUnit{\solarmass}{M_\odot}

\begin{document}

   \title{TS23/McDonald and FIES/NOT strike again}

   \subtitle{Two new warm Jupiter systems and outer companions in a third}

    \author{
        E.~Knudstrup\inst{\ref{sac}}\orcidlink{0000-0001-7880-594X} \and
        D.~Gandolfi\inst{\ref{torino}}\orcidlink{0000-0001-8627-9628} \and
        W.~D.~Cochran\inst{\ref{mcd}}\orcidlink{0000-0001-9662-3496} \and 
        M.~Endl\inst{\ref{mcd}} \and 
        P.~MacQueen\inst{\ref{mcd}} \and
        T.~Pugh\inst{\ref{mcd}} \and
        M.~S.~Lundkvist\inst{\ref{sac}}\orcidlink{0000-0002-8661-2571} \and
        C.~M.~Persson\inst{\ref{oso}} \and
        A.~Stokholm\inst{\ref{bir}} \and
        M.~L.~Winther\inst{\ref{sac}}\orcidlink{0000-0003-1687-3271} \and
        J.~R.~Larsen\inst{\ref{sac}}\orcidlink{0009-0006-0423-2353} \and
        J.~L.~R{\o}rsted\inst{\ref{sac}}\orcidlink{0000-0001-9234-430X}\and 
        S.~H.~Albrecht\inst{\ref{sac}}\orcidlink{0000-0003-1762-8235} \and
        K.~A.~Collins\inst{\ref{cfa}}\orcidlink{0000-0001-6588-9574} \and
        C.~Watkins\inst{\ref{bozeman}}\orcidlink{0000-0001-8621-6731} \and
        B.~Edwards\inst{\ref{sron},\ref{ucl}} \and
        D.~R.~Ciardi\inst{\ref{nes}}\orcidlink{0000-0002-5741-3047} \and
        C.~A.~Clark\inst{\ref{nes}}\orcidlink{0000-0002-2361-5812} \and
        A.~S.~Polanski\inst{\ref{lowell}}\orcidlink{0000-0001-7047-8681} \and
        C.~Ziegler\inst{\ref{nacog}} \and
        C.~Brice\~{n}o\inst{\ref{cerro}} \and
        A.~W.~Mann\inst{\ref{chapel}} \and
        M.~Everett\inst{\ref{nsf}}\orcidlink{0000-0002-0885-7215} \and
        S.~Howell\inst{\ref{ames}}\orcidlink{0000-0002-2532-2853} \and
        B.~Safonov\inst{\ref{stern}}\orcidlink{0000-0003-1713-3208} \and
        T.~Gan\inst{\ref{iac},\ref{ull}}\orcidlink{0000-0002-4503-9705} \and
        K.~Barkaoui\inst{\ref{iac},\ref{liege},\ref{mit}}\orcidlink{0000-0003-1464-9276} \and
        P.~Benni\inst{\ref{acton}}\orcidlink{0000-0001-6981-8722} \and
        R.~P.~Schwarz\inst{\ref{cfa}}\orcidlink{0000-0001-8227-1020} \and
        K.~Horne\inst{\ref{supa}}\orcidlink{0000-0003-1728-0304} \and
        R.~Sefako\inst{\ref{saao}}\orcidlink{0000-0003-3904-6754} \and
        D.~Muthukrishna\inst{\ref{mit2},\ref{aai}}\orcidlink{0000-0002-5788-9280}
    }
    \authorrunning{Knudstrup et~al.}
    
   \institute{
    astroCOALA, Department of Physics and Astronomy, Aarhus University, Ny Munkegade 120, DK-8000 Aarhus C, Denmark\label{sac}\email{emil@phys.au.dk}
    \and
    Dipartimento di Fisica, Universit\`{a} degli Studi di Torino, via Pietro Giuria 1, I-10125, Torino, Italy\label{torino} \and
    McDonald Observatory and Center for Planetary Systems Habitability, The University of Texas, Austin, Texas, USA\label{mcd} \and
    Department of Physics and Astronomy, Chalmers University of Technology, Onsala Space Observatory, SE-439 92 Onsala, Sweden\label{oso} \and
    School of Physics \& Astronomy, University of Birmingham, Edgbaston, Birmingham B15 2TT, UK \label{bir} \and
    Center for Astrophysics | Harvard \& Smithsonian, 60 Garden Street, Cambridge, MA 02138, USA\label{cfa} \and
    Bozeman, MT 59718, USA\label{bozeman} \and
    SRON, Netherlands Institute for Space Research, Niels Bohrweg 4, NL-2333 CA, Leiden, The Netherlands\label{sron} \and
    Department of Physics and Astronomy, University College London, Gower Street, London, WC1E 6BT, UK\label{ucl} \and
    NASA Exoplanet Science Institute-Caltech/IPAC, Pasadena, CA 91125, USA\label{nes} \and
    Lowell Observatory, 1400 West Mars Hill Road,Flagstaff, AZ86001, USA\label{lowell} \and
    Department of Physics, Engineering and Astronomy, Stephen F. Austin State University, 1936 North St, Nacogdoches, TX 75962, USA\label{nacog} \and
    Cerro Tololo Inter-American Observatory, Casilla 603, La Serena, Chile\label{cerro} \and
    Department of Physics and Astronomy, The University of North Carolina at Chapel Hill, Chapel Hill, NC 27599-3255, USA\label{chapel} \and
    NSF NOIRLab, 950 N. Cherry Ave., Tucson, AZ 85719, USA\label{nsf} \and
    NASA Ames Research Center, Moffett Field, CA 94035, USA\label{ames} \and 
    Sternberg Astronomical Institute Lomonosov Moscow State University, Universitetskii prospekt, 13, Moscow, Russia\label{stern} \and
    Instituto de Astrof\'{i}sica de Canarias (IAC), E-38205 La Laguna, Tenerife, Spain\label{iac} \and
    Departamento de Astrof\'isica, Universidad de La Laguna (ULL), E-38206 La Laguna, Tenerife, Spain\label{ull} \and
    Astrobiology Research Unit, Universit\'e de Li\`ege, 19C All\'ee du 6 Ao\^ut, 4000 Li\`ege, Belgium\label{liege} \and
    Department of Earth, Atmospheric and Planetary Science, Massachusetts Institute of Technology, 77 Massachusetts Avenue, Cambridge, MA 02139, USA\label{mit} \and
    Acton Sky Portal private observatory, Acton, MA, USA\label{acton} \and
    SUPA Physics and Astronomy, University of St. Andrews, Fife, KY16 9SS Scotland, UK \label{supa} \and
    South African Astronomical Observatory, P.O. Box 9, Observatory, Cape Town 7935, South Africa \label{saao} \and
    MIT Kavli Institute for Astrophysics \& Space Research, Massachusetts Institute of Technology, Cambridge, 02139, MA, USA \label{mit2} \and
    AstroAI, Center for Astrophysics | Harvard \& Smithsonian, 60 Garden Street, Cambridge, 02138, MA, USA\label{aai}
   }

   \date{Received September 15, 1996; accepted March 16, 1997}

 
  \abstract
   {
   Warm Jupiters (gas giant planets with orbital periods of $P\sim8$-$200$ d) 
   are thought to be largely unaffected by the tidal interactions that can erase information about the dynamical histories of hot Jupiters. 
   Their orbital eccentricities ($e$) and stellar obliquities are therefore expected to be preserved, 
   making them valuable probes of giant-planet formation and migration.
   }
   {Our aim is to identify 
   and characterise transiting warm Jupiters in order to expand the sample of well-characterised 
   systems and enable further studies of giant-planet formation and migration.
   }
   {We analyse TESS transit photometry 
   together with ground-based radial velocity measurements 
   and perform a joint fit of the photometric 
   and spectroscopic data from the TS23 and FIES spectrographs
   to determine the planetary and orbital parameters.}
   {We report on the discovery and characterisation of two new, transiting warm Jupiters; TOI-5120~b (\mpl~$=$~\mpfiveone~\mjup, \rpl~$=$~\radpfiveone~\rjup, $P=$~\perfiveone~d, $e=$~\eccfiveone) and TOI-5699~b (\mpl~$=$~\mpfivesix~\mjup, \rpl~$=$~\radpfivesix~\rjup, $P=$~\perfivesix~d, $e=$~\eccfivesix).
   We also refine the parameters of the known warm Jupiter TOI-2158~b (\mpl~$=$~\mptwoone~\mjup, \rpl~$=$~\radptwoone~\rjup, $P=$~\pertwoone~d, $e=$~\ecctwoone). In addition, 
   we detect an outer planetary companion, TOI-2158~c ($M_{\rm p}\sin i =$~\mpctwoone~\mjup, $P=$~\perctwoone~d, $e=$~\eccctwoone), 
   and identify a long-term radial velocity trend 
   that may indicate another planetary companion on a wide orbit.
   }
   {The diverse architectures of these systems, 
   in particular the presence and absence of outer companions,
   highlight the range of dynamical pathways leading to warm Jupiters.    
   }


   \maketitle
   \nolinenumbers
%

\section{Introduction}

Hot Jupiters are giant planets on orbits with small semi-major axis ($a\lesssim0.1$~AU),
whose existence and properties have posed significant challenges 
to theories on planet formation theory, 
making them key laboratories for studying planetary migration.
Found on slightly wider orbits,
but still interior to that of the Earth's,
the warm Jupiters reside.
As for hot Jupiters their origin remains debated, 
and several formation and migration pathways have been proposed, 
including in-situ formation, 
migration through the protoplanetary disc, and high-eccentricity tidal migration
\citep[see e.g.][for a review]{Dawson2018}. 
Each of these mechanisms is expected to leave distinct dynamical signatures 
in the resulting planetary systems. 
In particular, 
the orbital eccentricity ($e$) 
and stellar obliquity ($\psi$) can provide 
clues to the dynamical history of a system, as high-eccentricity (high-$e$) migration 
can excite both quantities to large values 
through processes such as planet–planet scattering \citep[e.g.][]{Rasio1996}
or secular interactions driven by outer companions \citep[e.g.][]{Wu2003},
whereas disc migration, for instance, 
is generally believed to keep eccentricities low \citep[e.g.][]{Duffell2015}.

In contrast to hot Jupiters, 
whose close-in orbits lead to strong tidal interactions with their host stars, 
warm Jupiters are expected to experience significantly weaker tidal dissipation. 
As a result, 
their orbital eccentricities 
and stellar obliquities may remain largely preserved over long timescales \citep[e.g.][]{Jackson2008,Albrecht2012}. 
Warm Jupiter systems may therefore provide a more direct 
record of the processes that shaped their orbital architectures 
and offer a valuable window into the formation and migration of giant planets.
Because tidal realignment is expected to be much less efficient for warm Jupiters than for hot Jupiters, 
stellar obliquities may also retain a record of their migration history. 
Recent obliquity measurements of warm Jupiters have so far revealed a tendency toward spin-orbit alignment, providing important constraints on their origin
\citep[e.g.][]{Rice2022,Wang2024}.
Another intriguing possibility is that some warm Jupiters represent planets caught in transition, 
occupying an intermediate stage between cold and hot Jupiters during high-$e$ tidal migration \citep[e.g.][]{Fabrycky2007,Socrates2012}. 
An often cited example is HD~80606~b, 
whose highly eccentric orbit suggests that it may be undergoing this process \citep{Naef2001,Wu2003}.

Observational studies of the giant-planet population have revealed several features that may be linked to these formation pathways. 
The eccentricity distribution of warm Jupiters appears to contain both low- and high-$e$ components \citep[e.g.][]{Petrovich2016}, 
suggesting that multiple migration channels may contribute to the observed population. 
In addition, 
the occurrence of warm Jupiters has been connected 
to the presence of outer companions and to host-star metallicity \citep[e.g.][]{Dong2014}, 
both of which may influence the dynamical evolution of planetary systems. 
These trends highlight the importance of expanding 
the sample of well-characterised warm Jupiter systems.

The Transiting Exoplanet Survey Satellite \citep[TESS;][]{Ricker2015}
has significantly increased 
the number of known transiting giant planets \citep[e.g.][]{Dong2021b} 
and provides an opportunity 
to discover and characterise new warm Jupiter systems. 
When combined with ground-based radial velocity (RV) follow-up, 
TESS discoveries allow precise measurements of planetary 
masses, radii, and orbital parameters, 
enabling detailed studies of their dynamical architectures.

In this work we
combine TESS photometry with
RVs from the 
Tull Coud\'{e} Spectrograph \citep[TS23:][]{Tull1995} and the FIbre-fed Echelle Spectrograph \citep[FIES;][]{Frandsen1999,Telting2014}
to report on the discovery and characterisation 
of two new transiting warm Jupiters, TOI-5120~b and TOI-5699~b.
We furthermore refine the parameters of the previously discovered warm Jupiter TOI-2158~b \citep{Knudstrup2022}, 
and
we also report the detection of an outer planetary companion in the TOI-2158 system and identify a long-term radial-velocity trend,
potentially caused by the presence of an additional substellar companion on a wide orbit. 
These systems add to the growing sample of well-characterised warm Jupiters 
and provide further insight into the diversity of architectures among giant-planet systems.

The paper is structured as follows. 
In Section~\ref{sec:obs} 
we present the various space-based and ground-based observations we have obtained.
Stellar parameters are derived from our
spectroscopic observations in Section~\ref{sec:stelpars},
and in Section~\ref{sec:fit}
we present our joint analysis of the photometric and spectroscopic data
along with the resulting orbital and planetary parameters.
In Section~\ref{sec:disc} we discuss our results before drawing our conclusions in
Section~\ref{sec:conc}.


\section{Observations}\label{sec:obs}

\subsection{Photometry}

The two new planet candidates, TOI-5120 and TOI-5699, were announced as TESS Objects of Interest (TOIs) by the TESS Science Office at MIT \citep{Guerrero2021}. This was done after transits were identified by the Massachusetts Institute of Technology (MIT) Quick Look Pipeline \citep[QLP;][]{Huang2020} through a box-least-squares \citep[BLS;][]{Kovacs2002,Hartman2016} search of the 10-min cadence Full Frame Images (FFIs). 

Subsequently the targets were placed on the 2-min cadence target list. The 2-min cadence data are processed by the Science Processing Operation Center \citep[SPOC;][]{Jenkins2016} team at the NASA Ames Research Center, where light curves are extracted through simple aperture photometry \citep[SAP;][]{Twicken2010,Morris2020} and processed using the Presearch Data Conditioning \citep[PDC;][]{Smith2012,Stumpe2012,Stumpe2014} algorithm.

For all three systems we downloaded and extracted the TESS observations using \texttt{lightkurve} \citep{lightkurve}. To detrend the light curves we iteratively performed fits to obtain good transit parameters to temporarily remove the transits from the photometric time series. After removing the transits, we detrended the light curve using Gaussian Process (GP) regression using a Mat\'ern-3/2 kernel from the \texttt{celerite} library \citep{celerite}. We then reinjected the transits in the time series and selected points in intervals 8 hours before and after the mid-transit times.

Ground-based time-series follow-up photometry was acquired for all three targets as part of the TESS Follow-up Observing Program \citep[TFOP;][]{Collins2019} with the aim of: (i) ruling out or identifying nearby eclipsing binaries; (ii) identifying transit-like events for the target to verify their depth, thereby determining the TESS photometric deblending factor; (iii) refining the ephemeris; (iv) imposing constraints on variations in transit depth across different optical filter bands. Transit observations were scheduled using the \texttt{TESS Transit Finder}, a customized version of the \texttt{Tapir} software package \citep{Jensen2013}.
For the ground-based observations, image calibration and photometric data extraction were performed using the \texttt{AstroImageJ} software package \citep{Collins2017}.

We observed TOI-5120 with the CHaracterising ExOPlanet Satellite \citep[CHEOPS;][]{Benz2021} on two separate dates as part of programme AO3-34 with the aim of refining the TESS ephemerides for targets that are potentially well-suited for atmospheric follow-up with the Atmospheric Remote-sensing Infrared Exoplanet Large-survey \citep[Ariel;][]{Tinetti2018} mission. The CHEOPS light curves were detrended using the \texttt{PYCHEOPS} package \citep{Maxted2022}. 

All the photometric observations for TOI-5120, TOI-5699, and TOI-2158 are shown in Figures~\ref{fig:time_5120}, \ref{fig:time_5699}, and \ref{fig:time_2158}, respectively. The space- and ground-based photometric observations are summarised in Tables~\ref{tab:space_phot} and \ref{tab:gb_phot}, respectively.

\begin{figure}
    \centering
    \includegraphics[width=\linewidth]{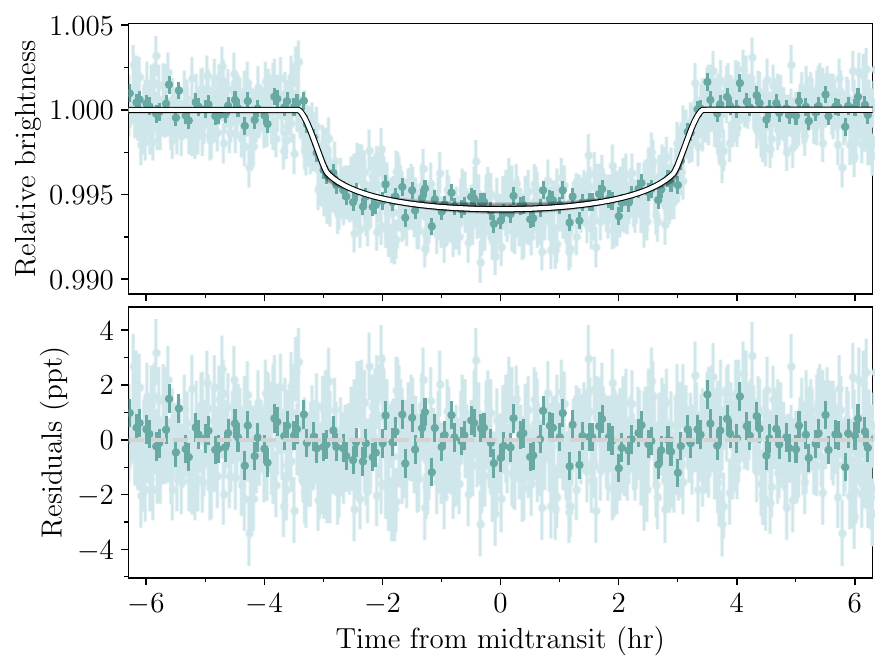}
    \caption{TESS light curve for TOI-5120~b. The dark green markers are TESS observations taken with a cadence of 10-min and the lighter points are 2-min cadence observations. Overplotted is the best-fitting transit model for TOI-5120~b.}
    \label{fig:lc_toi5120}
\end{figure}

\subsubsection{TOI-5120}

TOI-5120 was observed with a cadence of 10-min in TESS Sectors 44 and 45. 
The system was observed again in Sectors 71 and 72, this time in 2-min cadence. 
The TESS light curves centered around mid-transit for TOI-5120~b is shown in \fref{fig:lc_toi5120}.
The system was also observed in Sector 1751 
(special pointing to observe the interstellar comet 3I/ATLAS), but the seven days of observations did not include a transit.

\begin{figure}
    \centering
    \includegraphics[width=\linewidth]{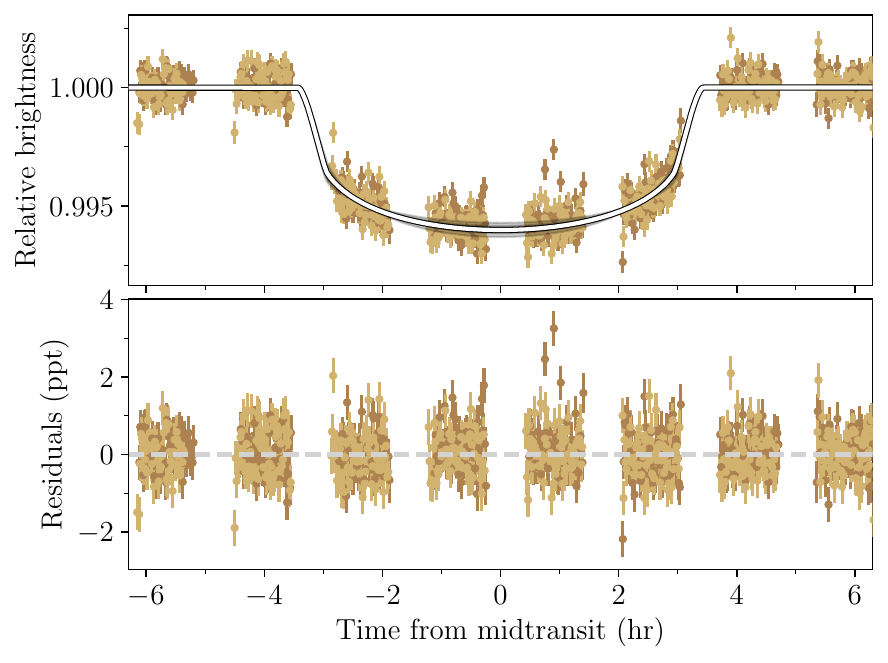}
    \caption{CHEOPS light curve for TOI-5120~b. The brown points are the CHEOPS observations from December 2022, while tan are observations from February 2024. As in \fref{fig:lc_toi5120} the best-fitting transit model is overplotted.}
    \label{fig:lc_toi5120_cheops}
\end{figure}

Two transits of TOI-5120~b were observed with CHEOPS; observations were carried out from (UTC) 02:49 to 20:07 on 2022-12-30 and from 17:50 on 2024-02-01 to 13:09 on 2024-02-02. 
The observations were taken with a cadence of 1~min, but with the characteristic gaps from the sun-synchronous orbit of the spacecraft. 
The CHEOPS observations are shown in \fref{fig:lc_toi5120_cheops}.

Ground-based photometric transit observations were made on four separate nights
with four different filters ($i^\prime$, $r^\prime$, $B$, $z$-short).
No full transits were observed, 
but the observations covered either ingress or egress 
plus a sizable segment ($>50\%$) of the transit.
The ground-based observations are shown in \fref{fig:lc_5120_gb}.

\subsubsection{TOI-5699}

\begin{figure}
    \centering
    \includegraphics[width=\linewidth]{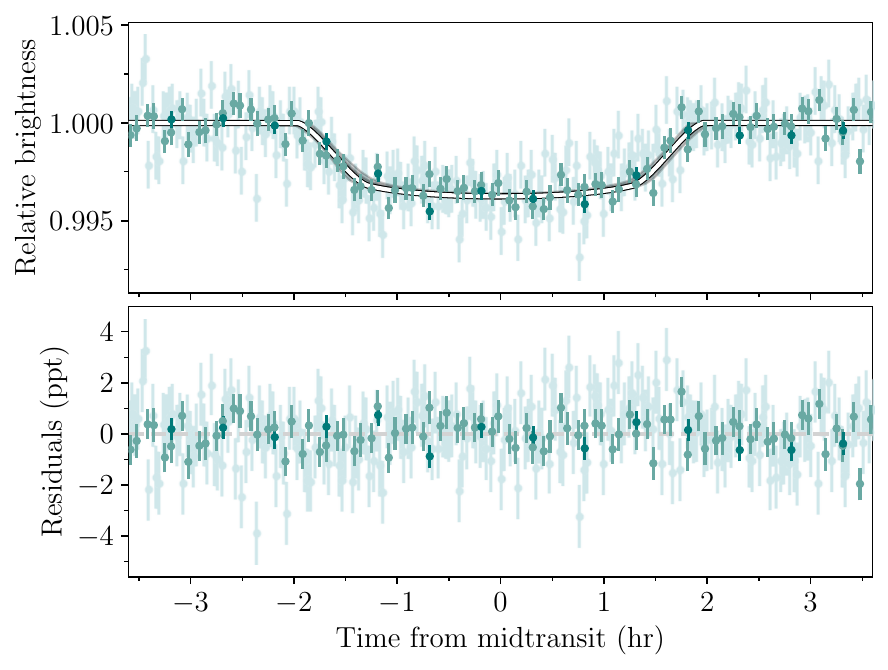}
    \caption{TESS light curve for TOI-5699~b. 
    Going from the darkest to the lightest colors the markers show the 30-min, 10-min, and 2-min TESS observations. 
    The best-fitting transit model for TOI-5699~b is overplotted.}
    \label{fig:lc_toi5699}
\end{figure}

TOI-5699 was observed by TESS in Sector 25 with 30-min cadence and in Sectors 51 and 52 at a 10-min cadence. 
In Sectors 78 and 79, TOI-5699 was observed with a cadence of 2~min, 
but given the long orbital period ($P\sim25$~d) no transit occurred during the Sector 78 observations.
The phasefolded TESS light curves of TOI-5699~b are shown in \fref{fig:lc_toi5699}. 

A full transit of TOI-5699~b was observed from ground in the $z$-short filter.
This transit is shown in \fref{fig:lc_5699_gb}.

\subsubsection{TOI-2158}

Two sectors from TESS were included in \citet{Knudstrup2022}, namely Sector 26 (30-minute cadence) and Sector 40 (2-minute cadence). TOI-2158 has since been observed by TESS in Sector 53, also with a cadence of 2~min.

\begin{figure}
    \centering
    \includegraphics[width=\linewidth]{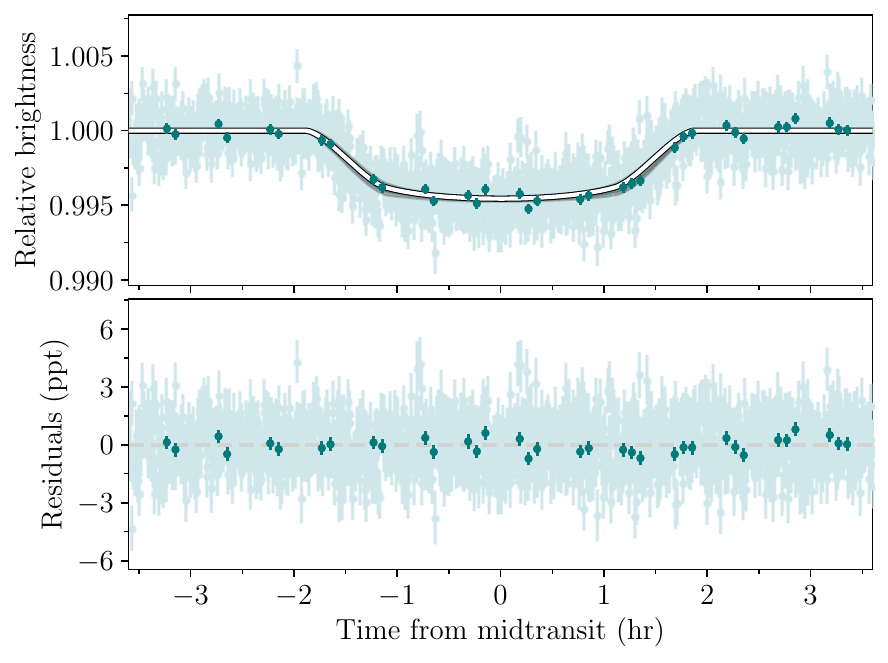}
    \caption{TESS light curve for TOI-2158~b. The darker green markers are TESS observations taken with a cadence of 30-min and the lighter points are 2-min cadence observations. Overplotted is the best-fitting transit model for TOI-2158~b.}
    \label{fig:lc_toi2158}
\end{figure}

In addition to the TESS photometry, 
we included the ground-based photometry from the Las Cumbres Observatory Global Telescope \citep[LOCGT;][]{Brown2013} presented in \citet{Knudstrup2022},
where three complete transits were observed (separately) in the $i^{\prime}$, $B$, and $z$-short filters. 
The TESS and ground-based observations centered around the mid-transit time are displayed in \fref{fig:lc_toi2158} and \fref{fig:lc_2158_gb}, respectively.

\subsection{Radial Velocity Follow-up}

Our RV monitoring was carried out using the TS23 spectrograph 
mounted at the 2.7~m Harlan J. Smith telescope at the McDonald Observatory, Texas, USA, and the FIES spectrograph
at the 2.6~m Nordic Optical Telescope \citep[NOT;][]{Djupvik2010} of the Roque de los Muchachos observatory, La Palma, Spain.

For the TS23 observations an $I_2$ gas cell was placed in the light beam to allow for the extraction of high precision RVs. 
All spectra were reduced and extracted through standard IRAF tasks, 
and RVs were extracted by making use of the Austral code \citep{Endl2000}.
TS23 Observations of each target were divided into three separate successive exposures 
to enable removal of energetic particle hits on the 
detector using the IRAF \texttt{crcombine} routine.

FIES spectra were reduced following the procedure outlined in \citet{Gandolfi2015}. 
This includes bias subtraction, flat fielding, order tracing, and extraction as well as wavelength calibration. 
As with the TS23 exposures, science exposures were divided into three sub-exposures (i.e. an exposure of 2700~s consisted of three exposures of 900~s) 
to mitigate the effects of energetic particles, 
and we performed a sigma clipping algorithm when combining the frames.
To trace the instrumental RV drift, we acquired long-exposed thorium-argon (ThAr) spectra immediately before and after each block of science exposures. 
Finally, the RVs were extracted from multi-order cross-correlations, 
where the first stellar spectrum was used as a template.
All RVs are tabulated in
\tref{tab:rvs_all}.

We monitored TOI-5120 with both TS23 and FIES starting 2022-02-10, and continuing until 2025-12-17, meaning a baseline of $1406$~d. We collected a total of 56 RVs using TS23 and 20 RVs using FIES, which are shown in \fref{fig:time_5120}.
We monitored TOI-5699 with TS23 between February 2023 and March 2026, collecting 86 RVs. 
Three epochs were discarded due to dome vignetting.
We furthermore collected four RVs 
using FIES starting observations in May and ending in June 2025. 
The baseline is thus $1128$~d.
The RV time series is shown in \fref{fig:time_5699}.

As mentioned, our RV follow-up for TOI-2158 
presented here 
is a continuation of the efforts from
\citet{Knudstrup2022}.
In addition to the 8~d signal from the orbital period of 
TOI-2158~b
a long-term quadratic trend was seen in the RVs. 
We therefore continued to monitor the system with both TS23 and FIES. 
The RV monitoring in \citet{Knudstrup2022} was carried out from
2021-03-03, 
to 2022-06-03, with 33 RVs from TS23 and 38 from FIES.
Here we have extended the monitoring until 2026-01-26, 
resulting in a total of 112 TS23 RVs and 53 FIES RVs 
with a baseline of $1764$~d.
As mentioned in \citet{Knudstrup2022},
the FIES spectrograph was refurbished on 2021-07-01,
which has introduced an RV offset between observations 
taken before
and after this date.
We have therefore treated our FIES observations 
as coming from two separate spectrographs;
FIES (before) and FIES+ (after).
All RVs for TOI-2158 are displayed in \fref{fig:time_2158}. 

\begin{figure*}
    \centering
    \includegraphics[width=\linewidth]{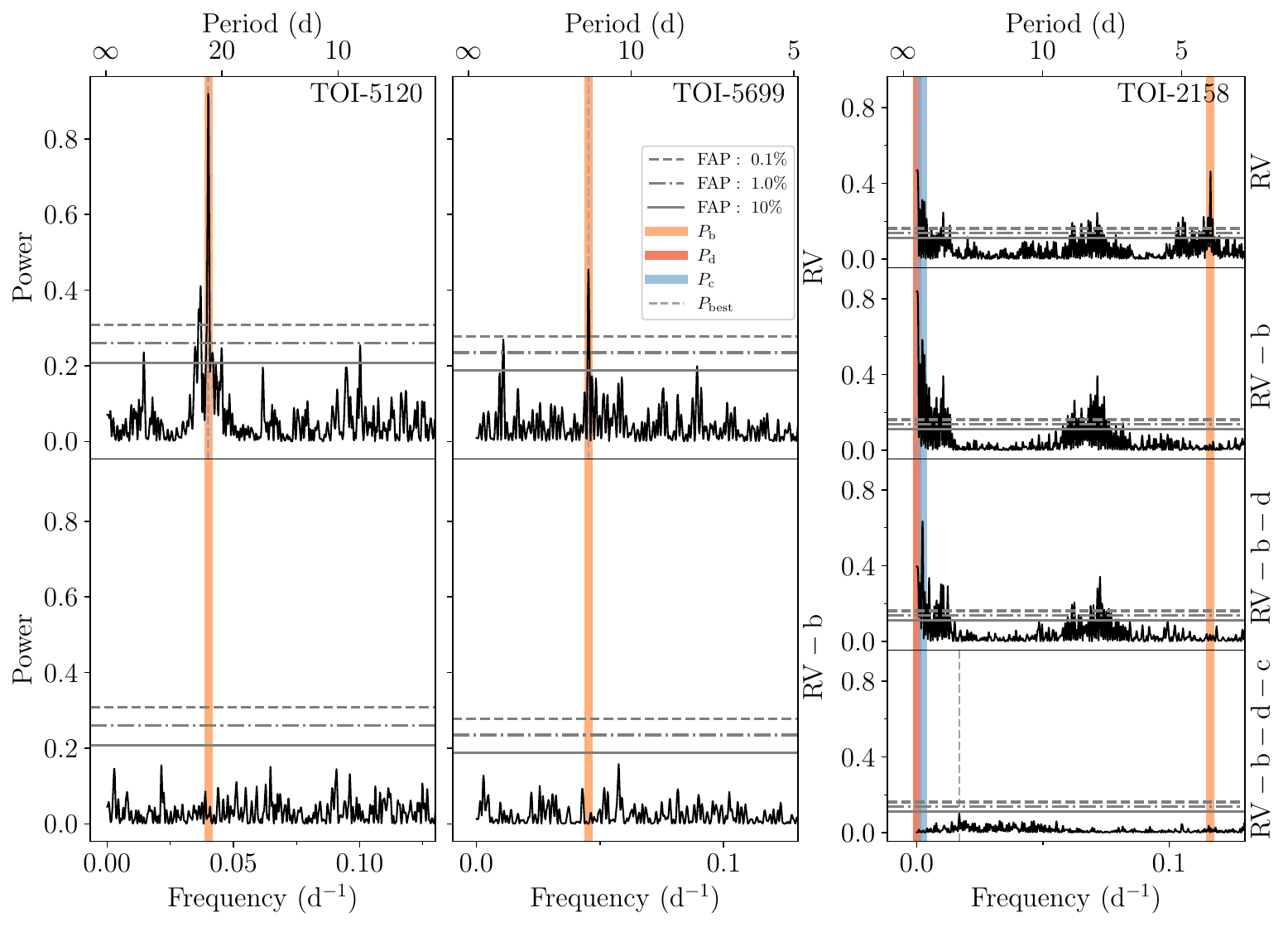}
    \caption{Lomb-Scargle periodogram created from our RVs. 
    The first two columns display the periodograms for the RVs obtained for TOI-5120 (left) and TOI-5699 (middle) with the raw RVs shown on top and after subtracting the best-fitting Keplerian orbit on the bottom (see Section~\ref{sec:fit}). 
    In the rightmost column we show the RVs for TOI-2158 again with the raw RVs on top and then subtracting the best-fitting Keplerian from planet b, then from planet (or signal) d in the panel below, and lastly planet c in the panel at the bottom. 
    The period of planet b is denoted with an orange line and planet c and d (TOI-2158) in blue and red, respectively. 
    The frequency of the highest peak in the periodogram is shown as the dashed line.
    We also give the false alarm probabilities (FAPs) at 0.1\%, 1.0\%, and 10\%.
    }
    \label{fig:lomb}
\end{figure*}

In \fref{fig:lomb} we show the Lomb-Scargle periodograms \citep{VanderPlas2012,VanderPlas2015} created from the RVs we have collected for all three systems. 
We also highlight the period of the transiting planet in orange 
and as indicated by the false alarm probabilities, 
the transiting planets are clearly detected in all three systems.
For TOI-2158 we also see additional peaks; 
one at around 5400~d and one at 440~d. 
The former of the two is not resolved 
as our baseline for TOI-2158 is only 1625~d
and is therefore only seen as a long-term trend,
potentially coming from a rather massive companion on a wide orbit.
For the latter we have covered roughly four cycles
and we interpret this as an additional planet in the system.

\subsection{High Resolution Imaging}
As part of the standard process for validating transiting exoplanets 
to assess the possible contamination of bound or unbound companions on the derived planetary radii \citep{Ciardi2015}, 
we observed TOI~5120 with optical speckle observations and TOI-5699 with optical speckle and near-infrared adaptive optics imaging. 
A faint ($\Delta I \approx 5.5$~mag) and potentially bound companion was detected in the TOI-5120 system. Below we summarise the high resolution imaging observations.

\subsubsection{Optical Speckle Imaging}

We searched for stellar companions to TOI-5120 with speckle imaging on the 4.1-m Southern Astrophysical Research (SOAR) telescope \citep{Tokovinin2018} on 2022-04-15, 
observing in Cousins I-band, a similar visible bandpass as TESS. 
This observation was sensitive to a 5.2-magnitude fainter star at an angular distance of 1\arcsec\, from the target. 
More details of the observations within the SOAR TESS survey are available in \citet{Ziegler2020}. 
The 5$\sigma$ detection sensitivity and speckle auto-correlation functions from the observations are shown in \fref{fig:soar_image}. 
No nearby stars were detected within 3\arcsec of TOI-5120 in the SOAR observations.

NESSI \citep[NN-EXPLORE Exoplanet and Stellar Speckle Imager;][]{Scott2018} is a speckle imager installed at the WIYN 3.5~m telescope on Kitt Peak. 
NESSI normally collects simultaneous speckle images in two filters, 
but for the night of 2022-04-18 when TOI-5120 was observed, only one beam was operational. 
NESSI data are therefore only available in one filter at $\lambda_c=832$~nm. 
The observation consisted of a set of 9000 speckle frames of 40~ms each. 
The field-of-view was confined using a readout region of $256\times256$ pixels or $4.6\times4.6$\arcsec. 
Additional observations of a single star near TOI-5120 were obtained for calibration of the underlying PSF.

To reduce the data, we used the pipeline described by \citet{Howell2011}.
Speckle data are best used to constrain the presence of additional sources out to a radius of $\sim1.2$\arcsec\, from the target star, 
but sources may be measured at greater radii with slightly reduced photometric accuracy. 
Pipeline data products include a reconstructed image of the field around each target and, measured from that, 
a contrast curve representing the relative magnitude limit as a function of angular separation from the star. 
There was a single, fainter companion star detected close to TOI-5120. 
The companion lies at an angular separation of $1.298$\arcsec\, 
and a position angle (PA), as measured from North through East, of 152.74$^\circ$
(corresponding to a physical separation of $\sim300$~AU). 
The magnitude contrast with TOI-5120 is 5.51 magnitudes in the 832~nm filter.

TOI-5120 and TOI-5699 were observed on 2022-12-24 and 2023-01-01, respectively, 
with the speckle polarimeter on the 2.5-m telescope at the Caucasian Observatory of Sternberg Astronomical Institute (SAI) of Lomonosov Moscow State University. 
A low-noise CMOS detector was used for the observations \citep{Strakhov2023}. 
The atmospheric dispersion compensator was active, which allowed using the $I_\mathrm{c}$ band. The respective angular resolution is 0.083\arcsec. 
The resulting sensitivity curves and autocorrelation functions (ACFs) are shown in 
\fref{fig:sai_toi5120} and \fref{fig:sai_toi5699} for TOI-5120 and TOI-5699, respectively.

For TOI-5120 we once again detected a stellar companion at a separation of $1.26\pm0.01$\arcsec, position angle $153.3\pm0.4^{\circ}$, and the magnitude difference is $\Delta I_\mathrm{c}=5.7\pm0.1$. 
We reobserved TOI-5120 on 2024-10-13. The companion was confirmed at a separation of $1.29\pm0.01$\arcsec, PA of $152.6\pm0.3^{\circ}$, and  $\Delta I_\mathrm{c}=6.0\pm0.1$.
The position of the component did not change significantly 
while from proper motion a change of 118.7~mas is expected on a time base of 1.8~yr for a background star. 
The component is likely to be physically bound. 
We note that the component is missing in the \emph{Gaia} DR3 catalogue despite the relatively large separation, 
although given the large brightness contrast the companion could go undetected at this separation \citep{ElBadry2021}.
Furthermore, the \emph{Gaia} renormalised unit weight error (RUWE) is 1.034,
indicating a clean single-source astrometric solution 
with no evidence that the companion measurably perturbed the astrometry.
No other companions were detected. 
The detection limits at distances $0.25$ and $1.0$\arcsec\, from the star are $\Delta I_\mathrm{c}=3.8^m$ and $6.5^m$ for TOI-5120 and $\Delta I_\mathrm{c}=3.8^m$ and $5.6^m$ for TOI-5699.

TOI-5699 was observed on 2024-05-22 at Gemini North 8-m telescope using the ‘Alopeke speckle imager.
The star has no close companions to within the angular (0.02 to 1.2\arcsec) and 5$\sigma$ magnitude contrast levels (5-8.5 mags) determined in each of two filters. 
The 5$\sigma$ contrast curves and the 832~nm reconstructed image are shown in \fref{fig:gemini_imaging}. 
No companion was detected.

For good measure, we further note that speckle polarimetry for TOI-2158 has also been obtained. 
No companions were detected in this system from these observations. 
Details and plots can be found in \citet{Knudstrup2022}.

\subsubsection{Near-Infrared AO Imaging}
    
Observations of TOI~5699 were made on 2023-06-07 with the PHARO instrument \citep{Hayward2001} 
on the Palomar Hale (5m) telescope behind the P3K natural guide star AO system \citep{Dekany2013} 
in the narrowband Br-gamma filter $(\lambda_o = 2.29; \Delta\lambda = 0.035~\mu$m). 
The PHARO pixel scale is $0.025\arcsec$ per pixel. 
A standard 5-point quincunx dither pattern with steps of 5\arcsec\, was performed three times with each repeat separated by 0.5\arcsec. 
The reduced science frames were combined into a single mosaiced image with a final resolution of 0.1\arcsec. 
The sensitivity of the final combined adaptive optics (AO) image were determined by injecting simulated sources azimuthally around the primary target every $20^\circ $ at separations of integer multiples of the central source's full width at half maximum \citep[FWHM;][]{Furlan2017}. 
The brightness of each injected source was scaled until standard aperture photometry detected it with 5$\sigma $ significance. 
The final 5$\sigma$ limit at each separation was determined from the average of all of the determined limits at that separation, 
and the uncertainty on the limit was set by the root-mean-square (rms) dispersion of the azimuthal slices at a given radial distance. 
The Palomar imaging and sensitivities are shown in \fref{fig:palomar_aoimaging}. 
The infrared imaging detected no stellar companions, 
in agreement with the speckle observations.

\begin{table}[]
    \centering
    \caption{Speckle observations of TOI-5120.}
    \begin{threeparttable}
        \begin{tabular}{c c c c}
        \toprule
            Date & Separation & PA & $\Delta m$  \\
            (UTC) & (\arcsec) & ($^\circ$) & (832~nm/$I_{\rm c}$) \\
        \midrule
           2022-04-18 & 1.298 & 152.74 & 5.51 \\
           2022-12-24 & $1.26\pm0.01$ & $153.3 \pm 0.4$ & $5.7\pm0.1$ \\
           2024-10-13 & $1.29\pm0.01$ & $152.6\pm0.3$ & $6.0\pm0.1$ \\
            
        \bottomrule
        \end{tabular}
        \begin{tablenotes}
            \item Separation, position angle (measured from North to East), and contrast of the companion in TOI-5120, which we note is not detected in \emph{Gaia}~DR3. The first observations is from NESSI and the two others are from SAI.
        \end{tablenotes}
    \end{threeparttable}
    \label{tab:speckle}
\end{table}



\section{Stellar Parameters}\label{sec:stelpars}

To derive stellar parameters for the three host stars, 
we used our high-resolution spectroscopic observations
for each target to obtain stellar parameters.
This was done by co-adding our FIES observations
to create a high signal-to-noise spectrum.

\begin{table*}
	\centering
	\caption{Stellar parameters.}
	\label{tab:stars}
	\begin{threeparttable}
		\begin{tabular}{c c c c c}
		\toprule
		
			& TESS Object of Interest & TOI-5120 & TOI-5699 & TOI-2158 \\
			& TESS Input Catalogue & TIC 58723861 & TIC 224328450 & TIC 342642208 \\
			& TYCHO-2 & TYC 1894-00102-1 & TYC 2582-02794-1 & TYC 1577-691-1 \\
		\midrule
			
		$ V $\tnote{(a)} & Tycho $V$ magnitude & \fiveoneVmag & \fivesixVmag & \twooneVmag \\
		$ B $\tnote{(a)} & Tycho $B$ magnitude & \fiveoneBmag & \fivesixBmag & \twooneBmag \\
		$ T $\tnote{(b)} & TESS magnitude & \fiveoneTmag & \fivesixTmag & \twooneTmag \\
		$ G $\tnote{(c)} & \emph{Gaia} $G$ magnitude & \fiveoneGAIAmag & \fivesixGAIAmag & \twooneGAIAmag \\
        $ G_{\rm BP} $\tnote{(c)} & \emph{Gaia} $G_{\rm BP}$ magnitude & \fiveonegaiabp & \fivesixgaiabp & \twoonegaiabp \\
        $ G_{\rm RP} $\tnote{(c)} & \emph{Gaia} $G_{\rm RP}$ magnitude & \fiveonegaiarp & \fivesixgaiarp & \twoonegaiarp \\

		\hdashline

		$ \alpha_{\rm J2000}\tnote{(c)} $ & Right ascension & 06:56:51.371 & 16:44:54.288 & 18:27:14.461 \\
		$ \delta_{\rm J2000}\tnote{(c)} $ & Declination & +22:35:44.627 & +31:12:39.317 & +20:31:36.671 \\
		$ \mu_\alpha $\tnote{(c)} & Proper motion in RA (mas~yr$^{-1}$) & \fiveonepmRA & \fivesixpmRA & \twoonepmRA \\
		$ \mu_\delta $\tnote{(c)} & Proper motion in Dec (mas~yr$^{-1}$) & \fiveonepmDEC & \fivesixpmDEC & \twoonepmDEC \\
		$ \varpi $\tnote{(c)} & Parallax (mas) & \fiveoneplx & \fivesixplx & \twooneplx \\
		$ \pi $\tnote{(c)} & Distance (pc) & \fiveoned & \fivesixd & \twooned \\
		$\gamma$\tnote{(c)} & Systemic velocity (km~s$^{-1}$) & \fiveoneRV & \fivesixRV & \twooneRV \\
		RUWE\tnote{(c)} & Renormalised unit weight error & \fiveoneRUWE & \fivesixRUWE & \twooneRUWE \\
		
		\hdashline

		$T_{\rm eff}$\tnote{(d)} & Effective temperature (K) & \fiveoneTeff & \fivesixTeff & \twooneTeff \\
		$\log g_\star$\tnote{(d)} & Surface gravity (cgs) & \fiveonelogg & \fivesixlogg & \twoonelogg \\
		$\rm [Fe/H]$\tnote{(d)} & Metallicity (dex) & \fiveoneFeH & \fivesixFeH & \twooneFeH \\
		$v \sin i_\star$\tnote{(d)} & Projected rotational velocity (km~s$^{-1}$) & \fiveonevsini & \fivesixvsini & \twoonevsini \\
		$\xi_\star$\tnote{(e)} & Micro-turbulent velocity (km~s$^{-1}$) & \fiveonevmic & \fivesixvmic & \twoonevmic \\
		$\zeta_\star$\tnote{(f)} & Macro-turbulent velocity (km~s$^{-1}$) & \fiveonevmac & \fivesixvmac & \twoonevmac \\
		
		\hdashline

		$R_\star$\tnote{(g)} & Radius (R$_\odot$) & \fiveoneradbasta & \fivesixradbasta & \twooneradbasta \\
		$M_\star$\tnote{(g)} & Mass (M$_\odot$) & \fiveonemassbasta & \fivesixmassbasta & \twoonemassbasta \\
		$\rho_\star$\tnote{(g)} & Density (g~cm$^{-3}$) & \fiveonerhobasta & \fivesixrhobasta & \twoonerhobasta \\
		$\tau$\tnote{(g)} & Age (Gyr) & \fiveoneagebasta & \fivesixagebasta & \twooneagebasta \\

		\bottomrule
		\end{tabular}
		\begin{tablenotes}
			\item[(a)] Tycho-2 \citep{Hog2000}.
			\item[(b)] TICv8.2 \citep{Stassun2019}.
			\item[(c)] \emph{Gaia} DR3.
			\item[(d)] This work using SME, except for TOI-2158 where the parameters are from \citet{Knudstrup2022}.
            \item[(e)] Empirical relation \citep{Bruntt2010}.
            \item[(f)] Empirical relation \citep{Doyle2014}.
            \item[(g)] This work using BASTA.
		\end{tablenotes}
	\end{threeparttable}

\end{table*}

\subsection{Spectral Analysis}\label{sec:spectral}

In the spectral analysis we used the empirical code
\texttt{SpecMatc-Emp} \citep{Yee2017}
to obtain initial values for the subsequent spectral modelling.
The primary quantities of interest in our spectral analysis 
are the effective temperature ($T_{\rm eff}$), 
the surface gravity ($\log g$), 
and the metallicity ($[\rm Fe/H]$).

Following the approach in \citet{Persson2018},
we made use of Spectroscopy Made Easy \citep[SME;][]{Valenti1996,Piskunov2017}
for the more detailed analysis.
SME fits observations to synthetic computed spectra, 
where we opted to use the MARCS grid of stellar atmospheres \citep{Gustafsson2008}
with retrieved atomic and molecular line data from the Vienna Atomic Line Database \citep[VALD;][]{Ryabchikova2015}.
In our analysis we varied one parameter at a time. 
We started out by estimating $T_{\rm eff}$ from the line wings of the H$_\alpha$ line
after which we determined $[\rm Fe/H]$, $[\rm Ca/H]$, and $[\rm Na/H]$ 
from clean and unblended lines between 6000 and 6500~\AA.
We then used the line wings of the Ca lines at 6102~\AA, 6122~\AA, 6162~\AA,  
as well as the Ca line at 6439~\AA \, to determine $\log g$.
We then iterated this process until the values had converged.
Micro- ($\xi_\star$) and macroturbulence ($\zeta_\star$)
were estimated at each step from the resulting stellar parameters 
using the the empirical relations from \citet{Bruntt2010} and \citet{Doyle2014}, respectively.
The resulting parameters are given in \tref{tab:stars}.

In \tref{tab:spec_app} we compare our spectroscopic values for all three systems with those from \emph{Gaia} Data Release~3 \citep{GaiaCollaboration2023} and the Exoplanet Follow-up Observation Program (ExoFOP) website. 
We generally find good agreement between these parameters. 
We furthermore list the calcium and sodium abundances there.

\subsection{Stellar Properties}\label{sec:phys}

For determination of the stellar properties, 
we used the BAyesian STellar Algorithm \citep[BASTA;][]{BASTA2015,BASTA2022}.
This is a Bayesian framework which computes the posterior distributions of each stellar property such as stellar mass, radius, and age given a set of observations and potentially a set of priors. 
For observational constraints, we choose to use our spectroscopic values ($T_{\rm eff},\log g,[\mathrm{Fe/H}]$
as derived with SME for TOI-5120 and TOI-5699, and as reported in \citealt{Knudstrup2022} for TOI-2158, see below),
along with the \emph{Gaia} apparent magnitudes and parallax (see \tref{tab:stars}). 
We corrected our parallaxes for the known parallax zero-point correction following \citet{Lindegren21}. 
When fitting with BASTA we used 
a custom grid computed with the Garching Stellar Evolution Code \citep[GARSTEC;][]{Weiss2008},
using the same input physics as described in \citet{Winther2023} and \citet{Larsen2025},
with the grid parameter ranges listed in \tref{tab:barbiegrid}.
For all three stars we got consistent results
when using BaSTI isochrones \citep[][opting for the models including convective core overshooting, microscopic diffusion, and mass loss]{BaSTI2018} instead of the custom grid.
As an additional check, we also made use of the ARIADNE\footnote{\url{https://github.com/jvines/astroARIADNE/tree/master}} \citep{Vines2022} package 
to perform an analysis of the broadband 
spectral energy distribution (SED) of the star in order to
derive an empirical measurement of the stellar radius. The resulting parameters are listed in \tref{tab:stars_app} in excellent agreement with those found using BASTA.

The initial modelling results of TOI-2158 (using the values from SME in \tref{tab:spec_app})
showed a stellar age consistent with $13.7$~Gyr. 
While this is physically permitted by the physics governing stellar evolution, 
it is difficult to reconcile with the star’s chemical composition in the broader context of Galactic evolution.
The chemical composition of the star is notably complex, exhibiting supersolar abundances of iron, calcium, and sodium.
Since the composition of a star broadly reflects that of the natal gas cloud from which it formed, Galactic chemical evolution models predict that stars born during the earliest epochs of the Milky Way should generally be metal-poor and deficient in these elements. 
This is because the nucleosynthetic channels responsible for enriching the interstellar medium in Fe, Ca, and Na require multiple generations of stellar evolution and supernova feedback before supersolar abundances can be reached.
Consequently, within the current understanding of the Milky Way's chemical enrichment history \citep{matteucci2021, spitoni2023}, it is highly implausible that a star with such enhanced abundances could have formed as early as $13.7$~Gyr ago.

Spectroscopic analysis of metal-rich stars is notoriously challenging 
due to the many degeneracies inherent in spectral fitting \citep[e.g.][]{Blanco2014,Slumstrup2019}.
Metal-rich spectra contain a large number of absorption lines, 
making it considerably more difficult to robustly determine the chemical composition compared to spectra of somewhat lower metallicity 
\citep[e.g.][]{Gray2005,Jofre2019}.
This ambiguity has direct consequences for the age inference: 
a higher metallicity implies more bound-bound and bound-free transitions within the stellar interior,
increasing the opacity and suppressing the outward energy flux.
A high-metallicity star is therefore intrinsically fainter than a lower-metallicity counterpart of the same mass and age, 
and to reproduce the observed luminosity, the best-fit stellar model must invoke a more evolved and consequently older star.

As a consistency check, we computed the Galactic orbit of all three stars using the astrometric information and line-of-sight velocity from \emph{Gaia} DR3 \citep{GaiaCollaboration2023}. 
We computed the Galactic orbital properties using \texttt{galpy} \citep{galpy}, adopting the axisymmetric gravitational potential from \citet{mcmillan2017} as a model for the Milky Way.
For the solar position and velocity, we assumed $(X_{\odot}, Y_{\odot}, Z_{\odot}) = (8.2, 0, 0.0208)$~\si{\kilo pc}, with a circular velocity of \SI{240}{\kilo\metre\per\second} \citep{schonrich2010,galpy,bennett2019,gravity2019}.
The orbits of all three stars are consistent with those of the Sun and other members of the Galactic disk
as shown in \fref{fig:toomre} and \fref{fig:gal_orbits}.
This kind of orbit is more typical of a younger, dynamically settled population than of a relic from the earliest epochs of the Milky Way, when the merger history of the Galaxy would have produced far more chaotic orbital configurations.

We therefore opted to use the spectroscopic parameters from \citet{Knudstrup2022}
derived from spectra from the Tillinghast Reflector Echelle Spectrograph \citep[TRES;][]{Furesz2008}
using the stellar parameter classification \citep[SPC;][]{Buchhave2012} tool.
Adopting this earlier solution as input to BASTA infers a significantly younger age off \twooneagebasta~Gyr, 
which is more consistent with our expectation from Galactic chemical enrichment models and with the derived Galacitic orbit.
The results for all three systems are given in \tref{tab:stars}.



\section{Joint Analysis}
\label{sec:fit}

The transit and orbital parameters
were estimated through a joint modelling of the transit photometry and RVs, where we sampled the posterior distributions through Monte Carlo Markov Chain (MCMC) sampling utilizing the \texttt{emcee} package \citep{Foreman2013}.

The light curves were modelled using the \texttt{batman} package \citep{Kreidberg2015} based on the formalism from \citet{Mandel2002}. For the photometric time series acquired with cadences of 10~min or more, we created evenly spaced model light curves over which we integrated to mimic a cadence of 2~min. We modelled the stellar limb darkening using a standard quadratic law. The coefficients were obtained from the tables by \citet{Claret2013} and \citet{Claret2018} using values appropriate for $T_{\rm eff}$, $\log g$, and $[\rm Fe/H]$ for each star (\tref{tab:stars}). During the MCMC sampling we were stepping in the sum of the coefficients applying a gaussian prior with a width of $0.1$, while keeping the difference fixed at the initial value. The limb darkening coefficients can be found in Table~\ref{tab:ld_all} in Appendix~\ref{sec:appB}.

The RVs were modelled as Keplerian orbits, where the systemic velocities ($\gamma$) were included for each instrument. 
Likewise, a jitter term ($\varsigma$) was added in quadrature to the RV errors stemming from a given instrument.
When fitting the eccentricity, we used the $h=\sqrt{e}\cos \omega$ and $k=\sqrt{e}\sin \omega$ parametrisation, instead of stepping in $e$ and $\omega$ (argument of periastron). 
For each system we also carried out runs assuming circular ($e=0$) orbits,
and below we discuss the preferred solution.

Our likelihood is defined as 

\begin{equation}
    \ln \mathcal{L} = -0.5 \sum_{i=1}^{N}  \left[ \frac{(O_i -C_i)^2}{\sigma_i^2}  + \ln 2 \pi \sigma_i^2 \right] + \sum_{j=1}^M \ln \mathcal{P}_j \, ,
\end{equation}
where $N$ denotes the total number of data points (photometric and RVs); 
$C_i$ the calculated model corresponding to the observed datum $O_i$
with associated uncertainty $\sigma_i$. 
$\mathcal{P}_j$ is the prior applied to the $j$th parameter, 
for which we use a uniform prior 
unless explicitly stated otherwise.

We initialised our MCMC in a 
`tight Gaussian ball' around the best-fitting
solution (obtained from a least-squares fit).
We assessed the convergence through visual
inspection of the chains and correlation plots
as well as 
by computing the ratio between the length
of the chains to the autocorrelation time.
Here we demanded that the minimum ratio was
more than 100.


\subsection{TOI-5120}

As pointed out by \citet{Lucy1971}, 
the positive-definite nature of the orbital eccentricity introduces a bias for eccentric solutions and to be 95\% confident that an eccentricity is non-zero, 
the result must be above 2.45$\sigma$. 
Allowing $e$ to vary in TOI-5120 resulted in an eccentricity of \eccfiveone\, -- a modest, but significant value. 

\begin{figure}
    \centering
    \includegraphics[width=\linewidth]{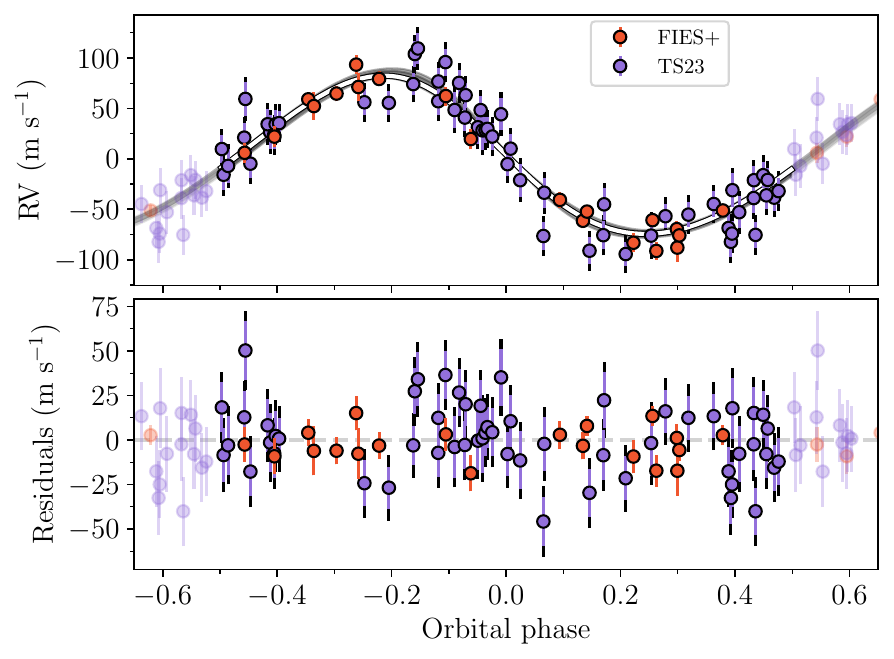}
    \caption{Radial velocity curve for TOI-5120~b. RVs from TS23 are shown with purple markers and FIES+ RVs are shown with orange. The best-fitting RV curve is overplotted. The grey shaded regions show the 1 and 2$\sigma$ confidence intervals.
    Low opacity markers are repeated data points shifted by one orbital phase.
    }
    \label{fig:rv_toi5120}
\end{figure}

To assess which model is preferred we calculated the Bayesian Information Criterion (BIC) from both of these runs. 
The difference in BIC ($\Delta$BIC) came out to be $\Delta \mathrm{BIC}=12.96$ in favour of the eccentric model,
which is our preferred model.
The best-fitting space-based light curves are shown in Figs.~\ref{fig:lc_toi5120} and \ref{fig:lc_toi5120_cheops} 
and in \fref{fig:lc_5120_gb} we show the ground-based light curves.
The best-fitting (eccentric) RV model is displayed in \fref{fig:rv_toi5120}. 
The best-fitting parameters are summarised in \tref{tab:pars}, 
and in \tref{tab:pars_circ} we list the best-fitting parameters
for the $e=0$ case.
Posterior distributions of key stepping parameters
for TOI-5120 can be found in \fref{fig:corner_toi5120}.

\subsection{TOI-5699}
\label{sec:dat5699}

As for TOI-5120 we investigated both a circular and an eccentric orbit for TOI-5699. 
When leaving $h=\sqrt{e}\cos \omega$ and $k=\sqrt{e}\sin \omega$ free to vary, 
we find an eccentricity of \eccfivesix, a little over a $3\sigma$ significance. However, calculating the BIC for both scenarios yields a $\Delta\mathrm{BIC}=3.20$ 
slightly in favour of the circular model. 
We also calculated the Akaike Information Criterion (AIC),
where $\Delta\mathrm{AIC}=-494.57$ favouring the eccentric solution,
signifying that the likelihood is greatly improved 
for an eccentric orbit, 
although the inclusion of two extra parameters are not justified
according to the $\Delta\mathrm{BIC}$.

To further assess the significance of the orbital eccentricity
and whether it is data-driven or instead a consequence of the $e\geq0$ boundary,
we performed an additional eccentric fit 
in which a prior was applied to the eccentricity.
We adopted a Gamma prior, $\Gamma (\alpha,\beta)$, 
using the values $\alpha=1.30$ and $\beta=5.88$ from
\citet[][]{Stevenson2025}.
As this distribution is designed to counteract 
the positive-definite bias of eccentricity near $e=0$,
the inferred eccentricity would be expected to collapse toward zero 
if it were purely a boundary effect. 
The resulting posterior yielded $e=0.26\pm0.09$,
indicating that the eccentricity survives a skeptical prior.
Applying the commonly used Beta prior from \citet[][$B(0.867,3.03)$]{Kipping2013} 
instead yielded $e=0.27_{-0.08}^{+0.11}$.

We performed 5,000 bootstrap realisations and computed the fraction of trials 
in which the fitted eccentricity was greater than or equal to the best-fitting value measured from the original data 
($e\geq0.32$). 
We find that this occurs in approximately 10\% of the realizations, 
indicating that an eccentricity of this magnitude can arise from noise alone with non-negligible probability under the assumption of a circular orbit. 
Thus, the bootstrap test provides only weak evidence for a non-zero eccentricity, 
and the RV data alone are insufficient to confidently reject a circular solution.

Finally, we tested what constraints on $e$ one can obtain from the photoeccentric effect
\citep{Seager2003}.
We performed an MCMC analysis of the transit photometry, 
imposing a prior on the stellar density ($\rho_\star=$~\fivesixrhobasta~g~cm$^{-3}$)
derived (independently) from our stellar characterization. 
The radial velocity data were excluded from this fit. 
This yielded an eccentricity of $e=0.16^{+0.08}_{-0.16}$.

\begin{figure}
    \centering
    \includegraphics[width=\linewidth]{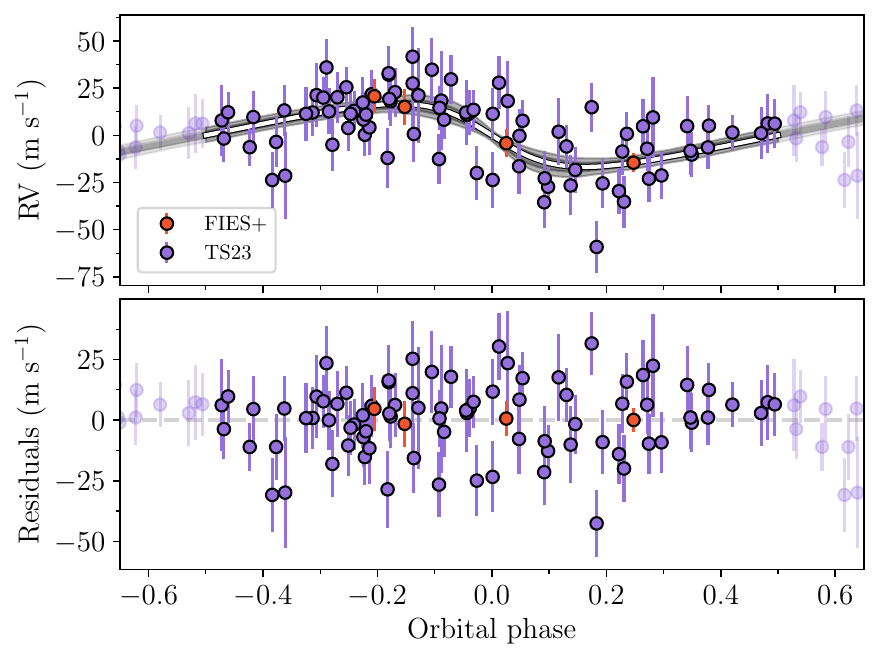}
    \caption{Radial velocity curve for TOI-5699~b. RVs from TS23 are shown with purple markers and FIES+ RVs with orange markers. The best-fitting RV curve is overplotted. The grey shaded regions show the 1 and 2$\sigma$ confidence intervals.}
    \label{fig:rv_toi5699}
\end{figure}

We therefore conclude that the data favour a mildly eccentric orbit, 
but the evidence is not strong enough to unambiguously rule out a circular solution. 
We further note that the tidal circularisation timescale is on the order of 140~Gyr
\citep[calculated using Eq. 2 in][assuming a tidal quality factor of $Q_{\rm p}=10^6$]{Adams2006},
meaning that an eccentric orbit is at least physically allowed.
We show the best-fitting model for the TESS light curve in \fref{fig:lc_toi5699} 
and the best-fitting eccentric RV model in \fref{fig:rv_toi5699}.
Posteriors from the eccentric run are summarised in \tref{tab:pars} 
and for the $e=0$ case in \tref{tab:pars_circ}.
The ground-based light curve can be found in \fref{fig:lc_5699_gb}.
Posterior distributions of key stepping parameters
for TOI-5699 can be found in \fref{fig:corner_toi5699}.

\subsection{TOI-2158}

As seen in the periodogram for TOI-2158 in \fref{fig:lomb} and as is clearly visible in the time series in \fref{fig:time_2158}, 
there seems to be multiple signals in the RVs. 
We therefore tested a suite of models starting from a model that only included just one Keplerian orbit and all the way up to three with linear and quadratic trends in between. 
For all of these runs the eccentricity was allowed to vary for planet b and c (when included), 
but for the three Keplerian case we fixed $e=0$ for planet d 
as we have not yet covered a full orbit,
and $e$ is therefore difficult to constrain.
The setup was otherwise identical to that outlined above, 
and we performed MCMC sampling for all of these.

As for TOI-5699 we calculated the $\Delta$BIC from each of these runs. 
The $\Delta$BIC between each of these and the preferred model are summarised in \tref{tab:trend_toi2158}. 
The preferred model is the 3 Keplerian model.

\begin{table}
    \centering
    \caption{The $\Delta$BIC between the setup preffered by the data (3 Keplerians) and the different tests for TOI-2158 system.}
    \begin{tabular}{l c}
    \toprule
        Model & $\Delta$BIC \\
    \midrule
        1 Keplerian & 248.91 \\ 
        1 Keplerian + lin. trend & 109.55 \\ 
        1 Keplerian + quad. trend & 107.45 \\ 
        2 Keplerians & 223.26 \\ 
        2 Keplerians + lin. trend & 59.87 \\ 
        2 Keplerians + quad. trend & 15.94 \\ 
        3 Keplerians & 0.00 \\ 
    \bottomrule
    \end{tabular}
    \label{tab:trend_toi2158}
\end{table}

In this preferred setup, the eccentricities for the orbits of planet b and c 
came out to $e_{\rm b}=$~\ecctwoone\, and $e_{\rm c}=$~\eccctwoone, respectively.
Planet c is on a quite eccentric orbit, 
whereas $e_{\rm b}$ is consistent with zero.
We further investigated whether a circular orbit for planet c
would be preferred in this setup, 
but the difference in BIC was $\Delta \mathrm{BIC}=52.94$ in favour of the eccentric solution.
We also tried to let the eccentricity of planet d vary,
while applying a Gamma prior (as we tried for TOI-5699).
We found that adding the two extra parameters ($e,\omega$)
was not preferred over fixing $e$ to zero.
Our final setup is therefore a three-planet model
in which the orbital eccentricities for planet b and c
are allowed to vary, 
while fixing $e=0$ for planet d.

Despite not covering a full cycle,
the orbital period of planet d 
is determined quite well, \perdtwoone~d,
although we note that the posteriors in \fref{fig:corner_toi2158d}
show long tails
towards longer orbital periods and larger $K$-amplitudes.
When allowing the eccentricity to vary,
these tails can be stretched out even further.
Furthermore, we gave relatively more weight to shorter orbital periods and $K$-amplitudes
by in stepping the logarithm of $P$ and $K$.
Finally, we note that the preference for a three planet Keplerian
over a two planet model including a quadratic trend came 
just after adding the latest TS23 datum acquired on 2026-01-26.
Further monitoring will establish 
if the proposed curvature is indeed accurate,
which ultimately affects the derived orbital period and minimum mass for this companion.

\begin{figure}
    \centering
    \includegraphics[width=\linewidth]{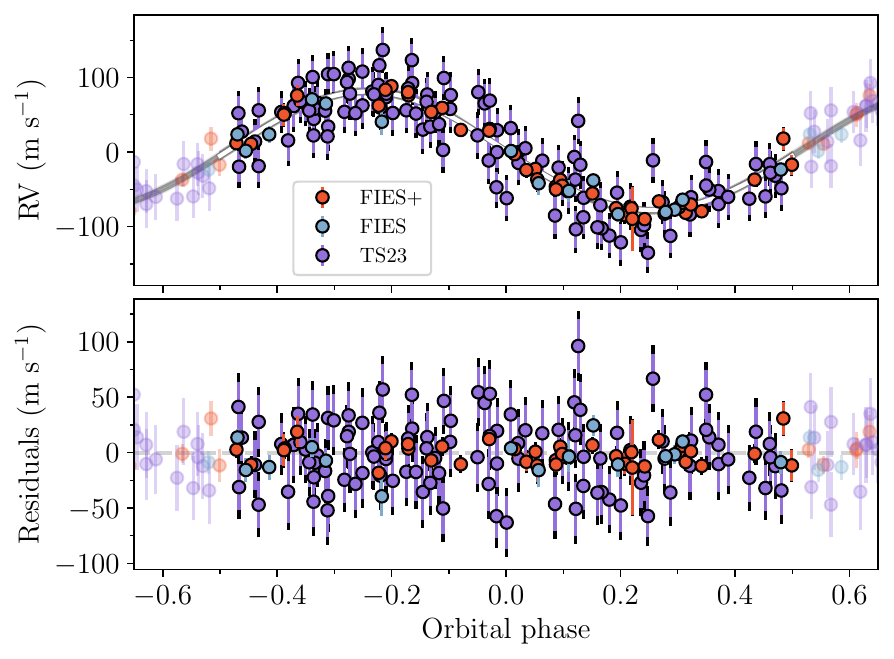}
    \caption{Radial velocity curve for TOI-2158~b. RVs from TS23 are shown with purple markers, FIES+ RVs are shown with orange, and light blue markers demarcate FIES observations. The best-fitting RV curve for TOI-2158~b is overplotted after subtracting the modulation from planet c and the long-term trend. The grey shaded regions show the 1 and 2$\sigma$ confidence intervals.}
    \label{fig:rv_toi2158_b}
\end{figure}

\begin{figure}
    \centering
    \includegraphics[width=\linewidth]{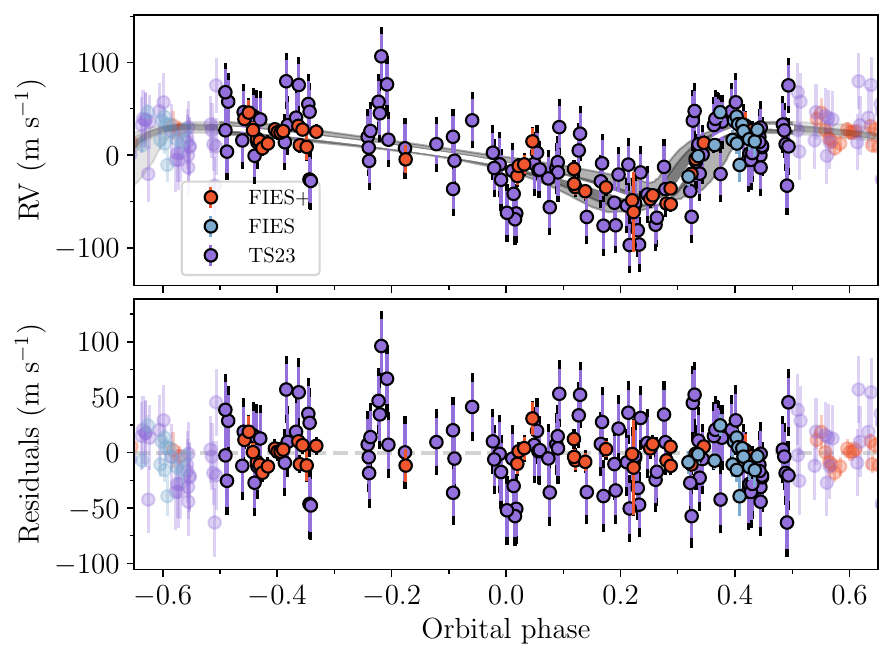}
    \caption{Radial velocity curve for TOI-2158~c. Same as in \fref{fig:rv_toi2158_b}, but for TOI-2158~c.}
    \label{fig:rv_toi2158_c}
\end{figure}

\begin{figure}
    \centering
    \includegraphics[width=\linewidth]{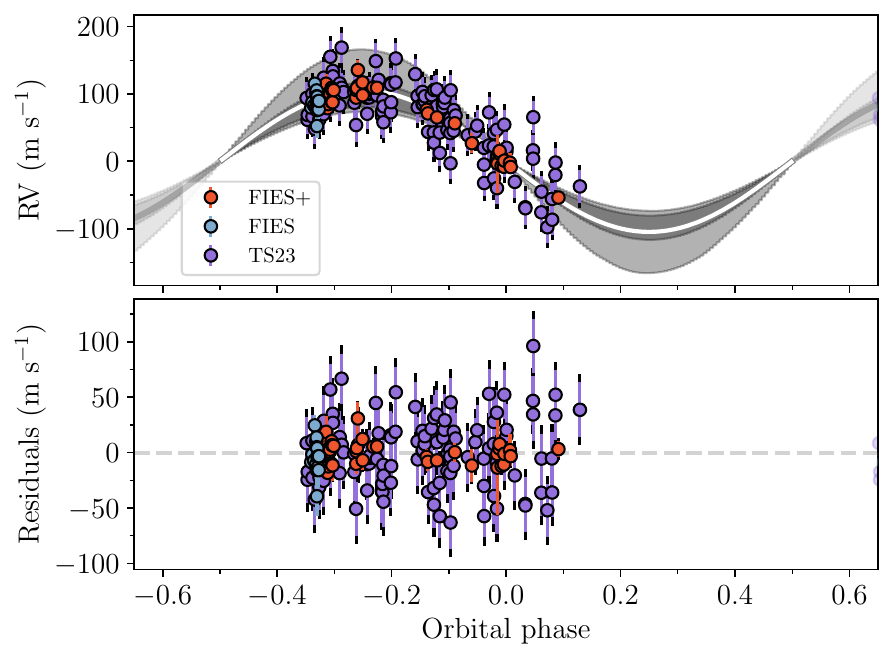}
    \caption{Radial velocity curve for TOI-2158~d. Same as in \fref{fig:rv_toi2158_b}, but for TOI-2158~d.}
    \label{fig:rv_toi2158_d}
\end{figure}

In \fref{fig:lc_toi2158} we show the phasefolded TESS light 
curve of planet b with the best-fitting model. 
The ground-based light curves can be seen in \fref{fig:lc_2158_gb}.
In \fref{fig:rv_toi2158_b}, \fref{fig:rv_toi2158_c}, and 
\fref{fig:rv_toi2158_d}
the best-fitting radial velocity models for planet b, c, and d, respectively,
are shown. 
Parameters for all planets are given in \tref{tab:pars}.
Posterior distributions of the orbital parameters
for each planet can be found in Figures~\ref{fig:corner_toi2158b}, \ref{fig:corner_toi2158c}, and \ref{fig:corner_toi2158d}.

\begin{table*}
	\centering
	\caption{
		Posterior values from the joint fit of the RVs and TESS photometry for 
		the three systems.
	}
	\label{tab:pars}
	\begin{threeparttable}
		\begin{tabular}{c c c c c}
		\toprule
		
			& TOI-5120 & TOI-5699 & \multicolumn{2}{c}{TOI-2158} \\
		
		\midrule
			
			& \multicolumn{4}{c}{Planet b} \\

		$ T_0 $~(BJD$_{\rm TDB}$) & \midfiveone & \midfivesix & \multicolumn{2}{c}{\midtwoone} \\
		$ P $~(d) & \perfiveone & \perfivesix & \multicolumn{2}{c}{\pertwoone} \\
		$ a/R_\star $ & \arfiveone & \arfivesix & \multicolumn{2}{c}{\artwoone} \\
		$ \cos i $ & \cosifiveone & \cosifivesix & \multicolumn{2}{c}{\cositwoone} \\
		$ R_{\rm p}/R_\star $ & \rpfiveone & \rpfivesix & \multicolumn{2}{c}{\rptwoone} \\
		$ K $~(m~s$^{-1}$) & \kampfiveone & \kampfivesix & \multicolumn{2}{c}{\kamptwoone} \\
		$ \sqrt{e} \cos \omega $ & \ecosfiveone & \ecosfivesix & \multicolumn{2}{c}{\ecostwoone} \\
		$ \sqrt{e} \sin \omega $ & \esinfiveone & \esinfivesix & \multicolumn{2}{c}{\esintwoone} \\
		
		\hdashline

		$ e $ & \eccfiveone & \eccfivesix & \multicolumn{2}{c}{\ecctwoone} \\
		$ \omega $~($^\circ$) & \wwfiveone & \wwfivesix & \multicolumn{2}{c}{\wwtwoone} \\
		$ i $~($^\circ$) & \incfiveone & \incfivesix & \multicolumn{2}{c}{\inctwoone} \\
		$ b $ & \impfiveone & \impfivesix & \multicolumn{2}{c}{\imptwoone} \\
        $ d $ & \impoccfiveone & \impoccfivesix & \multicolumn{2}{c}{\impocctwoone} \\
		$ T_{41} $~(hr) & \totalfiveone & \totalfivesix & \multicolumn{2}{c}{\totaltwoone} \\
		$ T_{21} $~(hr) & \gressfiveone & \gressfivesix & \multicolumn{2}{c}{\gresstwoone} \\
		$ R_{\rm p} $~(\rjup) & \radpfiveone & \radpfivesix & \multicolumn{2}{c}{\radptwoone} \\
		$ M_{\rm p} $~(\mjup) & \mpfiveone & \mpfivesix & \multicolumn{2}{c}{\mptwoone} \\
		$ T_{\rm eq} $~(K) & \teqfiveone & \teqfivesix & \multicolumn{2}{c}{\teqtwoone} \\
		$ a$~(AU) & \semifiveone & \semifivesix & \multicolumn{2}{c}{\semitwoone} \\
		$ F$~(F$_\oplus$) & \finsofiveone & \finsofivesix & \multicolumn{2}{c}{\finsotwoone} \\
        TSM & \tsmfiveone & \tsmfivesix & \multicolumn{2}{c}{\tsmtwoone} \\
        ESM & \esmfiveone & \esmfivesix & \multicolumn{2}{c}{\esmtwoone} \\

		\midrule
            & & & Planet c & Planet d \\

		$ T_0 $~(BJD$_{\rm TDB}$) & \midcfiveone  &  \midcfivesix & \midctwoone & \middtwoone \\
		$ P $~(d) & \percfiveone & \percfivesix & \perctwoone & \perdtwoone \\
		$ K $~(m~s$^{-1}$) & \kampcfiveone & \kampcfivesix & \kampctwoone & \kampdtwoone \\
		$ \sqrt{e} \cos \omega $ & \ecoscfiveone & \ecoscfivesix & \ecosctwoone & 0\\
		$ \sqrt{e} \sin \omega $ & \esincfiveone & \esincfivesix & \esinctwoone & 0 \\

		\hdashline

		$ e $ & \ecccfiveone & \ecccfivesix & \eccctwoone & 0 \\
		$ \omega $~($^\circ$) & \wwcfiveone & \wwcfivesix & \wwctwoone & 90\\
        $M_{\rm p} \sin i$~(\mjup) & \mpcfiveone & \mpcfivesix & \mpctwoone & \mpdtwoone \\
        $a \sin i$~(AU) & \semicfiveone & \semicfivesix & \semictwoone & \semidtwoone \\

		\midrule

			\multicolumn{5}{c}{System/instrument} \\
		
		$ \gamma_1 $~(m~s$^{-1}$) & \gammafiveone & \gammafivesix & \multicolumn{2}{c}{\gammatwoone} \\
		$ \gamma_2 $~(m~s$^{-1}$) & \gammatfiveone & \gammatfivesix & \multicolumn{2}{c}{\gammattwoone} \\
		$ \gamma_3 $~(m~s$^{-1}$) & \gammattfiveone & \gammattfivesix & \multicolumn{2}{c}{\gammatttwoone} \\
		$ \varsigma_{1} $~(m~s$^{-1}$) & \jitterfiveone & \jitterfivesix & \multicolumn{2}{c}{\jittertwoone} \\
		$ \varsigma_{2} $~(m~s$^{-1}$) & \jittertfiveone & \jittertfivesix & \multicolumn{2}{c}{\jitterttwoone} \\
		$ \varsigma_{3} $~(m~s$^{-1}$) & \jitterttfiveone & \jitterttfivesix & \multicolumn{2}{c}{\jittertttwoone} \\

		\bottomrule
		\end{tabular}
	\end{threeparttable}
	\begin{tablenotes}
		\item[] Values immediately below the dashed lines are derived from the posterior samples.
		\item[] $T_0$ is the mid-transit time, 
				$P$ is the orbital period, 
				$a/R_\star$ is the semi-major axis scaled by the stellar radius, 
				$i$ is the orbital inclination, 
				$R_{\rm p}/R_\star$ is the planet-to-star radius ratio, 
				$K$ is the RV semi-amplitude, 				
				$e$ is the eccentricity, 
				$\omega$ is the argument of periastron, 
				$b$ is the transit impact parameter,
                $d$ is the occultation impact parameter, 
				$T_{41}$ and $T_{21}$ are the full and ingress/egress durations, 
				$R_{\rm p}$ is the planet radius, 
				$M_{\rm p}$ is the planet mass, 
				$\rho_{\rm p}$ is the planet density, 
				$T_{\rm eq}$ is the equilibrium temperature, 
				$a$ is the semi-major axis, $F$ is the insolation.
                TSM and ESM are the transmission and emission spectroscopy metrics, respectively \citep[Eq. 1 and 4 in][]{Kempton2018}.
		\item[] $ \gamma $ is the RV offset and $ \varsigma $ is the RV jitter.
				The subscripts 1, 2, and 3 refer to TS23, FIES+, and FIES.
	\end{tablenotes}
\end{table*}




\section{Discussion}\label{sec:disc}

To place our systems in context, we compiled a comparison sample of known giant planets.
We queried the \texttt{pscomppars} table from the NASA Exoplanet Archive \citep[NEA][]{Akeson2013,Christiansen2025} on 2026-06-23. 
We replace eccentricities, whenever possible, with values from \citet{Bonomo2017}, 
who provided homogeneous eccentricity determinations 
for transiting giant planets based on a combined analysis of radial velocities and transit data.

\subsection{System Architectures and Properties}

The TOI-5120 system consists of a Jupiter-like planet 
(\rpl~$=$~\radpfiveone~\rjup, \mpl~$=$~\mpfiveone~\mjup)
on a wide orbit (\ars~$=$~\arfiveone, $a=$~\semifiveone~AU)
and a likely stellar companion, 
which we have detected 
at a separation of $1.26^{\prime\prime}$ ($\sim290$~AU). 
The orbit of TOI-5120~b is slightly eccentric at $e=$~\eccfiveone.

TOI-5699 is another warm Jupiter system (\ars~$=$~\arfivesix, $a=$~\semifivesix~AU, \rpl~$=$~\radpfivesix~\rjup, \mpl~$=$~\mpfivesix~\mjup), 
but unlike for the other two systems we detect no companions.
For the orbit of TOI-5699~b we measure a modest, 
but significant eccentricity of $e=$~\eccfivesix.
Although again we note that we cannot unambiguously rule out a circular orbit.

TOI-2158~b is a Jupiter-sized planet (\rpl~$=$~\radptwoone~\rjup, \mpl~$=$~\mptwoone~\mjup)
orbiting a slightly evolved G-type star in a system with potentially two outer companions.
The planet is at the border between being a warm Jupiter (\ars~$=$~\artwoone, $a=$~\semifivesix~AU)
if the distinction is made in \ars\, 
with \ars~$>10$ typically denoting warm Jupiters, 
or a hot Jupiter ($P=$~\pertwoone~d) 
when considering the orbital period,
where the distinction is typically made at $P=10$~d, although \citet{Dong2021b,Fairnington2026} use $P>8$~d for warm Jupiters.
We therefore refer to TOI-2158~b
as a warm Jupiter.
TOI-2158~c is likely a Jupiter-mass planet ($M_{\rm p}\sin i=$~\mpctwoone~\mjup) with an orbital period of \perctwoone~d. 
The orbit of planet c is quite eccentric ($e=$~\eccctwoone), 
whereas the orbit of planet b is consistent with being circular ($e=$~\ecctwoone).
For the outermost companion we have so far only detected a long-term trend 
and have not yet covered a full orbit.
Nonetheless, a three-planet model is statistically preferred over a two-planet model plus a quadratic trend.
Assuming a Keplerian model for this outer signal 
yields an orbital period of \perdtwoone~d 
and a $K$-amplitude of \kampdtwoone~m~s$^{-1}$, 
which would (most likely) make it a planetary object 
(with $M_{\rm p}\sin i=$~\mpdtwoone~\mjup\, assuming $e=0$).

\begin{figure}
    \centering
    \includegraphics[width=\columnwidth]{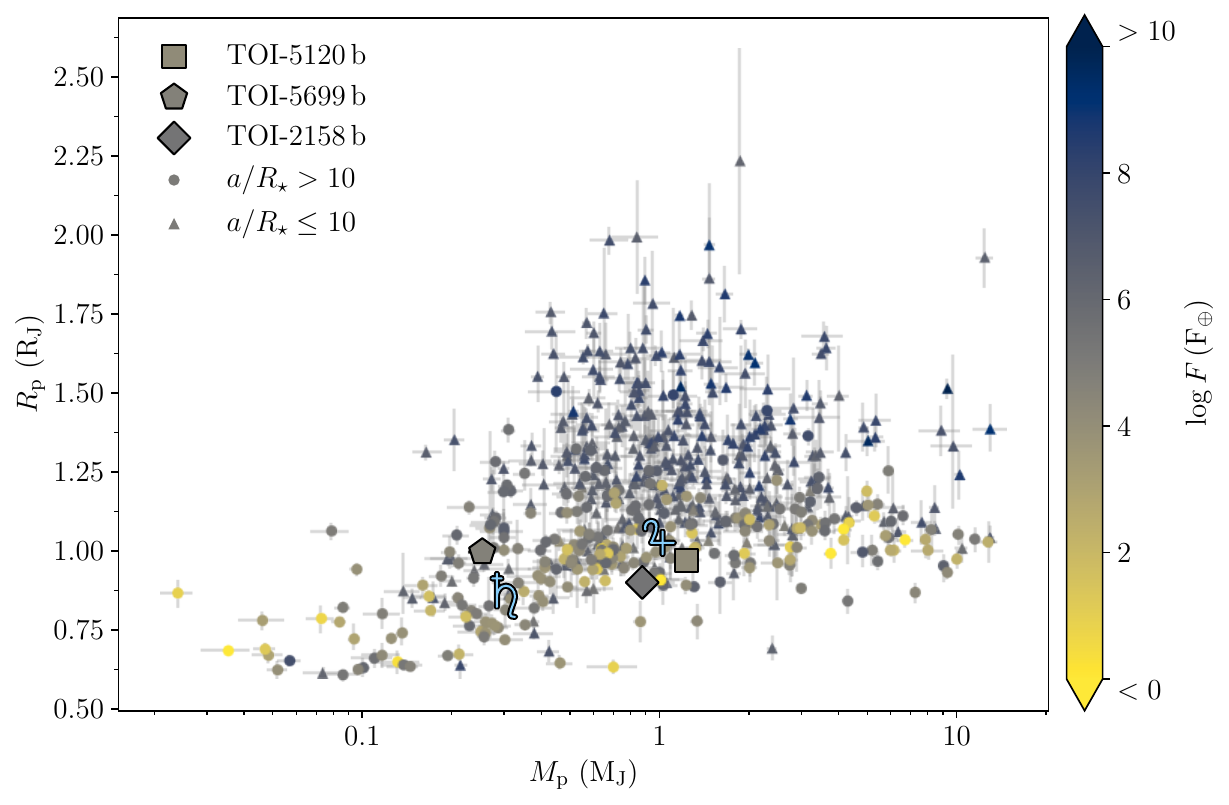}
    \caption{Mass-radius diagram for giant planets defined as 
    \mpl~$<13$~M$_{\rm J}$ and ( \mpl~$>0.3$~M$_{\rm J}$ or \rpl~$>0.6$~R$_{\rm J}$) 
    with mass and radius measurements more precise than 20\%.
    Triangles denote hot Jupiters
    and warm Jupiters are shown with circles.
    Markers are colour-coded according to the insolation.
    The locations of Jupiter and Saturn are shown with their respective symbols.
    }
    \label{fig:mr}
\end{figure}

In \fref{fig:mr} we show a radius-mass diagram of giant planets defined as planets
larger than $0.6$~\rjup\, or more massive than $0.3$~\mjup\, and in both cases less massive than $13$~\mjup.
We furthermore only show planets
with masses and radii determined to better than 20\%. 
The three transiting planets from this study are highlighted. 
In terms of size and mass, 
TOI-5120~b and TOI-2158~b are archetypical Jupiters with values very close to that of Jupiter.
While the size of TOI-5699~b is similar to that of Jupiter and the two other planets, 
it is significantly less massive and more comparable to Saturn in mass.
Dividing the planets into either hot or warm giants
using \ars\, below or above 10 in \fref{fig:mr} 
clearly illustrates 
the radius inflation often observed for hot Jupiters 

\begin{figure}
    \centering
    \includegraphics[width=\columnwidth]{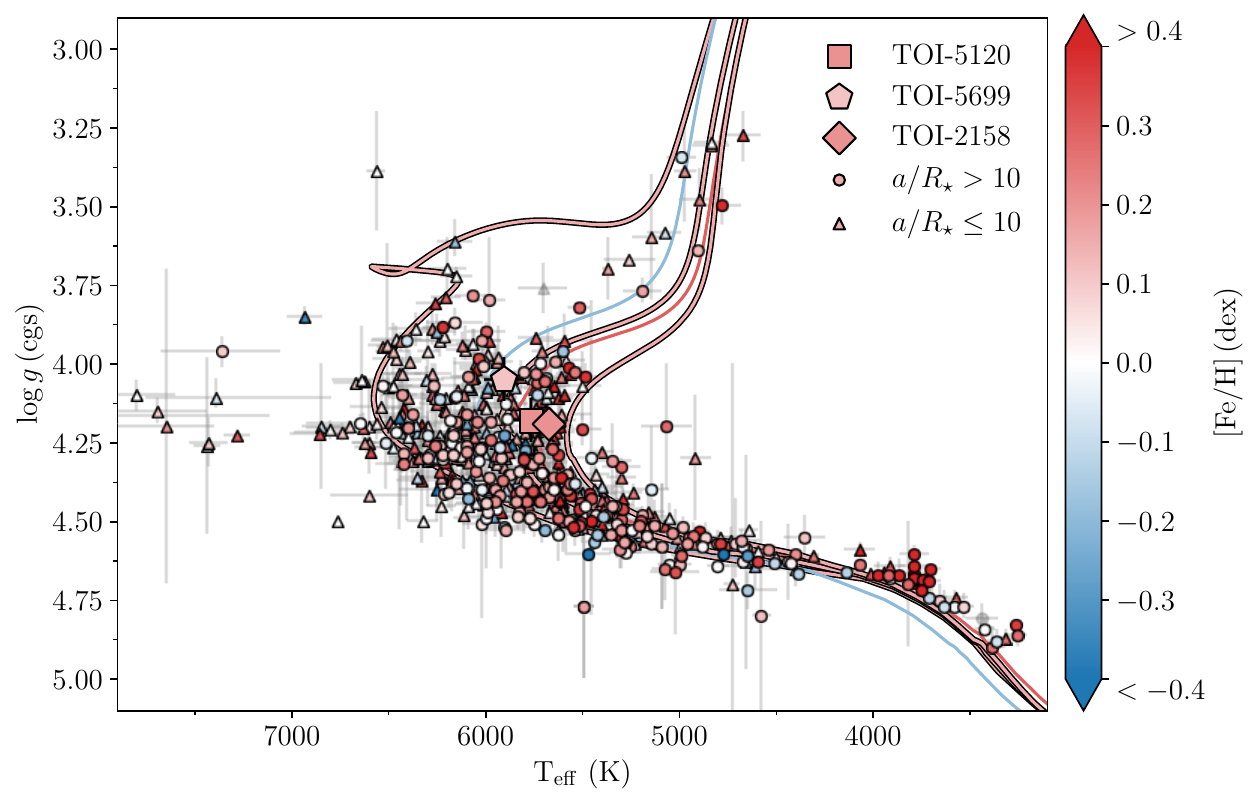}
    \caption{
    Kiel diagram for the host stars in \fref{fig:mr}.
    Warm Jupiter systems (\ars~$>10$) are shown with circles and hot Jupiter systems with (slightly transparent) squares.
    Stars are colour-coded according to their metallicity. 
    We show three BaSTI isochrones (with the inclusion of diffusion, mass loss, and overshooting) at $\rm [Fe/H]=0.15$ with ages (from left to right) of 2~Gyr, 7~Gyr, and 13~Gyr. We furthermore show two additional 7~Gyr isochrones with $\rm [Fe/H]=-0.08$ and $\rm [Fe/H]=0.3$. 
    }
    \label{fig:kiel}
\end{figure}

The three host stars are quite similar in terms of \teff, \logg, and \feh,
and they are all slightly evolved.
The Kiel diagram in \fref{fig:kiel} shows that among
stars hosting transiting, warm Jupiters,
these three do indeed belong to the more evolved segment.
In this context, it is of course also
worth noting that 
as a star evolves and expands, 
\ars\, shrinks (for constant/decreasing $a$) and
a warm Jupiter 
could transition to a hot Jupiter
(given the \ars\, distinction).

\fref{fig:kiel} also shows the well-known correlation 
that giant planets are more frequently found around stars 
with super solar metallicity \citep[e.g.][]{Fischer2005,Sousa2011} 
as is the case for the three systems investigated here.
This trend is commonly interpreted as evidence 
that metal-rich protoplanetary discs are more efficient 
at forming the solid cores required for the subsequent 
accretion of massive gaseous envelopes.

\subsection{Formation and migration}
The formation of a hot Jupiter is generally ascribed to one of three mechanisms; 
in-situ formation, disc migration, or high-eccentricity (high-$e$) tidal migration.
As the name implies, in-situ formation offers a straightforward explanation. 
The giant planet would form at the observed separation either
through gravitational instability or through core accretion.
However, both of these formation channels have their limitations
in forming 
these massive planets at these small orbital separation
\citep[e.g.][]{Rafikov2005,Schlichting2014,Dawson2018}.
If we do not believe these planets to be able form
at the observed locations,
they must naturally have been transported to their current location in some way.

In gas disc migration the semi-major axis shrinks
as the
planet exchanges angular momentum
with the gas and dust particles in the protoplanetary disc,
which implies that through this channel 
migration takes place within the lifetime of the disc (i.e. within the first $\sim 10$~Myr 
of the system).
Disc migration can effectively shrink the orbit from
several AU to just a few hundredths of an AU
\citep[e.g.][]{Lin1996}.
It is typically considered a quiet form of migration
in that both the eccentricity and obliquity remain low 
\citep[i.e. the planet is found on a close to circular orbit 
that is well-aligned with the stellar spin axis;][]{Duffell2015}.

\begin{figure*}
    \centering
    \includegraphics[width=\textwidth]{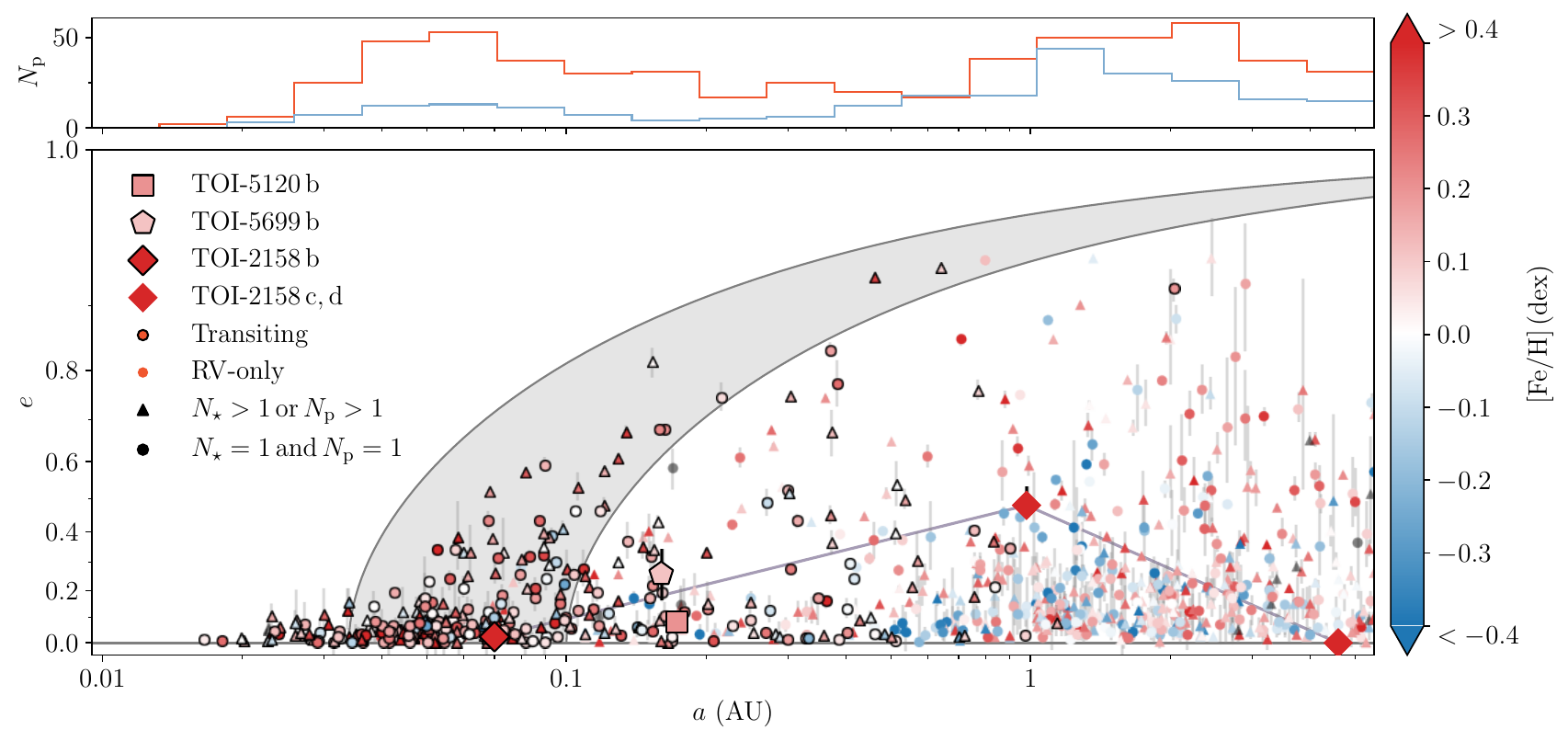}
    \caption{
    The eccentricity of giant planets plotted against semi-major axis.
    Markers with a black outline are transiting planets with the same criteria 
    as in \fref{fig:mr} and well-constrained eccentricities. 
    Markers without a black outline are giant planets discovered through RV in the same mass range and
    well-constrained eccentricities.
    Marker shapes denote whether the system is
    a multistar system ($N_\star >1$), 
    a multiplanet system ($N_{\rm p}>1$), 
    or both, 
    or whether it is a single star, single planet system (as given in the NEA).
    As \fref{fig:kiel} markers are colour-coded according to host star metallicity. 
    The grey shaded area shows where one might expect planets undergoing high-$e$ tidal migration.
    The $y$-axis is linear in $1-e^2$.
    Histograms for the number of planets in a given 
    semi-major axis bin
    are shown on top for host stars with super-solar metallicities ($\rm [Fe/H]>0.0$) in red and sub-solar in blue.
    Adapted from \citet{Dawson2018,Dong2021}.
    }
    \label{fig:highe}
\end{figure*}

In contrast, high-$e$ migration can excite both eccentricities and obliquities to large values.
This happens when a perturber reduces the orbital angular momentum of a Jupiter
by placing it on a highly elliptical orbit, 
either through planet-planet scattering
\citep[e.g.][]{Rasio1996}
or secular interactions
such as von Zeipel-Kozai-Lidov
\citep[ZKL;][]{Zeipel1910,Kozai1962,Lidov1962} oscillations
driven by outer planetary or stellar companions
\citep[e.g.][]{Wu2003}.
Orbital energy might then be removed 
as the planet is stretched by the star 
during the close passages during periapsis
\citep{Fabrycky2007}.
This tidal dissipation phase shrinks and circularises 
the orbit with a final semi-major axis given by
$a_{\rm final}=a (1 - e^2)$, which for $e \to 1$ becomes $a_{\rm final} \approx 2a (1 - e)$, 
meaning that the initial periastron is $\approx \frac{1}{2}a_{\rm final}$.

As high-$e$ migration can take place throughout the system's lifetime
that also implies that we could catch planets in transition, i.e., giant planets on wide orbits with extreme eccentricity undergoing tidal migration and circularisation.
The transition timescale for going from cold to hot, 
however, 
has been suggested to be short 
compared to the time spent as a cold Jupiter \citep{Anderson2017},
and catching planets en route might therefore be difficult.
Nonetheless, in \fref{fig:highe}
these proto hot Jupiters would be
found in the grey shaded region in \fref{fig:highe}, 
which is bounded by tracks of constant angular momentum.
As planets would be tidally disrupted 
as they pass inside the Roche limit,
$a_{\rm Roche}\approx f_{\rm p} R_{\rm p} (M_{\rm p}/M_\star)^{(1/3)}$  \citep[where $f_{\rm p}$ is a dimensionless coefficient we assume to be 2.7;][]{Guillochon2011}, 
$a_{\rm final}$ must therefore be greater than $2 a_{\rm Roche}$. 
For a $1$~\mjup,
$1.3$~\rjup\, planet orbiting a solar-mass star \citep[as used in][]{Dong2021}
this comes out to $0.034$~AU.
As tidal circularisation depends critically on the orbital separation, 
the lower limit is taken as $a_{\rm final}=0.1$~AU,
beyond which tidal migration is not expected 
on meaningful timescales \citep{Dawson2018}.

Some great examples of potential hot Jupiter progenitors are the transiting, warm Jupiters HD~80606~b \citep{Naef2001}, TOI-3362~b \citep{Dong2021}, and TIC~241249530~b \citep{Gupta2024} 
seen in \fref{fig:highe} as the three most eccentric planets in the grey region.
TOI-2158~b is found well within these boundaries 
on a close-in orbit with virtually no eccentricity. 
Considering that we have detected two outer companions, 
with the inner companion of those two found on a quite eccentric orbit,
high-$e$ tidal migration therefore provides a plausible evolutionary pathway for TOI-2158~b. 
However, the present-day architecture does not uniquely establish such a history, 
and alternative scenarios, 
such as migration through the protoplanetary disc followed by subsequent excitation of the outer planets, cannot be ruled out.
TOI-5120~b and TOI-5699~b are clearly not found within this region
and do therefore not appear to be hot Jupiters in the making.

In addition to the systems mentioned above, 
in \fref{fig:highe} we also show all transiting, giant planets
again defined as \mpl~$<13$~M$_{\rm J}$ and (\mpl~$>0.3$~M$_{\rm J}$ or \rpl~$>0.6$~R$_{\rm J}$), 
although we do not require a certain precision in mass or radius
as in \fref{fig:mr}. 
We do, however, only include 
planets with well-constrained eccentricities,
namely planets
with $e$ 
constrained to better than 50\% 
or with uncertainties less than $0.2$ 
for planets with $e<0.2$ 
\citep{Dong2021}.
Furthermore, 
we show planets detected through RV measurements
in the same mass range ($M_{\rm p}\sin i$ from $0.3$ to $13$~\mjup) 
and invoking the same eccentricity constraints.
The inclusion of these RV planets adds a lot of cold Jupiters to the sample,
which clearly illustrates a dip in the number of planets 
per log interval at warm Jupiter separations, 
the so-called period valley \citep[e.g.][]{Jones2003,Udry2003}.

\subsubsection{Warm Jupiter Eccentricity Distribution}

As for hot Jupiters, warm Jupiters are thought to form
through one of the three aforementioned mechanisms.
Another issue with invoking in situ formation
is that many warm Jupiters are observed with
relatively high eccentricities ($e\gtrsim0.2$).
This high-$e$ component for the distribution
also poses a problem in gas disc migration
as planet-planet
scattering at these separations
tends to lead to collisions rather than
eccentricity excitation \citep{Petrovich2014}.
On the other hand, the low-$e$ component
is difficult to account for if warm Jupiters
form through high-$e$ migration.
The eccentricity distribution of warm Jupiters 
thus comprises both low- and high-$e$ components,
both of which must be reproduced by theoretical models 
and properly considered when inferring the functional form of the observed distribution.

Owing to their flexibility with only two shape parameters,
the Gamma and Beta distributions (mentioned in Section~\ref{sec:fit}) 
are popular choices for modelling the exoplanet eccentricity distribution in general,
with the added benefit for the latter 
that its domain is restricted to $[0,1]$
\citep[e.g.][]{Kipping2013}.
The Beta distribution has been used extensively
as a prior when fitting the eccentricity,
especially in low signal-to-noise ratio cases,
or it has been used for populations syntheses,
for instance, as done in Section~\ref{sec:complete} below.

Physically motivated distributions 
include the mixture of the Exponential distribution and the Rayleigh distribution,
where the former should capture
the planets undergoing tidal dissipation
\citep{Rasio1996}
and the latter should reflect 
the outcome of planet-planet scattering \citep{Juric2008}.
In the high-$e$ tidal migration scenario,
this mixture model should therefore nicely account
for the observed distribution.
\citet[][$\mathcal{RE}$ in their notation]{Stevenson2025} found that this model
was the preferred distribution (among those investigated)
when considering all planets discovered with radial velocities.

Accordingly, we investigated the warm Jupiter eccentricity distribution using hierarchical Bayesian modelling (HBM).
We created the HBM and sampled the posterior
using \texttt{PyMC} \citep{pymc2023}.
We modelled each planet’s latent eccentricity 
as drawn from a two-component mixture on $[0,1]$: 
a truncated Exponential distribution and a truncated Rayleigh distribution, mixed with weight, $\alpha$, 
which can then be interpreted as the fraction of systems 
consistent with tidal damping versus dynamical excitation.
Truncation is needed \citep[as noted by][]{Kipping2013,Stevenson2025} since
a problem with the $\mathcal{RE}$
distribution is that
it assigns a non-zero probability to hyperbolic
orbits. 
We placed weakly informative hyperpriors 
on the parameters of mixture model:
$\alpha\sim\mathrm{Beta}(1,1)$, 
$\ln\sigma\sim\mathrm{Uniform}(-8,3)$, 
and 
$\ln\lambda\sim\mathrm{Uniform}(-5,5)$, 
where $\sigma=\exp(\ln\sigma)$ and 
$\lambda=\exp(\ln\lambda)$ 
are the shape parameters 
for the Rayleigh respectively Exponential distribution. 
Observed eccentricities are modelled with 
Gaussian measurement errors around latent eccentricities: $e_i^{\mathrm{obs}}\sim\mathcal{N}(e_i,s_i^2)$, 
where $s_i$ is the reported $1\sigma$ 
uncertainty (floored at $10^{-4}$ for numerical stability). 

\begin{figure}
    \centering
    \includegraphics[width=\columnwidth]{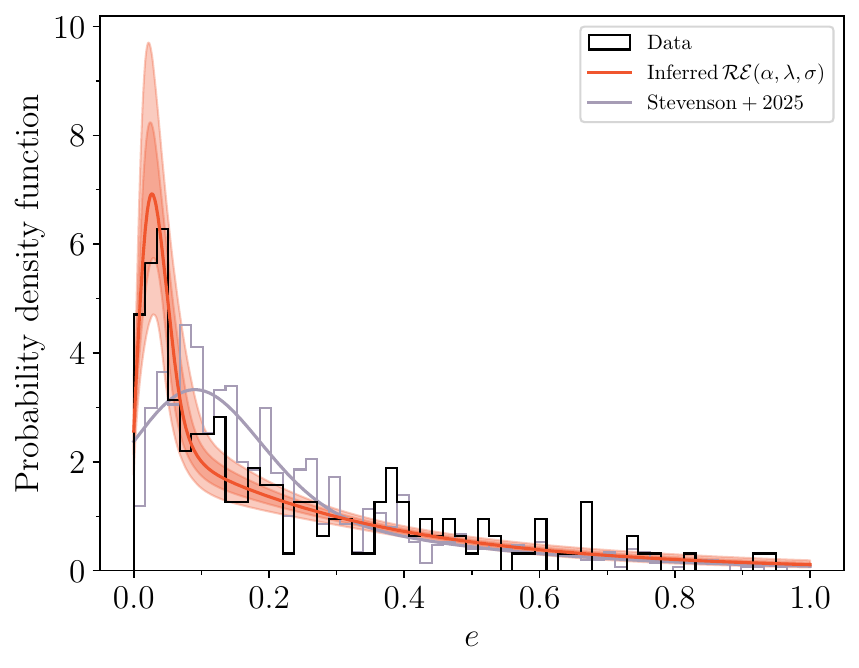}
    \caption{Shown in black is our sample of warm Jupiter eccentricities.
    In red we show the Rayleigh+Exponential
    distribution resulting from our HBM
    with parameters
    $\alpha=$~\hbmalpha, $\lambda=$~\hbmlambda, and
    $\sigma=$~\hbmsigma.
    In grey we show the RV eccentricity sample from \citet{Stevenson2025} along with their Rayleigh+Exponential
    distribution.
    }
    \label{fig:dist}
\end{figure}

We defined our sample of warm Jupiters as those planets 
meeting the criteria used to create \fref{fig:highe},
but only including planets with orbital periods between $8$ and $200$~d
as in 
\citep[][roughly between $0.08$ and $0.67$~AU]{Dong2021,Fairnington2026}, 
thereby also conveniently including TOI-2158~b.
A histogram for the eccentricities of these \Nhbm\, planets are shown in \fref{fig:dist}, 
where we also compare our sample to the ($N=891$)
sample of eccentricities from RV detected planets
used in \citet{Stevenson2025}.
Clearly, our sample has relatively more low-eccentricity planets compared to theirs,
possibly owing to the inclusion of more close-in planets,
which might have tidally circularised.
From our HBM we do indeed also get a slightly larger contribution from the Exponential population
with
$\alpha=$~\hbmalpha\,
compared to $\alpha=0.70\pm0.02$ in \citet{Stevenson2025}. 
For the shape parameters we get
$\lambda=$~\hbmlambda\, and
$\sigma=$~\hbmsigma\, \citep[$\lambda=3.39^{+0.09}_{-0.10}$,  $\sigma=0.111\pm0.003$ in][]{Stevenson2025}.
Although employing different models,  
by fitting a Beta distribution to the eccentricities of transiting warm Jupiters with orbit periods between 10 and 1000 days,
\citet{Gan2025} also found that most warm Jupiters exhibit circular orbits 
(shape parameters of $\alpha=0.64\pm0.07$ and $\beta=1.55\pm0.12$).

As mentioned, the overall occurrence rate of giant planets 
correlates with stellar metallicity,
and we might therefore also expect planet-planet interactions (through scattering or ZKL oscillations)
to occur more frequently in
such systems.
This in turn could thus be reflected in the relative contributions to the $\mathcal{RE}$ mixture model
when comparing the eccentricities of planets
around stars at either sub- or super-solar metallicity.
This effect might naturally also 
be visible more directly
when comparing the eccentricities of planets
with confirmed outer companions to those without.
Slicing and dicing the sample of eccentricities of warm Jupiters has been done in various ways, 
where, for instance,
\citet{Dong2014} reported  
that the fraction of warm Jupiters with outer companions (``jovian perturbers'') was an increasing function of eccentricity as opposed to those without companions,
and \citet{Morgan2026} recently found
that warm Jupiters orbiting metal-rich stars 
have higher eccentricities compared to those around metal-poor stars.

To see if the physically 
motivated distributions
might reveal more about the preferred migration channel,
we tried splitting our warm Jupiter sample into systems with companions,
with no distinction on the type of companion
($N_\star >1$ or $N_{\rm pl} >1$),
and those without ($N_\star =1$ and $N_{\rm pl} =1$).
From this we found $\alpha=$~\hbmalphacompanion\, 
for the systems with companions
and $\alpha=$~\hbmalphasingle\,
for those without, 
indicating that dynamically excited orbits are more prevalent in systems with companions, 
consistent with scattering or secular perturbations driving eccentricity growth.

We also tried to investigate the correlation
between metallicity and eccentricity,
but instead of dividing the sample into
two (or more) rather arbitrary bins,
for instance, a super- and sub-solar $\rm [Fe/H]$-population,
we modelled the mixture weight
as a function of host-star metallicity.
Specifically, for each planet $i$
we define the Rayleigh weight 
$\alpha_i$
via a logistic (sigmoid) link function
$\logit (\alpha_i) \equiv \ln \frac{\alpha_i}{1-\alpha_i} =\beta_0 + \beta_1 [\mathrm{Fe/H}]_i$,
meaning that 
\begin{equation}
    \alpha_i = \frac{1}{1+e^{-(\beta_0 + \beta_1 [\mathrm{Fe/H}]_i)}}
\end{equation}
where $\beta_0$ sets the intercept of the population-averaged weight at solar
metallicity, 
while the slope $\beta_1$ quantifies the metallicity dependence:
$\beta_1 > 0$ implies that metal-rich systems are more likely to have
eccentricities drawn from the Rayleigh distribution, 
consistent with dynamical
excitation in metal-rich systems hosting giant planets. 
We placed weakly
informative priors on $\beta_0, \beta_1 \sim \mathcal{N}(0, 2)$, 
which are broad
on the logit scale and allow $\alpha_i$ to span the full range $(0,1)$.
To enforce physically meaningful component shapes (with the Exponential representing
a near-circular population and the Rayleigh an excited one),
we restricted
$\lambda \in [3.3, 33]$, corresponding to a mean of $\langle e \rangle < 0.3$) and
$\sigma \in [0.20, 0.82]$, so that the Rayleigh mode $e = \sigma > 0.2$. 
The crossover metallicity 
$[\mathrm{Fe/H}]^{*} = -\beta_0/\beta_1$ marks where neither
component is preferred ($\alpha_i = 0.5$).
From this we found $\lambda=$~\hbmlambdalogit, $\sigma=$~\hbmsigmalogit, $\beta_0=$~\hbmbetazerologit, and $\beta_1=$~\hbmbetaonelogit, 
which corresponds to a cross-over metallicity
of $\rm [Fe/H]=$~\hbmcross,
meaning that the Rayleigh distribution
becomes dominant above $\rm [Fe/H]=0.2$.

We note that 
there are probably some 
uncertainties in the results presented here,
given how the samples were constructed.
First off, to properly investigate the influence
of companions one should assess the 
relevance
of these companions, meaning are they massive and/or nearby enough
(and sufficiently inclined), 
for the proposed migration mechanisms,
as well as the completeness for the systems without companions
(i.e. what range in mass and orbital separation can be excluded, see Section~\ref{sec:complete} below).
Secondly, we found that when modelling the mixture weight 
as a function of host-star metallicity using a logistic function,
it was necessary to enforce the priors as outlined, 
indicating that the results are perhaps not as 
robust as desired.
Furthermore, 
mixing samples of systems discovered
through RV and 
transit also means mixing selection functions
(e.g. the transit method 
is biased towards close-in and eccentric planets),
which we do not attempt to mitigate here.
We therefore do not necessarily expect our results
to reflect the true underlying distributions of the different samples; 
these are simply the distributions obtained 
when considering the currently available systems that meet our selection criteria.
Future constraints on outer companions 
and homogeneous samples will be key to determining
whether the eccentricity distribution can robustly disentangle 
the dominant migration pathways of warm Jupiters.

\subsubsection{Companions and Completeness}
\label{sec:complete}

The eccentricities for TOI-5120~b and TOI-5699~b 
appear to be too low for them to be undergoing high-$e$ tidal migration.
However, if (as of yet undetected) outer planetary companions reside in these systems, 
the inner planets might periodically
have their eccentricity excited
through secular interactions
and during this high-$e$ phase undergo tidal migration
as suggested by \citet{Petrovich2016}.
In this steady-state picture, 
the warm Jupiter spends most of its time
at relatively low eccentricities.
Moreover, if the eccentricity excitation is driven by ZKL cycles, 
the exchange between eccentricity and inclination implies that low-$e$ phases may coincide with high mutual inclinations,
which remain unconstrained for these systems. 
As a result, 
the currently observed low eccentricities do not necessarily rule out ongoing high-$e$ tidal migration.
As argued, although modest 
the eccentricity we do seem to measure for TOI-5699~b
(assuming it is significant cf. Section~\ref{sec:dat5699})
is difficult to account for in the in situ
or disc migration scenarios,
and as such this steady-state mechanism
might offer an explanation.

In TOI-5120 we have detected a stellar companion 
(physical separation $\sim300$~AU), 
which
might have triggered secular interactions
and caused the planet to migrate, 
although we currently know very little about this companion and thus its possible influence.
Furthermore, \citet{Petrovich2015a} found that stellar companions 
produced very few planets at separations intermediate to hot and cold Jupiters ($\sim0.1$ to 2~AU).

\begin{figure*}
    \centering
    \includegraphics[width=\textwidth]{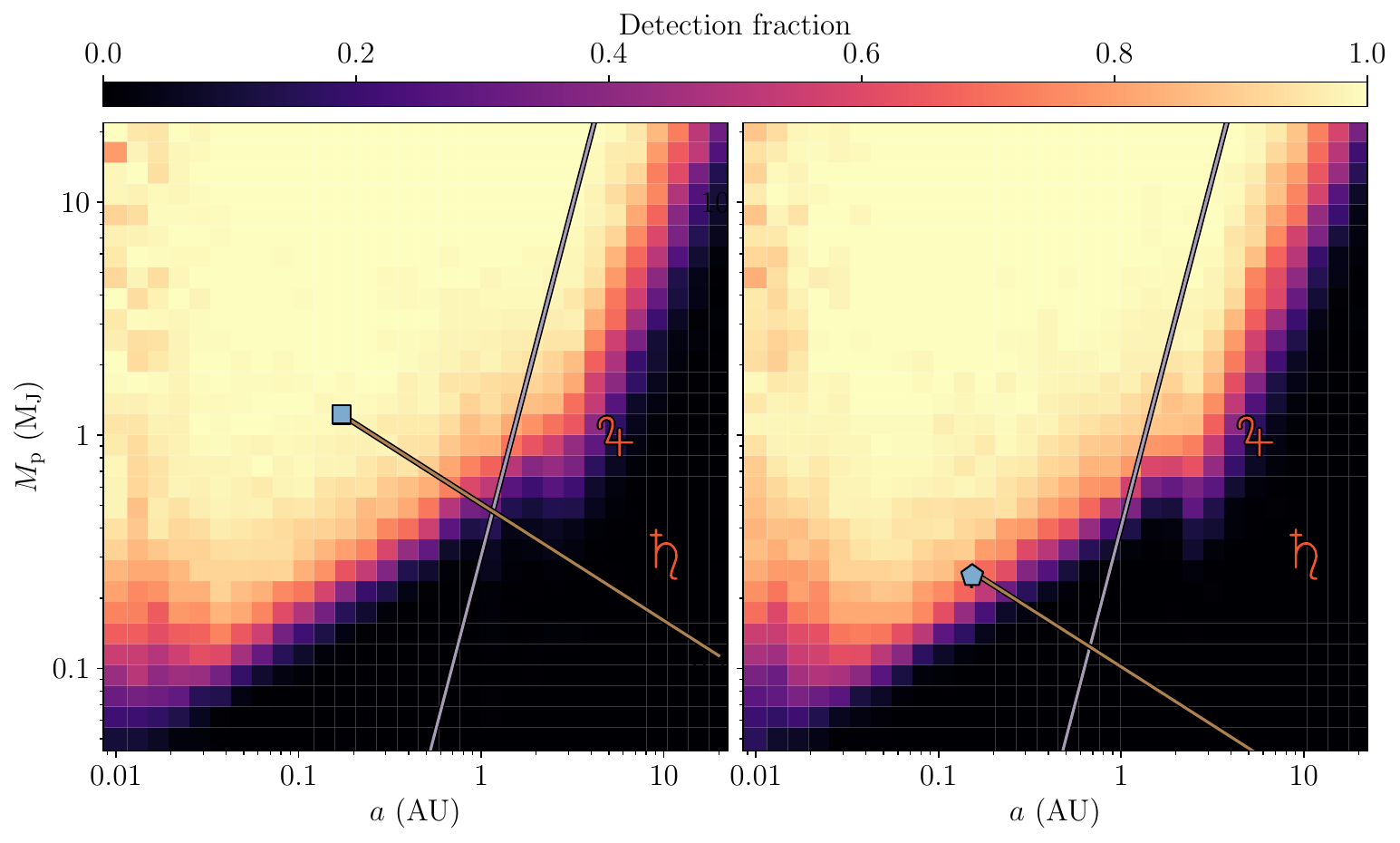}
    \caption{
    Completeness maps from our RV observations 
    of TOI-5120 (left) and TOI-5699 (right). 
    The position of the transiting planets are shown with the blue markers
    and the position of Jupiter and Saturn are shown for reference.
    Each cell represents 300 planets with values drawn uniformly in $\cos i$ from 0 to 1 and $\omega$ between 0 and $2\pi$ with $e$ drawn from a Beta distribution 
    \citep{Kipping2013}. 
    The cell is then colour-coded according to the fraction of significantly detected signals ($\Delta \mathrm{BIC}>10$).
    The grey line shows the minimum perturber mass required to overcome general-relativistic precession, 
    while the brown line indicates 
    where the outer companion dominates the system's angular momentum.
    }
    \label{fig:complete}
\end{figure*}

We therefore tried to place limits on 
potential planetary companions in these systems.
In \fref{fig:complete} we show the sensitivity limits 
in terms of mass and semi-major axis from our TS23 and FIES observations for TOI-5120 and TOI-5699.
These plots have been created through injection 
and recovery tests given our data
following the approaches in
\citet{Bryan2019,Bonomo2023}.
The plots are evidently quite similar, 
since we have comparable baselines, number of observations, and root mean squares.
Given our data, we are able to exclude 
Jupiter-mass planets out to a couple of AU.

In \fref{fig:complete} we also show an approximation for the mass and semi-major axis 
a perturber would need to have in order to overcome general-relativistic precession 
(Equation 14 in \citealt{Dawson2018}; see also \citealt{Dong2014}). 
A companion would therefore have to lie to the left of this line to excite the eccentricity of the inner planet. 
Efficient secular excitation further requires that the outer companion dominates the angular momentum budget of the system. 
In hierarchical systems the orbital angular momentum scales approximately as 
$L\propto M \sqrt{a}$, 
such that the ratio of inner to outer angular momentum is $L_{\rm in}/L_{\rm out} \sim
(M_{\rm p, in}/M_{\rm p, out})\sqrt{a_{\rm in}/a_{\rm out}}$
\citep[e.g.][]{Liu2015}. 
Large eccentricity excitation is therefore expected only 
when the outer orbit carries a comparable or larger share of the total angular momentum,
implying that perturbers capable of driving strong secular excitation 
are typically comparable to or more massive than the inner planet. 
Applying this constraint, as shown in \fref{fig:complete}, 
leaves only a narrow range in mass and semi-major axis for potential perturbers. 
While these constraints are admittedly somewhat ``hand-wavy'', 
they nevertheless illustrate that a companion capable of exciting the eccentricity of TOI-5120~b is unlikely. 
In contrast, for TOI-5699~b our current data still leave some room for a companion to remain undetected.

\subsection{Future Studies}

Continued RV monitoring of TOI-5120 and TOI-5699
will provide even stronger constraints on
the possible masses and orbital separations
of potential outer companions
or they might even reveal their presence.
Additional RVs would also help constrain the eccentricity of TOI-5699~b further
and determine whether the value we measure
($e=$~\eccfivesix) is not just (partly)
an artifact of the current data.
Continued RV monitoring of TOI-2158 is needed to cover a full orbit 
of the long-term trend 
and thereby determine the orbital period 
and minimum mass of the outer companion.
As TESS continues to monitor the sky,
more transits will be available for all three systems.
Currently, TOI-2158 is slated to be observed in two sectors in June and July 2027.

All three (transiting) planets are amenable targets for measurements of the
Rossiter-McLaughlin \citep[RM;][]{Rossiter1924,McLaughlin1924} effect.
Using the simple approximation
$0.7\sqrt{1-b^2}(R_{\rm p}/R_\star)^2 v\sin i_\star$, 
we calculated the amplitude for the RM anomaly to be
23~m~s$^{-1}$ for TOI-5120~b, 9~m~s$^{-1}$
for TOI-5699~b, and 10~m~s$^{-1}$ for TOI-2158 --
well-within reach for many spectrographs.
The relatively long duration of the transits 
($\gtrsim3.8$~h) helps ensure proper sampling.
As mentioned, dynamical interactions 
might also excite the obliquity of the host star,
measurements of the RM effect would 
allow us to determine the projected obliquity
and therefore
help shed light on the dynamical histories in these systems.
This is especially true for warm Jupiter systems
as the larger orbital separation would minimise
the effect of tidal realignment, 
which might erase this information \citep{Albrecht2012}.

We estimated the expected astrometric signal of TOI-2158~d 
following the standard reflex-motion relation. 
Given the minimum mass, orbital separation,  
and stellar mass (Tables~\ref{tab:pars} and \ref{tab:stars}), 
the induced stellar semi-major axis is
$a_\star = a (M_{\rm p}/M_\star) \sim 0.03 $~AU.
At a distance of \twooned~pc this translates to an astrometric signature of $\sim 151$~$\mu$as.
This is several times larger 
than \emph{Gaia}'s expected astrometric precision 
for stars with magnitudes of $G\sim 10-11$
\citep[tens of $\mu$as;][]{Perryman2014}.
Given the orbital period we find of \perdtwoone~d,
it might be possible to detect the wobble from TOI-2158~d in future \emph{Gaia} data releases.
Such an astrometric detection would provide the orbital inclination and true mass of TOI-2158~d,
and enable measurement of the mutual inclination between planets b and d.

Another piece of the puzzle could come from probing the atmospheres of these
warm Jupiters,
where, for instance, the C/O ratio might be able to trace 
the dynamical history and evolution \citep[e.g.][]{Madhusudhan2014}.
As discussed in relation to the CHEOPS observations of TOI-5120,
this planet has already been identified
as a potentially interesting target
for the Ariel mission,
although its transmission and emission spectroscopy metrics (TSM and ESM; Table~\ref{tab:pars})
are comparatively modest.
TOI-5699~b is the most favorable target for transmission spectroscopy, 
with a TSM of \tsmfivesix, 
owing primarily to its low mass and consequently large atmospheric scale height. 
In contrast, TOI-2158~b has the highest ESM (\esmtwoone), 
reflecting its higher equilibrium temperature 
and making it the most favorable target of the three for emission spectroscopy.



\section{Conclusion}\label{sec:conc}

We report on the discovery and characterisation of two new transiting warm Jupiters,
TOI-5120~b and TOI-5699~b, 
and refine the parameters of the previously discovered warm Jupiter TOI-2158~b.
All three planets orbit slightly evolved,
metal-rich host stars and have radii comparable to that of Jupiter, 
while spanning a range in masses and orbital architectures. 
In addition to the transiting planet, 
we detect an outer planetary companion in the TOI-2158 system 
and identify a long-term RV trend,
potentially indicating the presence of an additional 
planetary companion on a wide orbit.

The three systems illustrate the diversity of warm Jupiter architectures. 
TOI-5120~b resides in a system with a likely wide stellar companion, 
while TOI-5699~b currently shows no evidence 
for additional companions within the sensitivity of our RV observations. 
The TOI-2158 system, on the other hand, 
hosts multiple companions, 
including the eccentric outer planet TOI-2158~c and a possible additional distant body. 
These configurations highlight the range of dynamical environments in which warm Jupiters are found.

The modest eccentricities measured for TOI-5120~b and TOI-5699~b suggest that they are unlikely to be undergoing high-eccentricity tidal migration at present, although undetected companions could still excite their eccentricities through secular interactions. 
In contrast, the architecture of the TOI-2158 system, 
with an eccentric outer companion and indications of an additional perturber, 
is consistent with a dynamically active environment that could have influenced 
the migration history of the inner planet.

More broadly, these systems add to the growing sample of well-characterised warm Jupiters 
and contribute to population-level studies of their dynamical properties. 
As warm Jupiters are expected to retain signatures of their formation and migration histories, 
expanding this sample is essential for disentangling the relative roles of in situ formation, 
disc migration, 
and high-$e$ tidal migration. 
Continued RV monitoring and measurements of the RM effect 
for these systems will further constrain 
their dynamical architectures and provide additional insight into the origins of warm Jupiters.


\section*{Data availability}
Table \ref{tab:rvs_all} is available in electronic form at the CDS via anonymous ftp to cdsarc.u-strasbg.fr (130.79.128.5) or via \url{http://cdsweb.u-strasbg.fr/cgi-bin/qcat?J/A+A/}.


\begin{acknowledgements}
We thank the anonymous reviewer for valuable comments that helped improve the quality of the manuscript.
This work was supported by a research grant (42101) from VILLUM FONDEN.
DRC acknowledges partial support from NASA Grant 18-2XRP18\_2-0007. 
This paper includes data taken at The McDonald Observatory of The University of Texas at Austin supported by the NASA grant 80NSSC23K0429.
This study is based on observations made with the Nordic Optical Telescope, owned in collaboration by the University of Turku and Aarhus University, and operated jointly by Aarhus University, the University of Turku and the University of Oslo, representing Denmark, Finland and Norway, the University of Iceland and Stockholm University at the Observatorio del Roque de los Muchachos, La Palma, Spain, of the Instituto de Astrof\'{\i}sica de Canarias.
This paper includes data collected with the TESS mission, obtained from the MAST data archive at the Space Telescope Science Institute (STScI). Funding for the TESS mission is provided by the NASA Explorer Program. STScI is operated by the Association of Universities for Research in Astronomy, Inc., under NASA contract NAS 5–26555.
This research has made use of the Exoplanet Follow-up Observation Program (ExoFOP; DOI: 10.26134/ExoFOP5) website, which is operated by the California Institute of Technology, under contract with the National Aeronautics and Space Administration under the Exoplanet Exploration Program.
Based on observations obtained at the Hale Telescope, Palomar Observatory, as part of a collaborative agreement between the Caltech Optical Observatories and the Jet Propulsion Laboratory [operated by Caltech for NASA].
Some of the observations in the paper made use of the NN-EXPLORE Exoplanet and Stellar Speckle Imager (NESSI). NESSI was funded by the NASA Exoplanet Exploration Program and the NASA Ames Research Center. NESSI was built at the Ames Research Center by Steve B. Howell, Nic Scott, Elliott P. Horch, and Emmett Quigley.
This work has made use of data from the European Space Agency (ESA) mission
\emph{Gaia} (\url{https://www.cosmos.esa.int/gaia}), processed by the \emph{Gaia}
Data Processing and Analysis Consortium (DPAC,
\url{https://www.cosmos.esa.int/web/gaia/dpac/consortium}). Funding for the DPAC
has been provided by national institutions, in particular the institutions
participating in the \emph{Gaia} Multilateral Agreement. 
This work has made use of the VALD database, operated at Uppsala University, the Institute of Astronomy RAS in Moscow, and the University of Vienna.
This work makes use of observations from the LCOGT network. Part of the LCOGT telescope time was granted by NOIRLab through the Mid-Scale Innovations Program (MSIP). MSIP is funded by NSF.
The work of BS was conducted under the state assignment of Lomonosov Moscow State University.
T.G. acknowledges financial support from the Agencia Estatal de Investigaci\'on of the Ministerio de Ciencia e Innovaci\'on MCIN/AEI/10.13039/501100011033 and the ERDF “A way of making Europe” through projects PID2021-125627OB-C32 and PID2024-158486OB-C32. This work is supported by the European Union (ERC AdvG SPEAR, GA 101200674).
Funding for KB was provided by the European Union (ERC AdG SUBSTELLAR, GA 101054354).
%
%
This research has made use of the Exoplanet Follow-up Observation Program (ExoFOP; DOI: 10.26134/ExoFOP5) website, which is operated by the California Institute of Technology, under contract with the National Aeronautics and Space Administration under the Exoplanet Exploration Program. 
This research has made use of NASA’s Astrophysics Data System.
The TESS light curves were extracted using Lightkurve \citep{lightkurve}.
This work made use of the following \texttt{Python} packages; NumPy \citep{numpy}, SciPy \citep{scipy}, Matplotlib \citep{matplotlib}, dynesty \citep{Speagle2019}, Nested Sampling \citep{Skilling2004,Skilling2006}, corner \citep{corner}, kepler.py \citep{exoplanet:joss}, and Astropy \citep{astropy:2013, astropy:2018, astropy:2022}.
\end{acknowledgements}


%
%

\bibliographystyle{aa} 
\bibliography{myrefs} 

@ARTICLE{Adams2006,
       author = {{Adams}, Fred C. and {Laughlin}, Gregory},
        title = "{Long-Term Evolution of Close Planets Including the Effects of Secular Interactions}",
      journal = {\apj},
         year = 2006,
        month = oct,
       volume = {649},
       number = {2},
        pages = {1004-1009},
          doi = {10.1086/506145},
archivePrefix = {arXiv},
       eprint = {astro-ph/0606349},
 primaryClass = {astro-ph},
       adsurl = {https://ui.adsabs.harvard.edu/abs/2006ApJ...649.1004A}
}

@ARTICLE{Albrecht2012,
       author = {{Albrecht}, Simon and {Winn}, Joshua N. and {Johnson}, John A. and {Howard}, Andrew W. and {Marcy}, Geoffrey W. and {Butler}, R. Paul and {Arriagada}, Pamela and {Crane}, Jeffrey D. and {Shectman}, Stephen A. and {Thompson}, Ian B. and {Hirano}, Teruyuki and {Bakos}, Gaspar and {Hartman}, Joel D.},
        title = "{Obliquities of Hot Jupiter Host Stars: Evidence for Tidal Interactions and Primordial Misalignments}",
      journal = {\apj},
         year = 2012,
        month = sep,
       volume = {757},
       number = {1},
          eid = {18},
        pages = {18},
          doi = {10.1088/0004-637X/757/1/18},
archivePrefix = {arXiv},
       eprint = {1206.6105},
 primaryClass = {astro-ph.SR},
       adsurl = {https://ui.adsabs.harvard.edu/abs/2012ApJ...757...18A}
}

@ARTICLE{Akeson2013,
       author = {{Akeson}, R.~L. and {Chen}, X. and {Ciardi}, D. and {Crane}, M. and {Good}, J. and {Harbut}, M. and {Jackson}, E. and {Kane}, S.~R. and {Laity}, A.~C. and {Leifer}, S. and {Lynn}, M. and {McElroy}, D.~L. and {Papin}, M. and {Plavchan}, P. and {Ram{\'\i}rez}, S.~V. and {Rey}, R. and {von Braun}, K. and {Wittman}, M. and {Abajian}, M. and {Ali}, B. and {Beichman}, C. and {Beekley}, A. and {Berriman}, G.~B. and {Berukoff}, S. and {Bryden}, G. and {Chan}, B. and {Groom}, S. and {Lau}, C. and {Payne}, A.~N. and {Regelson}, M. and {Saucedo}, M. and {Schmitz}, M. and {Stauffer}, J. and {Wyatt}, P. and {Zhang}, A.},
        title = "{The NASA Exoplanet Archive: Data and Tools for Exoplanet Research}",
      journal = {\pasp},
         year = 2013,
        month = aug,
       volume = {125},
       number = {930},
        pages = {989},
          doi = {10.1086/672273},
archivePrefix = {arXiv},
       eprint = {1307.2944},
 primaryClass = {astro-ph.IM},
       adsurl = {https://ui.adsabs.harvard.edu/abs/2013PASP..125..989A}
}

@ARTICLE{Anderson2017,
       author = {{Anderson}, Kassandra R. and {Lai}, Dong},
        title = "{Moderately eccentric warm Jupiters from secular interactions with exterior companions}",
      journal = {\mnras},
         year = 2017,
        month = dec,
       volume = {472},
       number = {3},
        pages = {3692-3705},
          doi = {10.1093/mnras/stx2250},
archivePrefix = {arXiv},
       eprint = {1706.00084},
 primaryClass = {astro-ph.EP},
       adsurl = {https://ui.adsabs.harvard.edu/abs/2017MNRAS.472.3692A}
}

@ARTICLE{astropy:2013,
       author = {{Astropy Collaboration} and {Robitaille}, Thomas P. and {Tollerud}, Erik J. and {Greenfield}, Perry and {Droettboom}, Michael and {Bray}, Erik and {Aldcroft}, Tom and {Davis}, Matt and {Ginsburg}, Adam and {Price-Whelan}, Adrian M. and {Kerzendorf}, Wolfgang E. and {Conley}, Alexander and {Crighton}, Neil and {Barbary}, Kyle and {Muna}, Demitri and {Ferguson}, Henry and {Grollier}, Fr{\'e}d{\'e}ric and {Parikh}, Madhura M. and {Nair}, Prasanth H. and {Unther}, Hans M. and {Deil}, Christoph and {Woillez}, Julien and {Conseil}, Simon and {Kramer}, Roban and {Turner}, James E.~H. and {Singer}, Leo and {Fox}, Ryan and {Weaver}, Benjamin A. and {Zabalza}, Victor and {Edwards}, Zachary I. and {Azalee Bostroem}, K. and {Burke}, D.~J. and {Casey}, Andrew R. and {Crawford}, Steven M. and {Dencheva}, Nadia and {Ely}, Justin and {Jenness}, Tim and {Labrie}, Kathleen and {Lim}, Pey Lian and {Pierfederici}, Francesco and {Pontzen}, Andrew and {Ptak}, Andy and {Refsdal}, Brian and {Servillat}, Mathieu and {Streicher}, Ole},
        title = "{Astropy: A community Python package for astronomy}",
      journal = {\aap},
         year = 2013,
        month = oct,
       volume = {558},
          eid = {A33},
        pages = {A33},
          doi = {10.1051/0004-6361/201322068},
archivePrefix = {arXiv},
       eprint = {1307.6212},
 primaryClass = {astro-ph.IM},
       adsurl = {https://ui.adsabs.harvard.edu/abs/2013A&A...558A..33A}
}

@ARTICLE{astropy:2018,
       author = {{Astropy Collaboration} and {Price-Whelan}, A.~M. and {Sip{\H{o}}cz}, B.~M. and {G{\"u}nther}, H.~M. and {Lim}, P.~L. and {Crawford}, S.~M. and {Conseil}, S. and {Shupe}, D.~L. and {Craig}, M.~W. and {Dencheva}, N. and {Ginsburg}, A. and {VanderPlas}, J.~T. and {Bradley}, L.~D. and {P{\'e}rez-Su{\'a}rez}, D. and {de Val-Borro}, M. and {Aldcroft}, T.~L. and {Cruz}, K.~L. and {Robitaille}, T.~P. and {Tollerud}, E.~J. and {Ardelean}, C. and {Babej}, T. and {Bach}, Y.~P. and {Bachetti}, M. and {Bakanov}, A.~V. and {Bamford}, S.~P. and {Barentsen}, G. and {Barmby}, P. and {Baumbach}, A. and {Berry}, K.~L. and {Biscani}, F. and {Boquien}, M. and {Bostroem}, K.~A. and {Bouma}, L.~G. and {Brammer}, G.~B. and {Bray}, E.~M. and {Breytenbach}, H. and {Buddelmeijer}, H. and {Burke}, D.~J. and {Calderone}, G. and {Cano Rodr{\'\i}guez}, J.~L. and {Cara}, M. and {Cardoso}, J.~V.~M. and {Cheedella}, S. and {Copin}, Y. and {Corrales}, L. and {Crichton}, D. and {D'Avella}, D. and {Deil}, C. and {Depagne}, {\'E}. and {Dietrich}, J.~P. and {Donath}, A. and {Droettboom}, M. and {Earl}, N. and {Erben}, T. and {Fabbro}, S. and {Ferreira}, L.~A. and {Finethy}, T. and {Fox}, R.~T. and {Garrison}, L.~H. and {Gibbons}, S.~L.~J. and {Goldstein}, D.~A. and {Gommers}, R. and {Greco}, J.~P. and {Greenfield}, P. and {Groener}, A.~M. and {Grollier}, F. and {Hagen}, A. and {Hirst}, P. and {Homeier}, D. and {Horton}, A.~J. and {Hosseinzadeh}, G. and {Hu}, L. and {Hunkeler}, J.~S. and {Ivezi{\'c}}, {\v{Z}}. and {Jain}, A. and {Jenness}, T. and {Kanarek}, G. and {Kendrew}, S. and {Kern}, N.~S. and {Kerzendorf}, W.~E. and {Khvalko}, A. and {King}, J. and {Kirkby}, D. and {Kulkarni}, A.~M. and {Kumar}, A. and {Lee}, A. and {Lenz}, D. and {Littlefair}, S.~P. and {Ma}, Z. and {Macleod}, D.~M. and {Mastropietro}, M. and {McCully}, C. and {Montagnac}, S. and {Morris}, B.~M. and {Mueller}, M. and {Mumford}, S.~J. and {Muna}, D. and {Murphy}, N.~A. and {Nelson}, S. and {Nguyen}, G.~H. and {Ninan}, J.~P. and {N{\"o}the}, M. and {Ogaz}, S. and {Oh}, S. and {Parejko}, J.~K. and {Parley}, N. and {Pascual}, S. and {Patil}, R. and {Patil}, A.~A. and {Plunkett}, A.~L. and {Prochaska}, J.~X. and {Rastogi}, T. and {Reddy Janga}, V. and {Sabater}, J. and {Sakurikar}, P. and {Seifert}, M. and {Sherbert}, L.~E. and {Sherwood-Taylor}, H. and {Shih}, A.~Y. and {Sick}, J. and {Silbiger}, M.~T. and {Singanamalla}, S. and {Singer}, L.~P. and {Sladen}, P.~H. and {Sooley}, K.~A. and {Sornarajah}, S. and {Streicher}, O. and {Teuben}, P. and {Thomas}, S.~W. and {Tremblay}, G.~R. and {Turner}, J.~E.~H. and {Terr{\'o}n}, V. and {van Kerkwijk}, M.~H. and {de la Vega}, A. and {Watkins}, L.~L. and {Weaver}, B.~A. and {Whitmore}, J.~B. and {Woillez}, J. and {Zabalza}, V. and {Astropy Contributors}},
        title = "{The Astropy Project: Building an Open-science Project and Status of the v2.0 Core Package}",
      journal = {\aj},
         year = 2018,
        month = sep,
       volume = {156},
       number = {3},
          eid = {123},
        pages = {123},
          doi = {10.3847/1538-3881/aabc4f},
archivePrefix = {arXiv},
       eprint = {1801.02634},
 primaryClass = {astro-ph.IM},
       adsurl = {https://ui.adsabs.harvard.edu/abs/2018AJ....156..123A}
}

@ARTICLE{astropy:2022,
       author = {{Astropy Collaboration} and {Price-Whelan}, Adrian M. and {Lim}, Pey Lian and {Earl}, Nicholas and {Starkman}, Nathaniel and {Bradley}, Larry and {Shupe}, David L. and {Patil}, Aarya A. and {Corrales}, Lia and {Brasseur}, C.~E. and {N{\"o}the}, Maximilian and {Donath}, Axel and {Tollerud}, Erik and {Morris}, Brett M. and {Ginsburg}, Adam and {Vaher}, Eero and {Weaver}, Benjamin A. and {Tocknell}, James and {Jamieson}, William and {van Kerkwijk}, Marten H. and {Robitaille}, Thomas P. and {Merry}, Bruce and {Bachetti}, Matteo and {G{\"u}nther}, H. Moritz and {Aldcroft}, Thomas L. and {Alvarado-Montes}, Jaime A. and {Archibald}, Anne M. and {B{\'o}di}, Attila and {Bapat}, Shreyas and {Barentsen}, Geert and {Baz{\'a}n}, Juanjo and {Biswas}, Manish and {Boquien}, M{\'e}d{\'e}ric and {Burke}, D.~J. and {Cara}, Daria and {Cara}, Mihai and {Conroy}, Kyle E. and {Conseil}, Simon and {Craig}, Matthew W. and {Cross}, Robert M. and {Cruz}, Kelle L. and {D'Eugenio}, Francesco and {Dencheva}, Nadia and {Devillepoix}, Hadrien A.~R. and {Dietrich}, J{\"o}rg P. and {Eigenbrot}, Arthur Davis and {Erben}, Thomas and {Ferreira}, Leonardo and {Foreman-Mackey}, Daniel and {Fox}, Ryan and {Freij}, Nabil and {Garg}, Suyog and {Geda}, Robel and {Glattly}, Lauren and {Gondhalekar}, Yash and {Gordon}, Karl D. and {Grant}, David and {Greenfield}, Perry and {Groener}, Austen M. and {Guest}, Steve and {Gurovich}, Sebastian and {Handberg}, Rasmus and {Hart}, Akeem and {Hatfield-Dodds}, Zac and {Homeier}, Derek and {Hosseinzadeh}, Griffin and {Jenness}, Tim and {Jones}, Craig K. and {Joseph}, Prajwel and {Kalmbach}, J. Bryce and {Karamehmetoglu}, Emir and {Ka{\l}uszy{\'n}ski}, Miko{\l}aj and {Kelley}, Michael S.~P. and {Kern}, Nicholas and {Kerzendorf}, Wolfgang E. and {Koch}, Eric W. and {Kulumani}, Shankar and {Lee}, Antony and {Ly}, Chun and {Ma}, Zhiyuan and {MacBride}, Conor and {Maljaars}, Jakob M. and {Muna}, Demitri and {Murphy}, N.~A. and {Norman}, Henrik and {O'Steen}, Richard and {Oman}, Kyle A. and {Pacifici}, Camilla and {Pascual}, Sergio and {Pascual-Granado}, J. and {Patil}, Rohit R. and {Perren}, Gabriel I. and {Pickering}, Timothy E. and {Rastogi}, Tanuj and {Roulston}, Benjamin R. and {Ryan}, Daniel F. and {Rykoff}, Eli S. and {Sabater}, Jose and {Sakurikar}, Parikshit and {Salgado}, Jes{\'u}s and {Sanghi}, Aniket and {Saunders}, Nicholas and {Savchenko}, Volodymyr and {Schwardt}, Ludwig and {Seifert-Eckert}, Michael and {Shih}, Albert Y. and {Jain}, Anany Shrey and {Shukla}, Gyanendra and {Sick}, Jonathan and {Simpson}, Chris and {Singanamalla}, Sudheesh and {Singer}, Leo P. and {Singhal}, Jaladh and {Sinha}, Manodeep and {Sip{\H{o}}cz}, Brigitta M. and {Spitler}, Lee R. and {Stansby}, David and {Streicher}, Ole and {{\v{S}}umak}, Jani and {Swinbank}, John D. and {Taranu}, Dan S. and {Tewary}, Nikita and {Tremblay}, Grant R. and {de Val-Borro}, Miguel and {Van Kooten}, Samuel J. and {Vasovi{\'c}}, Zlatan and {Verma}, Shresth and {de Miranda Cardoso}, Jos{\'e} Vin{\'\i}cius and {Williams}, Peter K.~G. and {Wilson}, Tom J. and {Winkel}, Benjamin and {Wood-Vasey}, W.~M. and {Xue}, Rui and {Yoachim}, Peter and {Zhang}, Chen and {Zonca}, Andrea and {Astropy Project Contributors}},
        title = "{The Astropy Project: Sustaining and Growing a Community-oriented Open-source Project and the Latest Major Release (v5.0) of the Core Package}",
      journal = {\apj},
         year = 2022,
        month = aug,
       volume = {935},
       number = {2},
          eid = {167},
        pages = {167},
          doi = {10.3847/1538-4357/ac7c74},
archivePrefix = {arXiv},
       eprint = {2206.14220},
 primaryClass = {astro-ph.IM},
       adsurl = {https://ui.adsabs.harvard.edu/abs/2022ApJ...935..167A}
}

@ARTICLE{BASTA2015,
       author = {{Silva Aguirre}, V. and {Davies}, G.~R. and {Basu}, S. and {Christensen-Dalsgaard}, J. and {Creevey}, O. and {Metcalfe}, T.~S. and {Bedding}, T.~R. and {Casagrande}, L. and {Handberg}, R. and {Lund}, M.~N. and {Nissen}, P.~E. and {Chaplin}, W.~J. and {Huber}, D. and {Serenelli}, A.~M. and {Stello}, D. and {Van Eylen}, V. and {Campante}, T.~L. and {Elsworth}, Y. and {Gilliland}, R.~L. and {Hekker}, S. and {Karoff}, C. and {Kawaler}, S.~D. and {Kjeldsen}, H. and {Lundkvist}, M.~S.},
        title = "{Ages and fundamental properties of Kepler exoplanet host stars from asteroseismology}",
      journal = {\mnras},
         year = 2015,
        month = sep,
       volume = {452},
       number = {2},
        pages = {2127-2148},
          doi = {10.1093/mnras/stv1388},
archivePrefix = {arXiv},
       eprint = {1504.07992},
 primaryClass = {astro-ph.SR},
       adsurl = {https://ui.adsabs.harvard.edu/abs/2015MNRAS.452.2127S}
}

@ARTICLE{BASTA2022,
       author = {{Aguirre B{\o}rsen-Koch}, V. and {R{\o}rsted}, J.~L. and {Justesen}, A.~B. and {Stokholm}, A. and {Verma}, K. and {Winther}, M.~L. and {Knudstrup}, E. and {Nielsen}, K.~B. and {Sahlholdt}, C. and {Larsen}, J.~R. and {Cassisi}, S. and {Serenelli}, A.~M. and {Casagrande}, L. and {Christensen-Dalsgaard}, J. and {Davies}, G.~R. and {Ferguson}, J.~W. and {Lund}, M.~N. and {Weiss}, A. and {White}, T.~R.},
        title = "{The BAyesian STellar algorithm (BASTA): a fitting tool for stellar studies, asteroseismology, exoplanets, and Galactic archaeology}",
      journal = {\mnras},
         year = 2022,
        month = jan,
       volume = {509},
       number = {3},
        pages = {4344-4364},
          doi = {10.1093/mnras/stab2911},
archivePrefix = {arXiv},
       eprint = {2109.14622},
 primaryClass = {astro-ph.SR},
       adsurl = {https://ui.adsabs.harvard.edu/abs/2022MNRAS.509.4344A}
}

@ARTICLE{BaSTI2018,
       author = {{Hidalgo}, Sebastian L. and {Pietrinferni}, Adriano and {Cassisi}, Santi and {Salaris}, Maurizio and {Mucciarelli}, Alessio and {Savino}, Alessandro and {Aparicio}, Antonio and {Silva Aguirre}, Victor and {Verma}, Kuldeep},
        title = "{The Updated BaSTI Stellar Evolution Models and Isochrones. I. Solar-scaled Calculations}",
      journal = {\apj},
         year = 2018,
        month = apr,
       volume = {856},
       number = {2},
          eid = {125},
        pages = {125},
          doi = {10.3847/1538-4357/aab158},
archivePrefix = {arXiv},
       eprint = {1802.07319},
 primaryClass = {astro-ph.GA},
       adsurl = {https://ui.adsabs.harvard.edu/abs/2018ApJ...856..125H}
}

@ARTICLE{Benz2021,
       author = {{Benz}, W. and {Broeg}, C. and {Fortier}, A. and {Rando}, N. and {Beck}, T. and {Beck}, M. and {Queloz}, D. and {Ehrenreich}, D. and {Maxted}, P.~F.~L. and {Isaak}, K.~G. and {Billot}, N. and {Alibert}, Y. and {Alonso}, R. and {Ant{\'o}nio}, C. and {Asquier}, J. and {Bandy}, T. and {B{\'a}rczy}, T. and {Barrado}, D. and {Barros}, S.~C.~C. and {Baumjohann}, W. and {Bekkelien}, A. and {Bergomi}, M. and {Biondi}, F. and {Bonfils}, X. and {Borsato}, L. and {Brandeker}, A. and {Busch}, M. -D. and {Cabrera}, J. and {Cessa}, V. and {Charnoz}, S. and {Chazelas}, B. and {Collier Cameron}, A. and {Corral Van Damme}, C. and {Cortes}, D. and {Davies}, M.~B. and {Deleuil}, M. and {Deline}, A. and {Delrez}, L. and {Demangeon}, O. and {Demory}, B.~O. and {Erikson}, A. and {Farinato}, J. and {Fossati}, L. and {Fridlund}, M. and {Futyan}, D. and {Gandolfi}, D. and {Garcia Munoz}, A. and {Gillon}, M. and {Guterman}, P. and {Gutierrez}, A. and {Hasiba}, J. and {Heng}, K. and {Hernandez}, E. and {Hoyer}, S. and {Kiss}, L.~L. and {Kovacs}, Z. and {Kuntzer}, T. and {Laskar}, J. and {Lecavelier des Etangs}, A. and {Lendl}, M. and {L{\'o}pez}, A. and {Lora}, I. and {Lovis}, C. and {L{\"u}ftinger}, T. and {Magrin}, D. and {Malvasio}, L. and {Marafatto}, L. and {Michaelis}, H. and {de Miguel}, D. and {Modrego}, D. and {Munari}, M. and {Nascimbeni}, V. and {Olofsson}, G. and {Ottacher}, H. and {Ottensamer}, R. and {Pagano}, I. and {Palacios}, R. and {Pall{\'e}}, E. and {Peter}, G. and {Piazza}, D. and {Piotto}, G. and {Pizarro}, A. and {Pollaco}, D. and {Ragazzoni}, R. and {Ratti}, F. and {Rauer}, H. and {Ribas}, I. and {Rieder}, M. and {Rohlfs}, R. and {Safa}, F. and {Salatti}, M. and {Santos}, N.~C. and {Scandariato}, G. and {S{\'e}gransan}, D. and {Simon}, A.~E. and {Smith}, A.~M.~S. and {Sordet}, M. and {Sousa}, S.~G. and {Steller}, M. and {Szab{\'o}}, G.~M. and {Szoke}, J. and {Thomas}, N. and {Tschentscher}, M. and {Udry}, S. and {Van Grootel}, V. and {Viotto}, V. and {Walter}, I. and {Walton}, N.~A. and {Wildi}, F. and {Wolter}, D.},
        title = "{The CHEOPS mission}",
      journal = {Experimental Astronomy},
         year = 2021,
        month = feb,
       volume = {51},
       number = {1},
        pages = {109-151},
          doi = {10.1007/s10686-020-09679-4},
archivePrefix = {arXiv},
       eprint = {2009.11633},
 primaryClass = {astro-ph.IM},
       adsurl = {https://ui.adsabs.harvard.edu/abs/2021ExA....51..109B}
}

@ARTICLE{bennett2019,
       author = {{Bennett}, Morgan and {Bovy}, Jo},
        title = "{Vertical waves in the solar neighbourhood in Gaia DR2}",
      journal = {\mnras},
         year = 2019,
        month = jan,
       volume = {482},
       number = {1},
        pages = {1417-1425},
          doi = {10.1093/mnras/sty2813},
archivePrefix = {arXiv},
       eprint = {1809.03507},
 primaryClass = {astro-ph.GA},
       adsurl = {https://ui.adsabs.harvard.edu/abs/2019MNRAS.482.1417B}
}

@ARTICLE{Blanco2014,
       author = {{Blanco-Cuaresma}, S. and {Soubiran}, C. and {Heiter}, U. and {Jofr{\'e}}, P.},
        title = "{Determining stellar atmospheric parameters and chemical abundances of FGK stars with iSpec}",
      journal = {\aap},
         year = 2014,
        month = sep,
       volume = {569},
          eid = {A111},
        pages = {A111},
          doi = {10.1051/0004-6361/201423945},
archivePrefix = {arXiv},
       eprint = {1407.2608},
 primaryClass = {astro-ph.IM},
       adsurl = {https://ui.adsabs.harvard.edu/abs/2014A&A...569A.111B}
}

@ARTICLE{Bonomo2017,
       author = {{Bonomo}, A.~S. and {Desidera}, S. and {Benatti}, S. and {Borsa}, F. and {Crespi}, S. and {Damasso}, M. and {Lanza}, A.~F. and {Sozzetti}, A. and {Lodato}, G. and {Marzari}, F. and {Boccato}, C. and {Claudi}, R.~U. and {Cosentino}, R. and {Covino}, E. and {Gratton}, R. and {Maggio}, A. and {Micela}, G. and {Molinari}, E. and {Pagano}, I. and {Piotto}, G. and {Poretti}, E. and {Smareglia}, R. and {Affer}, L. and {Biazzo}, K. and {Bignamini}, A. and {Esposito}, M. and {Giacobbe}, P. and {H{\'e}brard}, G. and {Malavolta}, L. and {Maldonado}, J. and {Mancini}, L. and {Martinez Fiorenzano}, A. and {Masiero}, S. and {Nascimbeni}, V. and {Pedani}, M. and {Rainer}, M. and {Scandariato}, G.},
        title = "{The GAPS Programme with HARPS-N at TNG . XIV. Investigating giant planet migration history via improved eccentricity and mass determination for 231 transiting planets}",
      journal = {\aap},
         year = 2017,
        month = jun,
       volume = {602},
          eid = {A107},
        pages = {A107},
          doi = {10.1051/0004-6361/201629882},
archivePrefix = {arXiv},
       eprint = {1704.00373},
 primaryClass = {astro-ph.EP},
       adsurl = {https://ui.adsabs.harvard.edu/abs/2017A&A...602A.107B}
}

@ARTICLE{Bonomo2023,
       author = {{Bonomo}, A.~S. and {Dumusque}, X. and {Massa}, A. and {Mortier}, A. and {Bongiolatti}, R. and {Malavolta}, L. and {Sozzetti}, A. and {Buchhave}, L.~A. and {Damasso}, M. and {Haywood}, R.~D. and {Morbidelli}, A. and {Latham}, D.~W. and {Molinari}, E. and {Pepe}, F. and {Poretti}, E. and {Udry}, S. and {Affer}, L. and {Boschin}, W. and {Charbonneau}, D. and {Cosentino}, R. and {Cretignier}, M. and {Ghedina}, A. and {Lega}, E. and {L{\'o}pez-Morales}, M. and {Margini}, M. and {Mart{\'\i}nez Fiorenzano}, A.~F. and {Mayor}, M. and {Micela}, G. and {Pedani}, M. and {Pinamonti}, M. and {Rice}, K. and {Sasselov}, D. and {Tronsgaard}, R. and {Vanderburg}, A.},
        title = "{Cold Jupiters and improved masses in 38 Kepler and K2 small planet systems from 3661 HARPS-N radial velocities. No excess of cold Jupiters in small planet systems}",
      journal = {\aap},
         year = 2023,
        month = sep,
       volume = {677},
          eid = {A33},
        pages = {A33},
          doi = {10.1051/0004-6361/202346211},
archivePrefix = {arXiv},
       eprint = {2304.05773},
 primaryClass = {astro-ph.EP},
       adsurl = {https://ui.adsabs.harvard.edu/abs/2023A&A...677A..33B}
}

@ARTICLE{Brown2013,
       author = {{Brown}, T.~M. and {Baliber}, N. and {Bianco}, F.~B. and {Bowman}, M. and {Burleson}, B. and {Conway}, P. and {Crellin}, M. and {Depagne}, {\'E}. and {De Vera}, J. and {Dilday}, B. and {Dragomir}, D. and {Dubberley}, M. and {Eastman}, J.~D. and {Elphick}, M. and {Falarski}, M. and {Foale}, S. and {Ford}, M. and {Fulton}, B.~J. and {Garza}, J. and {Gomez}, E.~L. and {Graham}, M. and {Greene}, R. and {Haldeman}, B. and {Hawkins}, E. and {Haworth}, B. and {Haynes}, R. and {Hidas}, M. and {Hjelstrom}, A.~E. and {Howell}, D.~A. and {Hygelund}, J. and {Lister}, T.~A. and {Lobdill}, R. and {Martinez}, J. and {Mullins}, D.~S. and {Norbury}, M. and {Parrent}, J. and {Paulson}, R. and {Petry}, D.~L. and {Pickles}, A. and {Posner}, V. and {Rosing}, W.~E. and {Ross}, R. and {Sand}, D.~J. and {Saunders}, E.~S. and {Shobbrook}, J. and {Shporer}, A. and {Street}, R.~A. and {Thomas}, D. and {Tsapras}, Y. and {Tufts}, J.~R. and {Valenti}, S. and {Vander Horst}, K. and {Walker}, Z. and {White}, G. and {Willis}, M.},
        title = "{Las Cumbres Observatory Global Telescope Network}",
      journal = {\pasp},
         year = 2013,
        month = sep,
       volume = {125},
       number = {931},
        pages = {1031},
          doi = {10.1086/673168},
archivePrefix = {arXiv},
       eprint = {1305.2437},
 primaryClass = {astro-ph.IM},
       adsurl = {https://ui.adsabs.harvard.edu/abs/2013PASP..125.1031B}
}

@ARTICLE{Bruntt2010,
       author = {{Bruntt}, H. and {Bedding}, T.~R. and {Quirion}, P. -O. and {Lo Curto}, G. and {Carrier}, F. and {Smalley}, B. and {Dall}, T.~H. and {Arentoft}, T. and {Bazot}, M. and {Butler}, R.~P.},
        title = "{Accurate fundamental parameters for 23 bright solar-type stars}",
      journal = {\mnras},
         year = 2010,
        month = jul,
       volume = {405},
       number = {3},
        pages = {1907-1923},
          doi = {10.1111/j.1365-2966.2010.16575.x},
archivePrefix = {arXiv},
       eprint = {1002.4268},
 primaryClass = {astro-ph.SR},
       adsurl = {https://ui.adsabs.harvard.edu/abs/2010MNRAS.405.1907B}
}

@ARTICLE{Bryan2019,
       author = {{Bryan}, Marta L. and {Knutson}, Heather A. and {Lee}, Eve J. and {Fulton}, B.~J. and {Batygin}, Konstantin and {Ngo}, Henry and {Meshkat}, Tiffany},
        title = "{An Excess of Jupiter Analogs in Super-Earth Systems}",
      journal = {\aj},
         year = 2019,
        month = feb,
       volume = {157},
       number = {2},
          eid = {52},
        pages = {52},
          doi = {10.3847/1538-3881/aaf57f},
archivePrefix = {arXiv},
       eprint = {1806.08799},
 primaryClass = {astro-ph.EP},
       adsurl = {https://ui.adsabs.harvard.edu/abs/2019AJ....157...52B}
}

@ARTICLE{Buchhave2012,
       author = {{Buchhave}, Lars A. and {Latham}, David W. and {Johansen}, Anders and {Bizzarro}, Martin and {Torres}, Guillermo and {Rowe}, Jason F. and {Batalha}, Natalie M. and {Borucki}, William J. and {Brugamyer}, Erik and {Caldwell}, Caroline and {Bryson}, Stephen T. and {Ciardi}, David R. and {Cochran}, William D. and {Endl}, Michael and {Esquerdo}, Gilbert A. and {Ford}, Eric B. and {Geary}, John C. and {Gilliland}, Ronald L. and {Hansen}, Terese and {Isaacson}, Howard and {Laird}, John B. and {Lucas}, Philip W. and {Marcy}, Geoffrey W. and {Morse}, Jon A. and {Robertson}, Paul and {Shporer}, Avi and {Stefanik}, Robert P. and {Still}, Martin and {Quinn}, Samuel N.},
        title = "{An abundance of small exoplanets around stars with a wide range of metallicities}",
      journal = {\nat},
         year = 2012,
        month = jun,
       volume = {486},
       number = {7403},
        pages = {375-377},
          doi = {10.1038/nature11121},
       adsurl = {https://ui.adsabs.harvard.edu/abs/2012Natur.486..375B}
}

@ARTICLE{Ciardi2015,
       author = {{Ciardi}, David R. and {Beichman}, Charles A. and {Horch}, Elliott P. and {Howell}, Steve B.},
        title = "{Understanding the Effects of Stellar Multiplicity on the Derived Planet Radii from Transit Surveys: Implications for Kepler, K2, and TESS}",
      journal = {\apj},
         year = 2015,
        month = may,
       volume = {805},
       number = {1},
          eid = {16},
        pages = {16},
          doi = {10.1088/0004-637X/805/1/16},
archivePrefix = {arXiv},
       eprint = {1503.03516},
 primaryClass = {astro-ph.EP},
       adsurl = {https://ui.adsabs.harvard.edu/abs/2015ApJ...805...16C}
}

@ARTICLE{celerite,
       author = {{Foreman-Mackey}, Daniel and {Agol}, Eric and {Ambikasaran}, Sivaram and {Angus}, Ruth},
        title = "{Fast and Scalable Gaussian Process Modeling with Applications to Astronomical Time Series}",
      journal = {\aj},
         year = 2017,
        month = dec,
       volume = {154},
       number = {6},
          eid = {220},
        pages = {220},
          doi = {10.3847/1538-3881/aa9332},
archivePrefix = {arXiv},
       eprint = {1703.09710},
 primaryClass = {astro-ph.IM},
       adsurl = {https://ui.adsabs.harvard.edu/abs/2017AJ....154..220F}
}

@ARTICLE{Claret2013,
       author = {{Claret}, A. and {Hauschildt}, P.~H. and {Witte}, S.},
        title = "{New limb-darkening coefficients for Phoenix/1d model atmospheres. II. Calculations for 5000 K {\ensuremath{\leq}} T$_{eff}$ {\ensuremath{\leq}} 10 000 K Kepler, CoRot, Spitzer, uvby, UBVRIJHK, Sloan, and 2MASS photometric systems}",
      journal = {\aap},
         year = 2013,
        month = apr,
       volume = {552},
          eid = {A16},
        pages = {A16},
          doi = {10.1051/0004-6361/201220942},
       adsurl = {https://ui.adsabs.harvard.edu/abs/2013A&A...552A..16C}
}

@ARTICLE{Claret2018,
       author = {{Claret}, Antonio},
        title = "{A new method to compute limb-darkening coefficients for stellar atmosphere models with spherical symmetry: the space missions TESS, Kepler, CoRoT, and MOST}",
      journal = {\aap},
         year = 2018,
        month = oct,
       volume = {618},
          eid = {A20},
        pages = {A20},
          doi = {10.1051/0004-6361/201833060},
archivePrefix = {arXiv},
       eprint = {1804.10135},
 primaryClass = {astro-ph.SR},
       adsurl = {https://ui.adsabs.harvard.edu/abs/2018A&A...618A..20C}
}

@ARTICLE{Christiansen2025,
       author = {{Christiansen}, Jessie L. and {McElroy}, Douglas L. and {Harbut}, Marcy and {Ciardi}, David R. and {Crane}, Megan and {Good}, John and {Hardegree-Ullman}, Kevin K. and {Kesseli}, Aurora Y. and {Lund}, Michael B. and {Lynn}, Meca and {Muthiar}, Ananda and {Nilsson}, Ricky and {Oluyide}, Toba and {Papin}, Michael and {Rivera}, Amalia and {Swain}, Melanie and {Susemiehl}, Nicholas D. and {Tam}, Raymond and {van Eyken}, Julian and {Beichman}, Charles},
        title = "{The NASA Exoplanet Archive and Exoplanet Follow-up Observing Program: Data, Tools, and Usage}",
      journal = {The Planetary Science Journal},
         year = 2025,
        month = aug,
       volume = {6},
       number = {8},
          eid = {186},
        pages = {186},
          doi = {10.3847/PSJ/ade3c2},
archivePrefix = {arXiv},
       eprint = {2506.03299},
 primaryClass = {astro-ph.EP},
       adsurl = {https://ui.adsabs.harvard.edu/abs/2025PSJ.....6..186C}
}

@ARTICLE{Collins2017,
       author = {{Collins}, Karen A. and {Kielkopf}, John F. and {Stassun}, Keivan G. and {Hessman}, Frederic V.},
        title = "{AstroImageJ: Image Processing and Photometric Extraction for Ultra-precise Astronomical Light Curves}",
      journal = {\aj},
         year = 2017,
        month = feb,
       volume = {153},
       number = {2},
          eid = {77},
        pages = {77},
          doi = {10.3847/1538-3881/153/2/77},
archivePrefix = {arXiv},
       eprint = {1701.04817},
 primaryClass = {astro-ph.IM},
       adsurl = {https://ui.adsabs.harvard.edu/abs/2017AJ....153...77C}
}

@INPROCEEDINGS{Collins2019,
       author = {{Collins}, Karen},
        title = "{TESS Follow-up Observing Program Working Group (TFOP WG) Sub Group 1 (SG1): Ground-based Time-series Photometry}",
    booktitle = {American Astronomical Society Meeting Abstracts \#233},
         year = 2019,
       series = {American Astronomical Society Meeting Abstracts},
       volume = {233},
        month = jan,
          eid = {140.05},
        pages = {140.05},
       adsurl = {https://ui.adsabs.harvard.edu/abs/2019AAS...23314005C}
}

@article{corner,
      doi = {10.21105/joss.00024},
      url = {https://doi.org/10.21105/joss.00024},
      year  = {2016},
      month = {jun},
      publisher = {The Open Journal},
      volume = {1},
      number = {2},
      pages = {24},
      author = {Daniel Foreman-Mackey},
      title = {corner.py: Scatterplot matrices in Python},
      journal = {The Journal of Open Source Software}
}

@ARTICLE{Dawson2018,
       author = {{Dawson}, Rebekah I. and {Johnson}, John Asher},
        title = "{Origins of Hot Jupiters}",
      journal = {\araa},
         year = 2018,
        month = sep,
       volume = {56},
        pages = {175-221},
          doi = {10.1146/annurev-astro-081817-051853},
archivePrefix = {arXiv},
       eprint = {1801.06117},
 primaryClass = {astro-ph.EP},
       adsurl = {https://ui.adsabs.harvard.edu/abs/2018ARA&A..56..175D}
}

@ARTICLE{Dekany2013,
       author = {{Dekany}, Richard and {Roberts}, Jennifer and {Burruss}, Rick and {Bouchez}, Antonin and {Truong}, Tuan and {Baranec}, Christoph and {Guiwits}, Stephen and {Hale}, David and {Angione}, John and {Trinh}, Thang and {Zolkower}, Jeffry and {Shelton}, J. Christopher and {Palmer}, Dean and {Henning}, John and {Croner}, Ernest and {Troy}, Mitchell and {McKenna}, Dan and {Tesch}, Jonathan and {Hildebrandt}, Sergi and {Milburn}, Jennifer},
        title = "{PALM-3000: Exoplanet Adaptive Optics for the 5 m Hale Telescope}",
      journal = {\apj},
         year = 2013,
        month = oct,
       volume = {776},
       number = {2},
          eid = {130},
        pages = {130},
          doi = {10.1088/0004-637X/776/2/130},
archivePrefix = {arXiv},
       eprint = {1309.1216},
 primaryClass = {astro-ph.IM},
       adsurl = {https://ui.adsabs.harvard.edu/abs/2013ApJ...776..130D}
}

@INPROCEEDINGS{Djupvik2010,
       author = {{Djupvik}, Anlaug Amanda and {Andersen}, Johannes},
        title = "{The Nordic Optical Telescope}",
    booktitle = {Highlights of Spanish Astrophysics V},
         year = 2010,
       series = {Astrophysics and Space Science Proceedings},
       volume = {14},
        month = jan,
        pages = {211},
          doi = {10.1007/978-3-642-11250-8_21},
archivePrefix = {arXiv},
       eprint = {0901.4015},
 primaryClass = {astro-ph.IM},
       adsurl = {https://ui.adsabs.harvard.edu/abs/2010ASSP...14..211D}
}

@ARTICLE{Dong2014,
       author = {{Dong}, Subo and {Katz}, Boaz and {Socrates}, Aristotle},
        title = "{Warm Jupiters Need Close ``Friends'' for High-eccentricity Migration{\textemdash}a Stringent Upper Limit on the Perturber's Separation}",
      journal = {\apjl},
         year = 2014,
        month = jan,
       volume = {781},
       number = {1},
          eid = {L5},
        pages = {L5},
          doi = {10.1088/2041-8205/781/1/L5},
archivePrefix = {arXiv},
       eprint = {1309.0011},
 primaryClass = {astro-ph.EP},
       adsurl = {https://ui.adsabs.harvard.edu/abs/2014ApJ...781L...5D}
}

@ARTICLE{Dong2021b,
       author = {{Dong}, Jiayin and {Huang}, Chelsea X. and {Dawson}, Rebekah I. and {Foreman-Mackey}, Daniel and {Collins}, Karen A. and {Quinn}, Samuel N. and {Lissauer}, Jack J. and {Beatty}, Thomas and {Quarles}, Billy and {Sha}, Lizhou and {Shporer}, Avi and {Guo}, Zhao and {Kane}, Stephen R. and {Abe}, Lyu and {Barkaoui}, Khalid and {Benkhaldoun}, Zouhair and {Brahm}, Rafael and {Bouchy}, Fran{\c{c}}ois and {Carmichael}, Theron W. and {Collins}, Kevin I. and {Conti}, Dennis M. and {Crouzet}, Nicolas and {Dransfield}, Georgina and {Evans}, Phil and {Gan}, Tianjun and {Ghachoui}, Mourad and {Gillon}, Micha{\"e}l and {Grieves}, Nolan and {Guillot}, Tristan and {Hellier}, Coel and {Jehin}, Emmanu{\"e}l and {Jensen}, Eric L.~N. and {Jord{\'a}n}, Andres and {Kamler}, Jacob and {Kielkopf}, John F. and {M{\'e}karnia}, Djamel and {Nielsen}, Louise D. and {Pozuelos}, Francisco J. and {Radford}, Don J. and {Schmider}, Fran{\c{c}}ois-Xavier and {Schwarz}, Richard P. and {Stockdale}, Chris and {Tan}, Thiam-Guan and {Timmermans}, Mathilde and {Triaud}, Amaury H.~M.~J. and {Wang}, Gavin and {Ricker}, George and {Vanderspek}, Roland and {Latham}, David W. and {Seager}, Sara and {Winn}, Joshua N. and {Jenkins}, Jon M. and {Mireles}, Ismael and {Yahalomi}, Daniel A. and {Morgan}, Edward H. and {Vezie}, Michael and {Quintana}, Elisa V. and {Rose}, Mark E. and {Smith}, Jeffrey C. and {Shiao}, Bernie},
        title = "{Warm Jupiters in TESS Full-frame Images: A Catalog and Observed Eccentricity Distribution for Year 1}",
      journal = {\apjs},
         year = 2021,
        month = jul,
       volume = {255},
       number = {1},
          eid = {6},
        pages = {6},
          doi = {10.3847/1538-4365/abf73c},
archivePrefix = {arXiv},
       eprint = {2104.01970},
 primaryClass = {astro-ph.EP},
       adsurl = {https://ui.adsabs.harvard.edu/abs/2021ApJS..255....6D}
}

@ARTICLE{Dong2021,
       author = {{Dong}, Jiayin and {Huang}, Chelsea X. and {Zhou}, George and {Dawson}, Rebekah I. and {Rodriguez}, Joseph E. and {Eastman}, Jason D. and {Collins}, Karen A. and {Quinn}, Samuel N. and {Shporer}, Avi and {Triaud}, Amaury H.~M.~J. and {Wang}, Songhu and {Beatty}, Thomas and {Jackson}, Jonathon M. and {Collins}, Kevin I. and {Abe}, Lyu and {Suarez}, Olga and {Crouzet}, Nicolas and {M{\'e}karnia}, Djamel and {Dransfield}, Georgina and {Jensen}, Eric L.~N. and {Stockdale}, Chris and {Barkaoui}, Khalid and {Heitzmann}, Alexis and {Wright}, Duncan J. and {Addison}, Brett C. and {Wittenmyer}, Robert A. and {Okumura}, Jack and {Bowler}, Brendan P. and {Horner}, Jonathan and {Kane}, Stephen R. and {Kielkopf}, John and {Liu}, Huigen and {Plavchan}, Peter and {Mengel}, Matthew W. and {Ricker}, George R. and {Vanderspek}, Roland and {Latham}, David W. and {Seager}, S. and {Winn}, Joshua N. and {Jenkins}, Jon M. and {Christiansen}, Jessie L. and {Paegert}, Martin},
        title = "{TOI-3362b: A Proto Hot Jupiter Undergoing High-eccentricity Tidal Migration}",
      journal = {\apjl},
         year = 2021,
        month = oct,
       volume = {920},
       number = {1},
          eid = {L16},
        pages = {L16},
          doi = {10.3847/2041-8213/ac2600},
archivePrefix = {arXiv},
       eprint = {2109.03771},
 primaryClass = {astro-ph.EP},
       adsurl = {https://ui.adsabs.harvard.edu/abs/2021ApJ...920L..16D}
}

@ARTICLE{Doyle2014,
       author = {{Doyle}, Amanda P. and {Davies}, Guy R. and {Smalley}, Barry and {Chaplin}, William J. and {Elsworth}, Yvonne},
        title = "{Determining stellar macroturbulence using asteroseismic rotational velocities from Kepler}",
      journal = {\mnras},
         year = 2014,
        month = nov,
       volume = {444},
       number = {4},
        pages = {3592-3602},
          doi = {10.1093/mnras/stu1692},
archivePrefix = {arXiv},
       eprint = {1408.3988},
 primaryClass = {astro-ph.SR},
       adsurl = {https://ui.adsabs.harvard.edu/abs/2014MNRAS.444.3592D}
}

@ARTICLE{Duffell2015,
       author = {{Duffell}, Paul C. and {Chiang}, Eugene},
        title = "{Eccentric Jupiters via Disk-Planet Interactions}",
      journal = {\apj},
         year = 2015,
        month = oct,
       volume = {812},
       number = {2},
          eid = {94},
        pages = {94},
          doi = {10.1088/0004-637X/812/2/94},
archivePrefix = {arXiv},
       eprint = {1507.08667},
 primaryClass = {astro-ph.EP},
       adsurl = {https://ui.adsabs.harvard.edu/abs/2015ApJ...812...94D}
}

@ARTICLE{ElBadry2021,
       author = {{El-Badry}, Kareem and {Rix}, Hans-Walter and {Heintz}, Tyler M.},
        title = "{A million binaries from Gaia eDR3: sample selection and validation of Gaia parallax uncertainties}",
      journal = {\mnras},
         year = 2021,
        month = sep,
       volume = {506},
       number = {2},
        pages = {2269-2295},
          doi = {10.1093/mnras/stab323},
archivePrefix = {arXiv},
       eprint = {2101.05282},
 primaryClass = {astro-ph.SR},
       adsurl = {https://ui.adsabs.harvard.edu/abs/2021MNRAS.506.2269E}
}

@ARTICLE{emcee,
       author = {{Foreman-Mackey}, Daniel and {Hogg}, David W. and {Lang}, Dustin and {Goodman}, Jonathan},
        title = "{emcee: The MCMC Hammer}",
      journal = {\pasp},
         year = 2013,
        month = mar,
       volume = {125},
       number = {925},
        pages = {306},
          doi = {10.1086/670067},
archivePrefix = {arXiv},
       eprint = {1202.3665},
 primaryClass = {astro-ph.IM},
       adsurl = {https://ui.adsabs.harvard.edu/abs/2013PASP..125..306F}
}

@ARTICLE{Endl2000,
       author = {{Endl}, M. and {K{\"u}rster}, M. and {Els}, S.},
        title = "{The planet search program at the ESO Coud{\'e} Echelle spectrometer. I. Data modeling technique and radial velocity precision tests}",
      journal = {\aap},
         year = 2000,
        month = oct,
       volume = {362},
        pages = {585-594},
       adsurl = {https://ui.adsabs.harvard.edu/abs/2000A&A...362..585E}
}

@ARTICLE{exoplanet:joss,
       author = {{Foreman-Mackey}, Daniel and {Luger}, Rodrigo and {Agol}, Eric and {Barclay}, Thomas and {Bouma}, Luke and {Brandt}, Timothy and {Czekala}, Ian and {David}, Trevor and {Dong}, Jiayin and {Gilbert}, Emily and {Gordon}, Tyler and {Hedges}, Christina and {Hey}, Daniel and {Morris}, Brett and {Price-Whelan}, Adrian and {Savel}, Arjun},
        title = "{exoplanet: Gradient-based probabilistic inference for exoplanet data \& other astronomical time series}",
      journal = {The Journal of Open Source Software},
         year = 2021,
        month = jun,
       volume = {6},
       number = {62},
          eid = {3285},
        pages = {3285},
          doi = {10.21105/joss.03285},
archivePrefix = {arXiv},
       eprint = {2105.01994},
 primaryClass = {astro-ph.IM},
       adsurl = {https://ui.adsabs.harvard.edu/abs/2021JOSS....6.3285F}
}

@ARTICLE{Fabrycky2007,
       author = {{Fabrycky}, Daniel and {Tremaine}, Scott},
        title = "{Shrinking Binary and Planetary Orbits by Kozai Cycles with Tidal Friction}",
      journal = {\apj},
         year = 2007,
        month = nov,
       volume = {669},
       number = {2},
        pages = {1298-1315},
          doi = {10.1086/521702},
archivePrefix = {arXiv},
       eprint = {0705.4285},
 primaryClass = {astro-ph},
       adsurl = {https://ui.adsabs.harvard.edu/abs/2007ApJ...669.1298F}
}

@ARTICLE{Fairnington2026,
       author = {{Fairnington}, Tyler R. and {Dong}, Jiayin and {Huang}, Chelsea X. and {Nabbie}, Emma and {Zhou}, George and {Wright}, Duncan and {Collins}, Karen A. and {Ciardi}, David and {Jenkins}, Jon M. and {Latham}, David W. and {Ricker}, George and {Quinn}, Samuel N. and {Seager}, Sara and {Shporer}, Avi and {Vanderspek}, Roland and {Winn}, Joshua N. and {Barkaoui}, Khalid and {Bieryla}, Allyson and {Buchhave}, Lars and {Cheryasov}, Dmitry and {Christiansen}, Jessie and {Dressing}, Courtney and {Fukui}, Akihiko and {Garmash}, Alexey and {Giacalone}, Steven and {Hintz}, Eric G. and {Howell}, Steve B. and {Isogai}, Keisuke and {de Leon}, Jerome and {Lillo-Box}, Jorge and {Murgas}, Felipe and {Narita}, Norio and {Nielsen}, Louise D. and {Palle}, Enric and {Rabus}, Markus and {Rackham}, Benjamin V. and {Schwarz}, Richard P. and {Srdoc}, Gregor and {Stephens}, Denise C. and {Wang}, Gavin and {Watanabe}, Noriharu and {Wilkin}, Francis P. and {Williams}, Joe},
        title = "{The Orbital Eccentricity─Radius Distribution for Warm, Single Planets in TESS}",
      journal = {\apjl},
         year = 2026,
        month = sep,
       volume = {1008},
       number = {1},
          eid = {L6},
        pages = {L6},
          doi = {10.3847/2041-8213/ae9612},
archivePrefix = {arXiv},
       eprint = {2602.20015},
 primaryClass = {astro-ph.EP},
       adsurl = {https://ui.adsabs.harvard.edu/abs/2026ApJ..1008L...6F}
}

@ARTICLE{Fischer2005,
       author = {{Fischer}, Debra A. and {Valenti}, Jeff},
        title = "{The Planet-Metallicity Correlation}",
      journal = {\apj},
         year = 2005,
        month = apr,
       volume = {622},
       number = {2},
        pages = {1102-1117},
          doi = {10.1086/428383},
       adsurl = {https://ui.adsabs.harvard.edu/abs/2005ApJ...622.1102F}
}

@ARTICLE{Foreman2013,
       author = {{Foreman-Mackey}, Daniel and {Hogg}, David W. and {Lang}, Dustin and {Goodman}, Jonathan},
        title = "{emcee: The MCMC Hammer}",
      journal = {\pasp},
         year = 2013,
        month = mar,
       volume = {125},
       number = {925},
        pages = {306},
          doi = {10.1086/670067},
archivePrefix = {arXiv},
       eprint = {1202.3665},
 primaryClass = {astro-ph.IM},
       adsurl = {https://ui.adsabs.harvard.edu/abs/2013PASP..125..306F}
}

@INPROCEEDINGS{Frandsen1999,
       author = {{Frandsen}, S. and {Lindberg}, B.},
        title = "{FIES: A high resolution FIber fed Echelle Spectrograph for the NOT (poster)}",
    booktitle = {Astrophysics with the NOT},
         year = 1999,
       editor = {{Karttunen}, H. and {Piirola}, V.},
        month = jan,
        pages = {71},
       adsurl = {https://ui.adsabs.harvard.edu/abs/1999anot.conf...71F}
}

@phdthesis{Furesz2008,
  author  = "Fűrész, G.",
  title   = "Design and Application of High Resolution and Multiobject Spectrographs: Dynamical Studies of Open Clusters ",
  school  = "University of Szeged, Hungary",
  year    = "2008"
}

@ARTICLE{Furlan2017,
       author = {{Furlan}, E. and {Ciardi}, D.~R. and {Everett}, M.~E. and {Saylors}, M. and {Teske}, J.~K. and {Horch}, E.~P. and {Howell}, S.~B. and {van Belle}, G.~T. and {Hirsch}, L.~A. and {Gautier}, T.~N., III and {Adams}, E.~R. and {Barrado}, D. and {Cartier}, K.~M.~S. and {Dressing}, C.~D. and {Dupree}, A.~K. and {Gilliland}, R.~L. and {Lillo-Box}, J. and {Lucas}, P.~W. and {Wang}, J.},
        title = "{The Kepler Follow-up Observation Program. I. A Catalog of Companions to Kepler Stars from High-Resolution Imaging}",
      journal = {\aj},
         year = 2017,
        month = feb,
       volume = {153},
       number = {2},
          eid = {71},
        pages = {71},
          doi = {10.3847/1538-3881/153/2/71},
archivePrefix = {arXiv},
       eprint = {1612.02392},
 primaryClass = {astro-ph.SR},
       adsurl = {https://ui.adsabs.harvard.edu/abs/2017AJ....153...71F}
}

@ARTICLE{GaiaCollaboration2023,
       author = {{Gaia Collaboration} and {Vallenari}, A. and {Brown}, A.~G.~A. and {Prusti}, T. and {de Bruijne}, J.~H.~J. and {Arenou}, F. and {Babusiaux}, C. and {Biermann}, M. and {Creevey}, O.~L. and {Ducourant}, C. and {Evans}, D.~W. and {Eyer}, L. and {Guerra}, R. and {Hutton}, A. and {Jordi}, C. and {Klioner}, S.~A. and {Lammers}, U.~L. and {Lindegren}, L. and {Luri}, X. and {Mignard}, F. and {Panem}, C. and {Pourbaix}, D. and {Randich}, S. and {Sartoretti}, P. and {Soubiran}, C. and {Tanga}, P. and {Walton}, N.~A. and {Bailer-Jones}, C.~A.~L. and {Bastian}, U. and {Drimmel}, R. and {Jansen}, F. and {Katz}, D. and {Lattanzi}, M.~G. and {van Leeuwen}, F. and {Bakker}, J. and {Cacciari}, C. and {Casta{\~n}eda}, J. and {De Angeli}, F. and {Fabricius}, C. and {Fouesneau}, M. and {Fr{\'e}mat}, Y. and {Galluccio}, L. and {Guerrier}, A. and {Heiter}, U. and {Masana}, E. and {Messineo}, R. and {Mowlavi}, N. and {Nicolas}, C. and {Nienartowicz}, K. and {Pailler}, F. and {Panuzzo}, P. and {Riclet}, F. and {Roux}, W. and {Seabroke}, G.~M. and {Sordo}, R. and {Th{\'e}venin}, F. and {Gracia-Abril}, G. and {Portell}, J. and {Teyssier}, D. and {Altmann}, M. and {Andrae}, R. and {Audard}, M. and {Bellas-Velidis}, I. and {Benson}, K. and {Berthier}, J. and {Blomme}, R. and {Burgess}, P.~W. and {Busonero}, D. and {Busso}, G. and {C{\'a}novas}, H. and {Carry}, B. and {Cellino}, A. and {Cheek}, N. and {Clementini}, G. and {Damerdji}, Y. and {Davidson}, M. and {de Teodoro}, P. and {Nu{\~n}ez Campos}, M. and {Delchambre}, L. and {Dell'Oro}, A. and {Esquej}, P. and {Fern{\'a}ndez-Hern{\'a}ndez}, J. and {Fraile}, E. and {Garabato}, D. and {Garc{\'\i}a-Lario}, P. and {Gosset}, E. and {Haigron}, R. and {Halbwachs}, J.-L. and {Hambly}, N.~C. and {Harrison}, D.~L. and {Hern{\'a}ndez}, J. and {Hestroffer}, D. and {Hodgkin}, S.~T. and {Holl}, B. and {Jan{\ss}en}, K. and {Jevardat de Fombelle}, G. and {Jordan}, S. and {Krone-Martins}, A. and {Lanzafame}, A.~C. and {L{\"o}ffler}, W. and {Marchal}, O. and {Marrese}, P.~M. and {Moitinho}, A. and {Muinonen}, K. and {Osborne}, P. and {Pancino}, E. and {Pauwels}, T. and {Recio-Blanco}, A. and {Reyl{\'e}}, C. and {Riello}, M. and {Rimoldini}, L. and {Roegiers}, T. and {Rybizki}, J. and {Sarro}, L.~M. and {Siopis}, C. and {Smith}, M. and {Sozzetti}, A. and {Utrilla}, E. and {van Leeuwen}, M. and {Abbas}, U. and {{\'A}brah{\'a}m}, P. and {Abreu Aramburu}, A. and {Aerts}, C. and {Aguado}, J.~J. and {Ajaj}, M. and {Aldea-Montero}, F. and {Altavilla}, G. and {{\'A}lvarez}, M.~A. and {Alves}, J. and {Anders}, F. and {Anderson}, R.~I. and {Anglada Varela}, E. and {Antoja}, T. and {Baines}, D. and {Baker}, S.~G. and {Balaguer-N{\'u}{\~n}ez}, L. and {Balbinot}, E. and {Balog}, Z. and {Barache}, C. and {Barbato}, D. and {Barros}, M. and {Barstow}, M.~A. and {Bartolom{\'e}}, S. and {Bassilana}, J.-L. and {Bauchet}, N. and {Becciani}, U. and {Bellazzini}, M. and {Berihuete}, A. and {Bernet}, M. and {Bertone}, S. and {Bianchi}, L. and {Binnenfeld}, A. and {Blanco-Cuaresma}, S. and {Blazere}, A. and {Boch}, T. and {Bombrun}, A. and {Bossini}, D. and {Bouquillon}, S. and {Bragaglia}, A. and {Bramante}, L. and {Breedt}, E. and {Bressan}, A. and {Brouillet}, N. and {Brugaletta}, E. and {Bucciarelli}, B. and {Burlacu}, A. and {Butkevich}, A.~G. and {Buzzi}, R. and {Caffau}, E. and {Cancelliere}, R. and {Cantat-Gaudin}, T. and {Carballo}, R. and {Carlucci}, T. and {Carnerero}, M.~I. and {Carrasco}, J.~M. and {Casamiquela}, L. and {Castellani}, M. and {Castro-Ginard}, A. and {Chaoul}, L. and {Charlot}, P. and {Chemin}, L. and {Chiaramida}, V. and {Chiavassa}, A. and {Chornay}, N. and {Comoretto}, G. and {Contursi}, G. and {Cooper}, W.~J. and {Cornez}, T. and {Cowell}, S. and {Crifo}, F. and {Cropper}, M. and {Crosta}, M. and {Crowley}, C. and {Dafonte}, C. and {Dapergolas}, A. and {David}, M. and {David}, P. and {de Laverny}, P. and {De Luise}, F. and {De March}, R.},
        title = "{Gaia Data Release 3. Summary of the content and survey properties}",
      journal = {\aap},
         year = 2023,
        month = jun,
       volume = {674},
          eid = {A1},
        pages = {A1},
          doi = {10.1051/0004-6361/202243940},
archivePrefix = {arXiv},
       eprint = {2208.00211},
 primaryClass = {astro-ph.GA},
       adsurl = {https://ui.adsabs.harvard.edu/abs/2023A&A...674A...1G}
}

@ARTICLE{galpy,
       author = {{Bovy}, Jo},
        title = "{galpy: A python Library for Galactic Dynamics}",
      journal = {\apjs},
         year = 2015,
        month = feb,
       volume = {216},
       number = {2},
          eid = {29},
        pages = {29},
          doi = {10.1088/0067-0049/216/2/29},
archivePrefix = {arXiv},
       eprint = {1412.3451},
 primaryClass = {astro-ph.GA},
       adsurl = {https://ui.adsabs.harvard.edu/abs/2015ApJS..216...29B}
}

@ARTICLE{Gan2025,
       author = {{Gan}, Tianjun and {Cadieux}, Charles and {Ida}, Shigeru and {Wang}, Sharon X. and {Mao}, Shude and {Lin}, Zitao and {Stassun}, Keivan G. and {Burgasser}, Adam J. and {Howell}, Steve B. and {Clark}, Catherine A. and {Strakhov}, Ivan A. and {Benni}, Paul and {Ricker}, George R. and {Vanderspek}, Roland and {Latham}, David W. and {Seager}, Sara and {Winn}, Joshua N. and {Jenkins}, Jon M. and {Arnold}, Luc and {Artigau}, {\'E}tienne and {Charbonneau}, David and {Collins}, Karen A. and {Cook}, Neil J. and {de Beurs}, Zo{\"e} L. and {Deveny}, Sarah J. and {Doty}, John P. and {Doyon}, Ren{\'e} and {Littlefield}, Colin and {Pritchard}, Tyler and {Ross}, Gabrielle and {Shporer}, Avi and {Theissen}, Christopher R. and {Tofflemire}, Benjamin M. and {Vanderburg}, Andrew and {Watanabe}, David},
        title = "{A New Brown Dwarf Orbiting an M Star and an Investigation of the Eccentricity Distribution of Transiting Long-period Brown Dwarfs}",
      journal = {\apjl},
         year = 2025,
        month = aug,
       volume = {988},
       number = {2},
          eid = {L78},
        pages = {L78},
          doi = {10.3847/2041-8213/adef55},
archivePrefix = {arXiv},
       eprint = {2507.09461},
 primaryClass = {astro-ph.EP},
       adsurl = {https://ui.adsabs.harvard.edu/abs/2025ApJ...988L..78G}
}

@ARTICLE{Gandolfi2015,
       author = {{Gandolfi}, D. and {Parviainen}, H. and {Deeg}, H.~J. and {Lanza}, A.~F. and {Fridlund}, M. and {Prada Moroni}, P.~G. and {Alonso}, R. and {Augusteijn}, T. and {Cabrera}, J. and {Evans}, T. and {Geier}, S. and {Hatzes}, A.~P. and {Holczer}, T. and {Hoyer}, S. and {Kangas}, T. and {Mazeh}, T. and {Pagano}, I. and {Tal-Or}, L. and {Tingley}, B.},
        title = "{Kepler-423b: a half-Jupiter mass planet transiting a very old solar-like star}",
      journal = {\aap},
         year = 2015,
        month = apr,
       volume = {576},
          eid = {A11},
        pages = {A11},
          doi = {10.1051/0004-6361/201425062},
archivePrefix = {arXiv},
       eprint = {1409.8245},
 primaryClass = {astro-ph.EP},
       adsurl = {https://ui.adsabs.harvard.edu/abs/2015A&A...576A..11G}
}

@ARTICLE{gravity2019,
       author = {{Gravity Collaboration} and {Abuter}, R. and {Amorim}, A. and {Baub{\"o}ck}, M. and {Berger}, J.~P. and {Bonnet}, H. and {Brandner}, W. and {Cl{\'e}net}, Y. and {Coud{\'e} Du Foresto}, V. and {de Zeeuw}, P.~T. and {Dexter}, J. and {Duvert}, G. and {Eckart}, A. and {Eisenhauer}, F. and {F{\"o}rster Schreiber}, N.~M. and {Garcia}, P. and {Gao}, F. and {Gendron}, E. and {Genzel}, R. and {Gerhard}, O. and {Gillessen}, S. and {Habibi}, M. and {Haubois}, X. and {Henning}, T. and {Hippler}, S. and {Horrobin}, M. and {Jim{\'e}nez-Rosales}, A. and {Jocou}, L. and {Kervella}, P. and {Lacour}, S. and {Lapeyr{\`e}re}, V. and {Le Bouquin}, J. -B. and {L{\'e}na}, P. and {Ott}, T. and {Paumard}, T. and {Perraut}, K. and {Perrin}, G. and {Pfuhl}, O. and {Rabien}, S. and {Rodriguez Coira}, G. and {Rousset}, G. and {Scheithauer}, S. and {Sternberg}, A. and {Straub}, O. and {Straubmeier}, C. and {Sturm}, E. and {Tacconi}, L.~J. and {Vincent}, F. and {von Fellenberg}, S. and {Waisberg}, I. and {Widmann}, F. and {Wieprecht}, E. and {Wiezorrek}, E. and {Woillez}, J. and {Yazici}, S.},
        title = "{A geometric distance measurement to the Galactic center black hole with 0.3\% uncertainty}",
      journal = {\aap},
         year = 2019,
        month = may,
       volume = {625},
          eid = {L10},
        pages = {L10},
          doi = {10.1051/0004-6361/201935656},
archivePrefix = {arXiv},
       eprint = {1904.05721},
 primaryClass = {astro-ph.GA},
       adsurl = {https://ui.adsabs.harvard.edu/abs/2019A&A...625L..10G}
}

@BOOK{Gray2005,
       author = {{Gray}, David F.},
        title = "{The Observation and Analysis of Stellar Photospheres}",
         year = 2005,
       adsurl = {https://ui.adsabs.harvard.edu/abs/2005oasp.book.....G}
}

@ARTICLE{Guerrero2021,
       author = {{Guerrero}, Natalia M. and {Seager}, S. and {Huang}, Chelsea X. and {Vanderburg}, Andrew and {Garcia Soto}, Aylin and {Mireles}, Ismael and {Hesse}, Katharine and {Fong}, William and {Glidden}, Ana and {Shporer}, Avi and {Latham}, David W. and {Collins}, Karen A. and {Quinn}, Samuel N. and {Burt}, Jennifer and {Dragomir}, Diana and {Crossfield}, Ian and {Vanderspek}, Roland and {Fausnaugh}, Michael and {Burke}, Christopher J. and {Ricker}, George and {Daylan}, Tansu and {Essack}, Zahra and {G{\"u}nther}, Maximilian N. and {Osborn}, Hugh P. and {Pepper}, Joshua and {Rowden}, Pamela and {Sha}, Lizhou and {Villanueva}, Steven, Jr. and {Yahalomi}, Daniel A. and {Yu}, Liang and {Ballard}, Sarah and {Batalha}, Natalie M. and {Berardo}, David and {Chontos}, Ashley and {Dittmann}, Jason A. and {Esquerdo}, Gilbert A. and {Mikal-Evans}, Thomas and {Jayaraman}, Rahul and {Krishnamurthy}, Akshata and {Louie}, Dana R. and {Mehrle}, Nicholas and {Niraula}, Prajwal and {Rackham}, Benjamin V. and {Rodriguez}, Joseph E. and {Rowden}, Stephen J.~L. and {Sousa-Silva}, Clara and {Watanabe}, David and {Wong}, Ian and {Zhan}, Zhuchang and {Zivanovic}, Goran and {Christiansen}, Jessie L. and {Ciardi}, David R. and {Swain}, Melanie A. and {Lund}, Michael B. and {Mullally}, Susan E. and {Fleming}, Scott W. and {Rodriguez}, David R. and {Boyd}, Patricia T. and {Quintana}, Elisa V. and {Barclay}, Thomas and {Col{\'o}n}, Knicole D. and {Rinehart}, S.~A. and {Schlieder}, Joshua E. and {Clampin}, Mark and {Jenkins}, Jon M. and {Twicken}, Joseph D. and {Caldwell}, Douglas A. and {Coughlin}, Jeffrey L. and {Henze}, Chris and {Lissauer}, Jack J. and {Morris}, Robert L. and {Rose}, Mark E. and {Smith}, Jeffrey C. and {Tenenbaum}, Peter and {Ting}, Eric B. and {Wohler}, Bill and {Bakos}, G. {\'A}. and {Bean}, Jacob L. and {Berta-Thompson}, Zachory K. and {Bieryla}, Allyson and {Bouma}, Luke G. and {Buchhave}, Lars A. and {Butler}, Nathaniel and {Charbonneau}, David and {Doty}, John P. and {Ge}, Jian and {Holman}, Matthew J. and {Howard}, Andrew W. and {Kaltenegger}, Lisa and {Kane}, Stephen R. and {Kjeldsen}, Hans and {Kreidberg}, Laura and {Lin}, Douglas N.~C. and {Minsky}, Charlotte and {Narita}, Norio and {Paegert}, Martin and {P{\'a}l}, Andr{\'a}s and {Palle}, Enric and {Sasselov}, Dimitar D. and {Spencer}, Alton and {Sozzetti}, Alessandro and {Stassun}, Keivan G. and {Torres}, Guillermo and {Udry}, Stephane and {Winn}, Joshua N.},
        title = "{The TESS Objects of Interest Catalog from the TESS Prime Mission}",
      journal = {\apjs},
         year = 2021,
        month = jun,
       volume = {254},
       number = {2},
          eid = {39},
        pages = {39},
          doi = {10.3847/1538-4365/abefe1},
archivePrefix = {arXiv},
       eprint = {2103.12538},
 primaryClass = {astro-ph.EP},
       adsurl = {https://ui.adsabs.harvard.edu/abs/2021ApJS..254...39G}
}

@ARTICLE{Guillochon2011,
       author = {{Guillochon}, James and {Ramirez-Ruiz}, Enrico and {Lin}, Douglas},
        title = "{Consequences of the Ejection and Disruption of Giant Planets}",
      journal = {\apj},
         year = 2011,
        month = may,
       volume = {732},
       number = {2},
          eid = {74},
        pages = {74},
          doi = {10.1088/0004-637X/732/2/74},
archivePrefix = {arXiv},
       eprint = {1012.2382},
 primaryClass = {astro-ph.EP},
       adsurl = {https://ui.adsabs.harvard.edu/abs/2011ApJ...732...74G}
}

@ARTICLE{Gupta2024,
       author = {{Gupta}, Arvind F. and {Millholland}, Sarah C. and {Im}, Haedam and {Dong}, Jiayin and {Jackson}, Jonathan M. and {Carleo}, Ilaria and {Libby-Roberts}, Jessica and {Delamer}, Megan and {Giovinazzi}, Mark R. and {Lin}, Andrea S.~J. and {Kanodia}, Shubham and {Wang}, Xian-Yu and {Stassun}, Keivan and {Masseron}, Thomas and {Dragomir}, Diana and {Mahadevan}, Suvrath and {Wright}, Jason and {Alvarado-Montes}, Jaime A. and {Bender}, Chad and {Blake}, Cullen H. and {Caldwell}, Douglas and {Ca{\~n}as}, Caleb I. and {Cochran}, William D. and {Dalba}, Paul and {Everett}, Mark E. and {Fernandez}, Pipa and {Golub}, Eli and {Guillet}, Bruno and {Halverson}, Samuel and {Hebb}, Leslie and {Higuera}, Jesus and {Huang}, Chelsea X. and {Klusmeyer}, Jessica and {Knight}, Rachel and {Leroux}, Liouba and {Logsdon}, Sarah E. and {Loose}, Margaret and {McElwain}, Michael W. and {Monson}, Andrew and {Ninan}, Joe P. and {Nowak}, Grzegorz and {Palle}, Enric and {Patel}, Yatrik and {Pepper}, Joshua and {Primm}, Michael and {Rajagopal}, Jayadev and {Robertson}, Paul and {Roy}, Arpita and {Schneider}, Donald P. and {Schwab}, Christian and {Schweiker}, Heidi and {Sgro}, Lauren and {Shimizu}, Masao and {Simard}, Georges and {Stef{\'a}nsson}, Gudmundur and {Stevens}, Daniel J. and {Villanueva}, Steven and {Wisniewski}, John and {Will}, Stefan and {Ziegler}, Carl},
        title = "{A hot-Jupiter progenitor on a super-eccentric retrograde orbit}",
      journal = {\nat},
         year = 2024,
        month = aug,
       volume = {632},
       number = {8023},
        pages = {50-54},
          doi = {10.1038/s41586-024-07688-3},
       adsurl = {https://ui.adsabs.harvard.edu/abs/2024Natur.632...50G}
}

@ARTICLE{Gustafsson2008,
       author = {{Gustafsson}, B. and {Edvardsson}, B. and {Eriksson}, K. and {J{\o}rgensen}, U.~G. and {Nordlund}, {\r{A}}. and {Plez}, B.},
        title = "{A grid of MARCS model atmospheres for late-type stars. I. Methods and general properties}",
      journal = {\aap},
         year = 2008,
        month = aug,
       volume = {486},
       number = {3},
        pages = {951-970},
          doi = {10.1051/0004-6361:200809724},
archivePrefix = {arXiv},
       eprint = {0805.0554},
 primaryClass = {astro-ph},
       adsurl = {https://ui.adsabs.harvard.edu/abs/2008A&A...486..951G}
}

@ARTICLE{Hartman2016,
       author = {{Hartman}, J.~D. and {Bakos}, G. {\'A}.},
        title = "{VARTOOLS: A program for analyzing astronomical time-series data}",
      journal = {Astronomy and Computing},
         year = 2016,
        month = oct,
       volume = {17},
        pages = {1-72},
          doi = {10.1016/j.ascom.2016.05.006},
archivePrefix = {arXiv},
       eprint = {1605.06811},
 primaryClass = {astro-ph.IM},
       adsurl = {https://ui.adsabs.harvard.edu/abs/2016A&C....17....1H}
}

@ARTICLE{Hayward2001,
       author = {{Hayward}, T.~L. and {Brandl}, B. and {Pirger}, B. and {Blacken}, C. and {Gull}, G.~E. and {Schoenwald}, J. and {Houck}, J.~R.},
        title = "{PHARO: A Near-Infrared Camera for the Palomar Adaptive Optics System}",
      journal = {\pasp},
         year = 2001,
        month = jan,
       volume = {113},
       number = {779},
        pages = {105-118},
          doi = {10.1086/317969},
       adsurl = {https://ui.adsabs.harvard.edu/abs/2001PASP..113..105H}
}

@ARTICLE{Hog2000,
       author = {{H{\o}g}, E. and {Fabricius}, C. and {Makarov}, V.~V. and {Urban}, S. and {Corbin}, T. and {Wycoff}, G. and {Bastian}, U. and {Schwekendiek}, P. and {Wicenec}, A.},
        title = "{The Tycho-2 catalogue of the 2.5 million brightest stars}",
      journal = {\aap},
         year = 2000,
        month = mar,
       volume = {355},
        pages = {L27-L30},
       adsurl = {https://ui.adsabs.harvard.edu/abs/2000A&A...355L..27H}
}

@ARTICLE{Howell2011,
       author = {{Howell}, Steve B. and {Everett}, Mark E. and {Sherry}, William and {Horch}, Elliott and {Ciardi}, David R.},
        title = "{Speckle Camera Observations for the NASA Kepler Mission Follow-up Program}",
      journal = {\aj},
         year = 2011,
        month = jul,
       volume = {142},
       number = {1},
          eid = {19},
        pages = {19},
          doi = {10.1088/0004-6256/142/1/19},
       adsurl = {https://ui.adsabs.harvard.edu/abs/2011AJ....142...19H}
}

@ARTICLE{Huang2020,
       author = {{Huang}, Chelsea X. and {Vanderburg}, Andrew and {P{\'a}l}, Andras and {Sha}, Lizhou and {Yu}, Liang and {Fong}, Willie and {Fausnaugh}, Michael and {Shporer}, Avi and {Guerrero}, Natalia and {Vanderspek}, Roland and {Ricker}, George},
        title = "{Photometry of 10 Million Stars from the First Two Years of TESS Full Frame Images: Part II}",
      journal = {Research Notes of the American Astronomical Society},
         year = 2020,
        month = nov,
       volume = {4},
       number = {11},
          eid = {206},
        pages = {206},
          doi = {10.3847/2515-5172/abca2d},
       adsurl = {https://ui.adsabs.harvard.edu/abs/2020RNAAS...4..206H}
}

@ARTICLE{Jackson2008,
       author = {{Jackson}, Brian and {Greenberg}, Richard and {Barnes}, Rory},
        title = "{Tidal Evolution of Close-in Extrasolar Planets}",
      journal = {\apj},
         year = 2008,
        month = may,
       volume = {678},
       number = {2},
        pages = {1396-1406},
          doi = {10.1086/529187},
archivePrefix = {arXiv},
       eprint = {0802.1543},
 primaryClass = {astro-ph},
       adsurl = {https://ui.adsabs.harvard.edu/abs/2008ApJ...678.1396J}
}

@INPROCEEDINGS{Jenkins2016,
   author = {{Jenkins}, J.~M. and {Twicken}, J.~D. and {McCauliff}, S. and 
	{Campbell}, J. and {Sanderfer}, D. and {Lung}, D. and {Mansouri-Samani}, M. and 
	{Girouard}, F. and {Tenenbaum}, P. and {Klaus}, T. and {Smith}, J.~C. and 
	{Caldwell}, D.~A. and {Chacon}, A.~D. and {Henze}, C. and {Heiges}, C. and 
	{Latham}, D.~W. and {Morgan}, E. and {Swade}, D. and {Rinehart}, S. and 
	{Vanderspek}, R.},
    title = "{The TESS science processing operations center}",
booktitle = {Software and Cyberinfrastructure for Astronomy IV},
     year = 2016,
   series = {\procspie},
   volume = 9913,
    month = aug,
      eid = {99133E},
    pages = {99133E},
      doi = {10.1117/12.2233418}
}

@misc{Jensen2013,
       author = {{Jensen}, Eric},
        title = "{Tapir: A web interface for transit/eclipse observability}",
 howpublished = {Astrophysics Source Code Library, record ascl:1306.007},
         year = 2013,
        month = jun,
          eid = {ascl:1306.007},
       adsurl = {https://ui.adsabs.harvard.edu/abs/2013ascl.soft06007J}
}

@ARTICLE{Jofre2019,
       author = {{Jofr{\'e}}, Paula and {Heiter}, Ulrike and {Soubiran}, Caroline},
        title = "{Accuracy and Precision of Industrial Stellar Abundances}",
      journal = {\araa},
         year = 2019,
        month = aug,
       volume = {57},
        pages = {571-616},
          doi = {10.1146/annurev-astro-091918-104509},
archivePrefix = {arXiv},
       eprint = {1811.08041},
 primaryClass = {astro-ph.SR},
       adsurl = {https://ui.adsabs.harvard.edu/abs/2019ARA&A..57..571J}
}

@ARTICLE{Jones2003,
       author = {{Jones}, Hugh R.~A. and {Butler}, R. Paul and {Tinney}, C.~G. and {Marcy}, Geoffrey W. and {Penny}, Alan J. and {McCarthy}, Chris and {Carter}, Brad D.},
        title = "{An exoplanet in orbit around {\ensuremath{\tau}}$^{1}$ Gruis}",
      journal = {\mnras},
         year = 2003,
        month = may,
       volume = {341},
       number = {3},
        pages = {948-952},
          doi = {10.1046/j.1365-8711.2003.06481.x},
archivePrefix = {arXiv},
       eprint = {astro-ph/0209302},
 primaryClass = {astro-ph},
       adsurl = {https://ui.adsabs.harvard.edu/abs/2003MNRAS.341..948J}
}

@ARTICLE{Juric2008,
       author = {{Juri{\'c}}, Mario and {Tremaine}, Scott},
        title = "{Dynamical Origin of Extrasolar Planet Eccentricity Distribution}",
      journal = {\apj},
         year = 2008,
        month = oct,
       volume = {686},
       number = {1},
        pages = {603-620},
          doi = {10.1086/590047},
archivePrefix = {arXiv},
       eprint = {astro-ph/0703160},
 primaryClass = {astro-ph},
       adsurl = {https://ui.adsabs.harvard.edu/abs/2008ApJ...686..603J}
}

@ARTICLE{Kempton2018,
       author = {{Kempton}, Eliza M.-R. and {Bean}, Jacob L. and {Louie}, Dana R. and {Deming}, Drake and {Koll}, Daniel D.~B. and {Mansfield}, Megan and {Christiansen}, Jessie L. and {L{\'o}pez-Morales}, Mercedes and {Swain}, Mark R. and {Zellem}, Robert T. and {Ballard}, Sarah and {Barclay}, Thomas and {Barstow}, Joanna K. and {Batalha}, Natasha E. and {Beatty}, Thomas G. and {Berta-Thompson}, Zach and {Birkby}, Jayne and {Buchhave}, Lars A. and {Charbonneau}, David and {Cowan}, Nicolas B. and {Crossfield}, Ian and {de Val-Borro}, Miguel and {Doyon}, Ren{\'e} and {Dragomir}, Diana and {Gaidos}, Eric and {Heng}, Kevin and {Hu}, Renyu and {Kane}, Stephen R. and {Kreidberg}, Laura and {Mallonn}, Matthias and {Morley}, Caroline V. and {Narita}, Norio and {Nascimbeni}, Valerio and {Pall{\'e}}, Enric and {Quintana}, Elisa V. and {Rauscher}, Emily and {Seager}, Sara and {Shkolnik}, Evgenya L. and {Sing}, David K. and {Sozzetti}, Alessandro and {Stassun}, Keivan G. and {Valenti}, Jeff A. and {von Essen}, Carolina},
        title = "{A Framework for Prioritizing the TESS Planetary Candidates Most Amenable to Atmospheric Characterization}",
      journal = {\pasp},
         year = 2018,
        month = nov,
       volume = {130},
       number = {993},
        pages = {114401},
          doi = {10.1088/1538-3873/aadf6f},
archivePrefix = {arXiv},
       eprint = {1805.03671},
 primaryClass = {astro-ph.EP},
       adsurl = {https://ui.adsabs.harvard.edu/abs/2018PASP..130k4401K}
}

@ARTICLE{Kipping2013,
       author = {{Kipping}, D.~M.},
        title = "{Parametrizing the exoplanet eccentricity distribution with the beta  distribution.}",
      journal = {\mnras},
         year = 2013,
        month = jul,
       volume = {434},
        pages = {L51-L55},
          doi = {10.1093/mnrasl/slt075},
archivePrefix = {arXiv},
       eprint = {1306.4982},
 primaryClass = {astro-ph.EP},
       adsurl = {https://ui.adsabs.harvard.edu/abs/2013MNRAS.434L..51K}
}

@ARTICLE{Knudstrup2022,
       author = {{Knudstrup}, Emil and {Serrano}, Luisa M. and {Gandolfi}, Davide and {Albrecht}, Simon H. and {Cochran}, William D. and {Endl}, Michael and {MacQueen}, Phillip and {Tronsgaard}, Ren{\'e} and {Bieryla}, Allyson and {Buchhave}, Lars A. and {Stassun}, Keivan and {Collins}, Karen A. and {Nowak}, Grzegorz and {Deeg}, Hans J. and {Barkaoui}, Khalid and {Safonov}, Boris S. and {Strakhov}, Ivan A. and {Belinski}, Alexandre A. and {Twicken}, Joseph D. and {Jenkins}, Jon M. and {Howard}, Andrew W. and {Isaacson}, Howard and {Winn}, Joshua N. and {Collins}, Kevin I. and {Conti}, Dennis M. and {Furesz}, Gabor and {Gan}, Tianjun and {Kielkopf}, John F. and {Massey}, Bob and {Murgas}, Felipe and {Murphy}, Lauren G. and {Palle}, Enric and {Quinn}, Samuel N. and {Reed}, Phillip A. and {Ricker}, George R. and {Seager}, Sara and {Shiao}, Bernie and {Schwarz}, Richard P. and {Srdoc}, Gregor and {Watanabe}, David},
        title = "{Confirmation and characterisation of three giant planets detected by TESS from the FIES/NOT and Tull/McDonald spectrographs}",
      journal = {\aap},
         year = 2022,
        month = nov,
       volume = {667},
          eid = {A22},
        pages = {A22},
          doi = {10.1051/0004-6361/202243656},
archivePrefix = {arXiv},
       eprint = {2204.13956},
 primaryClass = {astro-ph.EP},
       adsurl = {https://ui.adsabs.harvard.edu/abs/2022A&A...667A..22K}
}

@ARTICLE{Kovacs2002,
       author = {{Kov{\'a}cs}, G. and {Zucker}, S. and {Mazeh}, T.},
        title = "{A box-fitting algorithm in the search for periodic transits}",
      journal = {\aap},
         year = 2002,
        month = aug,
       volume = {391},
        pages = {369-377},
          doi = {10.1051/0004-6361:20020802},
archivePrefix = {arXiv},
       eprint = {astro-ph/0206099},
 primaryClass = {astro-ph},
       adsurl = {https://ui.adsabs.harvard.edu/abs/2002A&A...391..369K}
}

@ARTICLE{Kozai1962,
       author = {{Kozai}, Yoshihide},
        title = "{Secular perturbations of asteroids with high inclination and eccentricity}",
      journal = {\aj},
         year = 1962,
        month = nov,
       volume = {67},
        pages = {591-598},
          doi = {10.1086/108790},
       adsurl = {https://ui.adsabs.harvard.edu/abs/1962AJ.....67..591K}
}

@ARTICLE{Kreidberg2015,
       author = {{Kreidberg}, Laura},
        title = "{batman: BAsic Transit Model cAlculatioN in Python}",
      journal = {\pasp},
         year = 2015,
        month = nov,
       volume = {127},
       number = {957},
        pages = {1161},
          doi = {10.1086/683602},
archivePrefix = {arXiv},
       eprint = {1507.08285},
 primaryClass = {astro-ph.EP},
       adsurl = {https://ui.adsabs.harvard.edu/abs/2015PASP..127.1161K}
}

@ARTICLE{Larsen2025,
       author = {{Larsen}, J.~R. and {R{\o}rsted}, J.~L. and {Aguirre B{\o}rsen-Koch}, V. and {Lundkvist}, M.~S. and {Christensen-Dalsgaard}, J. and {Winther}, M.~L. and {Stokholm}, A. and {Li}, Y. and {Slumstrup}, D. and {Kjeldsen}, H. and {Corsaro}, E. and {Benomar}, O. and {Dhanpal}, S. and {Weiss}, A. and {Mosser}, B. and {Hekker}, S. and {Stello}, D. and {Korn}, A.~J. and {Jendreieck}, A. and {Elsworth}, Y. and {Handberg}, R. and {Kallinger}, T. and {Jiang}, C. and {Ruchti}, G.},
        title = "{Pushing the boundaries of asteroseismic individual frequency modelling: Unveiling two evolved very low-metallicity red giants}",
      journal = {\aap},
         year = 2025,
        month = may,
       volume = {697},
          eid = {A153},
        pages = {A153},
          doi = {10.1051/0004-6361/202453554},
archivePrefix = {arXiv},
       eprint = {2503.23063},
 primaryClass = {astro-ph.SR},
       adsurl = {https://ui.adsabs.harvard.edu/abs/2025A&A...697A.153L}
}

@ARTICLE{Lidov1962,
       author = {{Lidov}, M.~L.},
        title = "{The evolution of orbits of artificial satellites of planets under the action of gravitational perturbations of external bodies}",
      journal = {\planss},
         year = 1962,
        month = oct,
       volume = {9},
       number = {10},
        pages = {719-759},
          doi = {10.1016/0032-0633(62)90129-0},
       adsurl = {https://ui.adsabs.harvard.edu/abs/1962P&SS....9..719L}
}

@MISC{lightkurve,
   author = {{Lightkurve Collaboration} and {Cardoso}, J.~V.~d.~M. and
             {Hedges}, C. and {Gully-Santiago}, M. and {Saunders}, N. and
             {Cody}, A.~M. and {Barclay}, T. and {Hall}, O. and
             {Sagear}, S. and {Turtelboom}, E. and {Zhang}, J. and
             {Tzanidakis}, A. and {Mighell}, K. and {Coughlin}, J. and
             {Bell}, K. and {Berta-Thompson}, Z. and {Williams}, P. and
             {Dotson}, J. and {Barentsen}, G.},
    title = "{Lightkurve: Kepler and TESS time series analysis in Python}",
howpublished = {Astrophysics Source Code Library},
     year = 2018,
    month = dec,
archivePrefix = "ascl",
   eprint = {1812.013},
   adsurl = {http://adsabs.harvard.edu/abs/2018ascl.soft12013L},
}

@ARTICLE{Lin1996,
       author = {{Lin}, D.~N.~C. and {Bodenheimer}, P. and {Richardson}, D.~C.},
        title = "{Orbital migration of the planetary companion of 51 Pegasi to its present location}",
      journal = {\nat},
         year = 1996,
        month = apr,
       volume = {380},
       number = {6575},
        pages = {606-607},
          doi = {10.1038/380606a0},
       adsurl = {https://ui.adsabs.harvard.edu/abs/1996Natur.380..606L}
}

@ARTICLE{Lindegren21,
       author = {{Lindegren}, L. and {Bastian}, U. and {Biermann}, M. and {Bombrun}, A. and {de Torres}, A. and {Gerlach}, E. and {Geyer}, R. and {Hern{\'a}ndez}, J. and {Hilger}, T. and {Hobbs}, D. and {Klioner}, S.~A. and {Lammers}, U. and {McMillan}, P.~J. and {Ramos-Lerate}, M. and {Steidelm{\"u}ller}, H. and {Stephenson}, C.~A. and {van Leeuwen}, F.},
        title = "{Gaia Early Data Release 3. Parallax bias versus magnitude, colour, and position}",
      journal = {\aap},
         year = 2021,
        month = may,
       volume = {649},
          eid = {A4},
        pages = {A4},
          doi = {10.1051/0004-6361/202039653},
archivePrefix = {arXiv},
       eprint = {2012.01742},
 primaryClass = {astro-ph.IM},
       adsurl = {https://ui.adsabs.harvard.edu/abs/2021A&A...649A...4L}
}

@ARTICLE{Liu2015,
       author = {{Liu}, Bin and {Mu{\~n}oz}, Diego J. and {Lai}, Dong},
        title = "{Suppression of extreme orbital evolution in triple systems with short-range forces}",
      journal = {\mnras},
         year = 2015,
        month = feb,
       volume = {447},
       number = {1},
        pages = {747-764},
          doi = {10.1093/mnras/stu2396},
archivePrefix = {arXiv},
       eprint = {1409.6717},
 primaryClass = {astro-ph.EP},
       adsurl = {https://ui.adsabs.harvard.edu/abs/2015MNRAS.447..747L}
}

@ARTICLE{Lucy1971,
       author = {{Lucy}, L.~B. and {Sweeney}, M.~A.},
        title = "{Spectroscopic binaries with circular orbits.}",
      journal = {\aj},
         year = 1971,
        month = aug,
       volume = {76},
        pages = {544-556},
          doi = {10.1086/111159},
       adsurl = {https://ui.adsabs.harvard.edu/abs/1971AJ.....76..544L}
}

@ARTICLE{Madhusudhan2014,
       author = {{Madhusudhan}, Nikku and {Amin}, Mustafa A. and {Kennedy}, Grant M.},
        title = "{Toward Chemical Constraints on Hot Jupiter Migration}",
      journal = {\apjl},
         year = 2014,
        month = oct,
       volume = {794},
       number = {1},
          eid = {L12},
        pages = {L12},
          doi = {10.1088/2041-8205/794/1/L12},
archivePrefix = {arXiv},
       eprint = {1408.3668},
 primaryClass = {astro-ph.EP},
       adsurl = {https://ui.adsabs.harvard.edu/abs/2014ApJ...794L..12M}
}

@ARTICLE{Mandel2002,
       author = {{Mandel}, Kaisey and {Agol}, Eric},
        title = "{Analytic Light Curves for Planetary Transit Searches}",
      journal = {\apjl},
         year = 2002,
        month = dec,
       volume = {580},
       number = {2},
        pages = {L171-L175},
          doi = {10.1086/345520},
archivePrefix = {arXiv},
       eprint = {astro-ph/0210099},
 primaryClass = {astro-ph},
       adsurl = {https://ui.adsabs.harvard.edu/abs/2002ApJ...580L.171M}
}

@Article{matplotlib,
  Author    = {Hunter, J. D.},
  Title     = {Matplotlib: A 2D graphics environment},
  Journal   = {Computing in Science \& Engineering},
  Volume    = {9},
  Number    = {3},
  Pages     = {90--95},
  publisher = {IEEE COMPUTER SOC},
  doi       = {10.1109/MCSE.2007.55},
  year      = 2007
}

@ARTICLE{matteucci2021,
       author = {{Matteucci}, Francesca},
        title = "{Modelling the chemical evolution of the Milky Way}",
      journal = {\aapr},
         year = 2021,
        month = dec,
       volume = {29},
       number = {1},
          eid = {5},
        pages = {5},
          doi = {10.1007/s00159-021-00133-8},
archivePrefix = {arXiv},
       eprint = {2106.13145},
 primaryClass = {astro-ph.GA},
       adsurl = {https://ui.adsabs.harvard.edu/abs/2021A&ARv..29....5M}
}

@ARTICLE{Maxted2022,
       author = {{Maxted}, P.~F.~L. and {Ehrenreich}, D. and {Wilson}, T.~G. and {Alibert}, Y. and {Cameron}, A. Collier and {Hoyer}, S. and {Sousa}, S.~G. and {Olofsson}, G. and {Bekkelien}, A. and {Deline}, A. and {Delrez}, L. and {Bonfanti}, A. and {Borsato}, L. and {Alonso}, R. and {Anglada Escud{\'e}}, G. and {Barrado}, D. and {Barros}, S.~C.~C. and {Baumjohann}, W. and {Beck}, M. and {Beck}, T. and {Benz}, W. and {Billot}, N. and {Biondi}, F. and {Bonfils}, X. and {Brandeker}, A. and {Broeg}, C. and {B{\'a}rczy}, T. and {Cabrera}, J. and {Charnoz}, S. and {Corral Van Damme}, C. and {Csizmadia}, Sz and {Davies}, M.~B. and {Deleuil}, M. and {Demangeon}, O.~D.~S. and {Demory}, B. -O. and {Erikson}, A. and {Flor{\'e}n}, H.~G. and {Fortier}, A. and {Fossati}, L. and {Fridlund}, M. and {Futyan}, D. and {Gandolfi}, D. and {Gillon}, M. and {Guedel}, M. and {Guterman}, P. and {Heng}, K. and {Isaak}, K.~G. and {Kiss}, L. and {Laskar}, J. and {Lecavelier des Etangs}, A. and {Lendl}, M. and {Lovis}, C. and {Magrin}, D. and {Nascimbeni}, V. and {Ottensamer}, R. and {Pagano}, I. and {Pall{\'e}}, E. and {Peter}, G. and {Piotto}, G. and {Pollacco}, D. and {Pozuelos}, F.~J. and {Queloz}, D. and {Ragazzoni}, R. and {Rando}, N. and {Rauer}, H. and {Reimers}, C. and {Ribas}, I. and {Salmon}, S. and {Santos}, N.~C. and {Scandariato}, G. and {Simon}, A.~E. and {Smith}, A.~M.~S. and {Steller}, M. and {Swayne}, M.~I. and {Szab{\'o}}, Gy M. and {S{\'e}gransan}, D. and {Thomas}, N. and {Udry}, S. and {Van Grootel}, V. and {Walton}, N.~A.},
        title = "{Analysis of Early Science observations with the CHaracterising ExOPlanets Satellite (CHEOPS) using PYCHEOPS}",
      journal = {\mnras},
         year = 2022,
        month = jul,
       volume = {514},
       number = {1},
        pages = {77-104},
          doi = {10.1093/mnras/stab3371},
archivePrefix = {arXiv},
       eprint = {2111.08828},
 primaryClass = {astro-ph.EP},
       adsurl = {https://ui.adsabs.harvard.edu/abs/2022MNRAS.514...77M}
}

@ARTICLE{McLaughlin1924,
       author = {{McLaughlin}, D.~B.},
        title = "{Some results of a spectrographic study of the Algol system.}",
      journal = {\apj},
         year = 1924,
        month = jul,
       volume = {60},
        pages = {22-31},
          doi = {10.1086/142826},
       adsurl = {https://ui.adsabs.harvard.edu/abs/1924ApJ....60...22M}
}

@ARTICLE{mcmillan2017,
       author = {{McMillan}, Paul J.},
        title = "{The mass distribution and gravitational potential of the Milky Way}",
      journal = {\mnras},
         year = 2017,
        month = feb,
       volume = {465},
       number = {1},
        pages = {76-94},
          doi = {10.1093/mnras/stw2759},
archivePrefix = {arXiv},
       eprint = {1608.00971},
 primaryClass = {astro-ph.GA},
       adsurl = {https://ui.adsabs.harvard.edu/abs/2017MNRAS.465...76M}
}

@ARTICLE{Morgan2026,
       author = {{Morgan}, Marvin and {Bowler}, Brendan P. and {Tran}, Quang H.},
        title = "{Exploring Warm Jupiter Migration Pathways with Eccentricities. II. Correlations with Host Star Properties and Orbital Separation}",
      journal = {\aj},
         year = 2026,
        month = jan,
       volume = {171},
       number = {1},
          eid = {19},
        pages = {19},
          doi = {10.3847/1538-3881/ae0e16},
archivePrefix = {arXiv},
       eprint = {2510.02591},
 primaryClass = {astro-ph.EP},
       adsurl = {https://ui.adsabs.harvard.edu/abs/2026AJ....171...19M}
}

@MISC{Morris2020,
       author = {{Morris}, Robert L. and {Twicken}, Joseph D. and {Smith}, Jeffrey C. and {Clarke}, Bruce D. and {Jenkins}, Jon M. and {Bryson}, Stephen T. and {Girouard}, Forrest and {Klaus}, Todd C.},
        title = "{Kepler Data Processing Handbook: Photometric Analysis}",
 howpublished = {Kepler Science Document KSCI-19081-003},
         year = 2020,
        month = mar,
          eid = {6},
        pages = {6},
       adsurl = {https://ui.adsabs.harvard.edu/abs/2020ksci.rept....6M}
}

@ARTICLE{Naef2001,
       author = {{Naef}, D. and {Latham}, D.~W. and {Mayor}, M. and {Mazeh}, T. and {Beuzit}, J.~L. and {Drukier}, G.~A. and {Perrier-Bellet}, C. and {Queloz}, D. and {Sivan}, J.~P. and {Torres}, G. and {Udry}, S. and {Zucker}, S.},
        title = "{HD 80606 b, a planet on an extremely elongated orbit}",
      journal = {\aap},
         year = 2001,
        month = aug,
       volume = {375},
        pages = {L27-L30},
          doi = {10.1051/0004-6361:20010853},
archivePrefix = {arXiv},
       eprint = {astro-ph/0106256},
 primaryClass = {astro-ph},
       adsurl = {https://ui.adsabs.harvard.edu/abs/2001A&A...375L..27N}
}

@Article{numpy,
 title         = {Array programming with {NumPy}},
 author        = {Charles R. Harris and K. Jarrod Millman and St{\'{e}}fan J.
                 van der Walt and Ralf Gommers and Pauli Virtanen and David
                 Cournapeau and Eric Wieser and Julian Taylor and Sebastian
                 Berg and Nathaniel J. Smith and Robert Kern and Matti Picus
                 and Stephan Hoyer and Marten H. van Kerkwijk and Matthew
                 Brett and Allan Haldane and Jaime Fern{\'{a}}ndez del
                 R{\'{i}}o and Mark Wiebe and Pearu Peterson and Pierre
                 G{\'{e}}rard-Marchant and Kevin Sheppard and Tyler Reddy and
                 Warren Weckesser and Hameer Abbasi and Christoph Gohlke and
                 Travis E. Oliphant},
 year          = {2020},
 month         = sep,
 journal       = {Nature},
 volume        = {585},
 number        = {7825},
 pages         = {357--362},
 doi           = {10.1038/s41586-020-2649-2},
 publisher     = {Springer Science and Business Media {LLC}},
 url           = {https://doi.org/10.1038/s41586-020-2649-2}
}

@ARTICLE{Perryman2014,
       author = {{Perryman}, Michael and {Hartman}, Joel and {Bakos}, G{\'a}sp{\'a}r {\'A}. and {Lindegren}, Lennart},
        title = "{Astrometric Exoplanet Detection with Gaia}",
      journal = {\apj},
         year = 2014,
        month = dec,
       volume = {797},
       number = {1},
          eid = {14},
        pages = {14},
          doi = {10.1088/0004-637X/797/1/14},
archivePrefix = {arXiv},
       eprint = {1411.1173},
 primaryClass = {astro-ph.EP},
       adsurl = {https://ui.adsabs.harvard.edu/abs/2014ApJ...797...14P}
}

@ARTICLE{Persson2018,
       author = {{Persson}, C.~M. and {Fridlund}, M. and {Barrag{\'a}n}, O. and {Dai}, F. and {Gandolfi}, D. and {Hatzes}, A.~P. and {Hirano}, T. and {Grziwa}, S. and {Korth}, J. and {Prieto-Arranz}, J. and {Fossati}, L. and {Van Eylen}, V. and {Justesen}, A.~B. and {Livingston}, J. and {Kubyshkina}, D. and {Deeg}, H.~J. and {Guenther}, E.~W. and {Nowak}, G. and {Cabrera}, J. and {Eigm{\"u}ller}, Ph. and {Csizmadia}, Sz. and {Smith}, A.~M.~S. and {Erikson}, A. and {Albrecht}, S. and {Sobrino}, Alonso and {Cochran}, W.~D. and {Endl}, M. and {Esposito}, M. and {Fukui}, A. and {Heeren}, P. and {Hidalgo}, D. and {Hjorth}, M. and {Kuzuhara}, M. and {Narita}, N. and {Nespral}, D. and {Palle}, E. and {P{\"a}tzold}, M. and {Rauer}, H. and {Rodler}, F. and {Winn}, J.~N.},
        title = "{Super-Earth of 8 M$_{{\ensuremath{\oplus}}}$ in a 2.2-day orbit around the K5V star K2-216}",
      journal = {\aap},
         year = 2018,
        month = oct,
       volume = {618},
          eid = {A33},
        pages = {A33},
          doi = {10.1051/0004-6361/201832867},
archivePrefix = {arXiv},
       eprint = {1805.04774},
 primaryClass = {astro-ph.EP},
       adsurl = {https://ui.adsabs.harvard.edu/abs/2018A&A...618A..33P}
}

@ARTICLE{Petrovich2014,
       author = {{Petrovich}, Cristobal and {Tremaine}, Scott and {Rafikov}, Roman},
        title = "{Scattering Outcomes of Close-in Planets: Constraints on Planet Migration}",
      journal = {\apj},
         year = 2014,
        month = may,
       volume = {786},
       number = {2},
          eid = {101},
        pages = {101},
          doi = {10.1088/0004-637X/786/2/101},
archivePrefix = {arXiv},
       eprint = {1401.4457},
 primaryClass = {astro-ph.EP},
       adsurl = {https://ui.adsabs.harvard.edu/abs/2014ApJ...786..101P}
}

@ARTICLE{Petrovich2015a,
       author = {{Petrovich}, Cristobal},
        title = "{Steady-state Planet Migration by the Kozai-Lidov Mechanism in Stellar Binaries}",
      journal = {\apj},
         year = 2015,
        month = jan,
       volume = {799},
       number = {1},
          eid = {27},
        pages = {27},
          doi = {10.1088/0004-637X/799/1/27},
archivePrefix = {arXiv},
       eprint = {1405.0280},
 primaryClass = {astro-ph.EP},
       adsurl = {https://ui.adsabs.harvard.edu/abs/2015ApJ...799...27P}
}

@ARTICLE{Petrovich2016,
       author = {{Petrovich}, Cristobal and {Tremaine}, Scott},
        title = "{Warm Jupiters from Secular Planet-Planet Interactions}",
      journal = {\apj},
         year = 2016,
        month = oct,
       volume = {829},
       number = {2},
          eid = {132},
        pages = {132},
          doi = {10.3847/0004-637X/829/2/132},
archivePrefix = {arXiv},
       eprint = {1604.00010},
 primaryClass = {astro-ph.EP},
       adsurl = {https://ui.adsabs.harvard.edu/abs/2016ApJ...829..132P}
}

@ARTICLE{Piskunov2017,
       author = {{Piskunov}, Nikolai and {Valenti}, Jeff A.},
        title = "{Spectroscopy Made Easy: Evolution}",
      journal = {\aap},
         year = 2017,
        month = jan,
       volume = {597},
          eid = {A16},
        pages = {A16},
          doi = {10.1051/0004-6361/201629124},
archivePrefix = {arXiv},
       eprint = {1606.06073},
 primaryClass = {astro-ph.IM},
       adsurl = {https://ui.adsabs.harvard.edu/abs/2017A&A...597A..16P}
}

@article{pymc2023,
  title = {{PyMC}: A Modern and Comprehensive Probabilistic Programming Framework in {P}ython},
  author = {Oriol Abril-Pla and Virgile Andreani and Colin Carroll and Larry Dong and Christopher J. Fonnesbeck and Maxim Kochurov and Ravin Kumar and Junpeng Lao and Christian C. Luhmann and Osvaldo A. Martin and Michael Osthege and Ricardo Vieira and Thomas Wiecki and Robert Zinkov },
  journal = {{PeerJ} Computer Science},
  volume = {9},
  number = {e1516},
  doi = {10.7717/peerj-cs.1516},
  year = {2023}
}

@ARTICLE{Rafikov2005,
       author = {{Rafikov}, Roman R.},
        title = "{Can Giant Planets Form by Direct Gravitational Instability?}",
      journal = {\apjl},
         year = 2005,
        month = mar,
       volume = {621},
       number = {1},
        pages = {L69-L72},
          doi = {10.1086/428899},
archivePrefix = {arXiv},
       eprint = {astro-ph/0406469},
 primaryClass = {astro-ph},
       adsurl = {https://ui.adsabs.harvard.edu/abs/2005ApJ...621L..69R}
}

@ARTICLE{Rasio1996,
       author = {{Rasio}, Frederic A. and {Ford}, Eric B.},
        title = "{Dynamical instabilities and the formation of extrasolar planetary systems}",
      journal = {Science},
         year = 1996,
        month = nov,
       volume = {274},
        pages = {954-956},
          doi = {10.1126/science.274.5289.954},
       adsurl = {https://ui.adsabs.harvard.edu/abs/1996Sci...274..954R}
}

@ARTICLE{Rice2022,
       author = {{Rice}, Malena and {Wang}, Songhu and {Wang}, Xian-Yu and {Stef{\'a}nsson}, Gu{\dj}mundur and {Isaacson}, Howard and {Howard}, Andrew W. and {Logsdon}, Sarah E. and {Schweiker}, Heidi and {Dai}, Fei and {Brinkman}, Casey and {Giacalone}, Steven and {Holcomb}, Rae},
        title = "{A Tendency Toward Alignment in Single-star Warm-Jupiter Systems}",
      journal = {\aj},
         year = 2022,
        month = sep,
       volume = {164},
       number = {3},
          eid = {104},
        pages = {104},
          doi = {10.3847/1538-3881/ac8153},
archivePrefix = {arXiv},
       eprint = {2207.06511},
 primaryClass = {astro-ph.EP},
       adsurl = {https://ui.adsabs.harvard.edu/abs/2022AJ....164..104R}
}

@ARTICLE{Ricker2015,
       author = {{Ricker}, George R. and {Winn}, Joshua N. and {Vanderspek}, Roland and {Latham}, David W. and {Bakos}, G{\'a}sp{\'a}r {\'A}. and {Bean}, Jacob L. and {Berta-Thompson}, Zachory K. and {Brown}, Timothy M. and {Buchhave}, Lars and {Butler}, Nathaniel R. and {Butler}, R. Paul and {Chaplin}, William J. and {Charbonneau}, David and {Christensen-Dalsgaard}, J{\o}rgen and {Clampin}, Mark and {Deming}, Drake and {Doty}, John and {De Lee}, Nathan and {Dressing}, Courtney and {Dunham}, Edward W. and {Endl}, Michael and {Fressin}, Francois and {Ge}, Jian and {Henning}, Thomas and {Holman}, Matthew J. and {Howard}, Andrew W. and {Ida}, Shigeru and {Jenkins}, Jon M. and {Jernigan}, Garrett and {Johnson}, John Asher and {Kaltenegger}, Lisa and {Kawai}, Nobuyuki and {Kjeldsen}, Hans and {Laughlin}, Gregory and {Levine}, Alan M. and {Lin}, Douglas and {Lissauer}, Jack J. and {MacQueen}, Phillip and {Marcy}, Geoffrey and {McCullough}, Peter R. and {Morton}, Timothy D. and {Narita}, Norio and {Paegert}, Martin and {Palle}, Enric and {Pepe}, Francesco and {Pepper}, Joshua and {Quirrenbach}, Andreas and {Rinehart}, Stephen A. and {Sasselov}, Dimitar and {Sato}, Bun'ei and {Seager}, Sara and {Sozzetti}, Alessandro and {Stassun}, Keivan G. and {Sullivan}, Peter and {Szentgyorgyi}, Andrew and {Torres}, Guillermo and {Udry}, Stephane and {Villasenor}, Joel},
        title = "{Transiting Exoplanet Survey Satellite (TESS)}",
      journal = {Journal of Astronomical Telescopes, Instruments, and Systems},
         year = 2015,
        month = jan,
       volume = {1},
          eid = {014003},
        pages = {014003},
          doi = {10.1117/1.JATIS.1.1.014003},
       adsurl = {https://ui.adsabs.harvard.edu/abs/2015JATIS...1a4003R}
}

@ARTICLE{Rossiter1924,
       author = {{Rossiter}, R.~A.},
        title = "{On the detection of an effect of rotation during eclipse in the velocity of the brighter component of beta Lyrae, and on the constancy of velocity of this system.}",
      journal = {\apj},
         year = 1924,
        month = jul,
       volume = {60},
        pages = {15-21},
          doi = {10.1086/142825},
       adsurl = {https://ui.adsabs.harvard.edu/abs/1924ApJ....60...15R}
}

@ARTICLE{Ryabchikova2015,
       author = {{Ryabchikova}, T. and {Piskunov}, N. and {Kurucz}, R.~L. and {Stempels}, H.~C. and {Heiter}, U. and {Pakhomov}, Yu and {Barklem}, P.~S.},
        title = "{A major upgrade of the VALD database}",
      journal = {\physscr},
         year = 2015,
        month = may,
       volume = {90},
       number = {5},
          eid = {054005},
        pages = {054005},
          doi = {10.1088/0031-8949/90/5/054005},
       adsurl = {https://ui.adsabs.harvard.edu/abs/2015PhyS...90e4005R}
}

@ARTICLE{Schlichting2014,
       author = {{Schlichting}, Hilke E.},
        title = "{Formation of Close in Super-Earths and Mini-Neptunes: Required Disk Masses and their Implications}",
      journal = {\apjl},
         year = 2014,
        month = nov,
       volume = {795},
       number = {1},
          eid = {L15},
        pages = {L15},
          doi = {10.1088/2041-8205/795/1/L15},
archivePrefix = {arXiv},
       eprint = {1410.1060},
 primaryClass = {astro-ph.EP},
       adsurl = {https://ui.adsabs.harvard.edu/abs/2014ApJ...795L..15S}
}

@ARTICLE{scipy,
  author  = {Virtanen, Pauli and Gommers, Ralf and Oliphant, Travis E. and
            Haberland, Matt and Reddy, Tyler and Cournapeau, David and
            Burovski, Evgeni and Peterson, Pearu and Weckesser, Warren and
            Bright, Jonathan and {van der Walt}, St{\'e}fan J. and
            Brett, Matthew and Wilson, Joshua and Millman, K. Jarrod and
            Mayorov, Nikolay and Nelson, Andrew R. J. and Jones, Eric and
            Kern, Robert and Larson, Eric and Carey, C J and
            Polat, {\.I}lhan and Feng, Yu and Moore, Eric W. and
            {VanderPlas}, Jake and Laxalde, Denis and Perktold, Josef and
            Cimrman, Robert and Henriksen, Ian and Quintero, E. A. and
            Harris, Charles R. and Archibald, Anne M. and
            Ribeiro, Ant{\^o}nio H. and Pedregosa, Fabian and
            {van Mulbregt}, Paul and {SciPy 1.0 Contributors}},
  title   = {{{SciPy} 1.0: Fundamental Algorithms for Scientific
            Computing in Python}},
  journal = {Nature Methods},
  year    = {2020},
  volume  = {17},
  pages   = {261--272},
  adsurl  = {https://rdcu.be/b08Wh},
  doi     = {10.1038/s41592-019-0686-2},
}

@ARTICLE{schonrich2010,
   author = {{Sch{\"o}nrich}, R. and {Binney}, J. and {Dehnen}, W.},
    title = "{Local kinematics and the local standard of rest}",
  journal = {\mnras},
archivePrefix = "arXiv",
   eprint = {0912.3693},
     year = 2010,
    month = apr,
   volume = 403,
    pages = {1829-1833},
      doi = {10.1111/j.1365-2966.2010.16253.x},
   adsurl = {http://adsabs.harvard.edu/abs/2010MNRAS.403.1829S}
}

@ARTICLE{Scott2018,
       author = {{Scott}, Nicholas J. and {Howell}, Steve B. and {Horch}, Elliott P. and {Everett}, Mark E.},
        title = "{The NN-explore Exoplanet Stellar Speckle Imager: Instrument Description and Preliminary Results}",
      journal = {\pasp},
         year = 2018,
        month = may,
       volume = {130},
       number = {987},
        pages = {054502},
          doi = {10.1088/1538-3873/aab484},
       adsurl = {https://ui.adsabs.harvard.edu/abs/2018PASP..130e4502S}
}

@ARTICLE{Seager2003,
       author = {{Seager}, S. and {Mall{\'e}n-Ornelas}, G.},
        title = "{A Unique Solution of Planet and Star Parameters from an Extrasolar Planet Transit Light Curve}",
      journal = {\apj},
         year = 2003,
        month = mar,
       volume = {585},
       number = {2},
        pages = {1038-1055},
          doi = {10.1086/346105},
archivePrefix = {arXiv},
       eprint = {astro-ph/0206228},
 primaryClass = {astro-ph},
       adsurl = {https://ui.adsabs.harvard.edu/abs/2003ApJ...585.1038S}
}

@ARTICLE{Slumstrup2019,
       author = {{Slumstrup}, D. and {Grundahl}, F. and {Silva Aguirre}, V. and {Brogaard}, K.},
        title = "{Systematic differences in the spectroscopic analysis of red giants}",
      journal = {\aap},
         year = 2019,
        month = feb,
       volume = {622},
          eid = {A111},
        pages = {A111},
          doi = {10.1051/0004-6361/201833739},
archivePrefix = {arXiv},
       eprint = {1812.05630},
 primaryClass = {astro-ph.SR},
       adsurl = {https://ui.adsabs.harvard.edu/abs/2019A&A...622A.111S}
}

@ARTICLE{Smith2012,
   author = {{Smith}, J.~C. and {Stumpe}, M.~C. and {Van Cleve}, J.~E. and 
	{Jenkins}, J.~M. and {Barclay}, T.~S. and {Fanelli}, M.~N. and 
	{Girouard}, F.~R. and {Kolodziejczak}, J.~J. and {McCauliff}, S.~D. and 
	{Morris}, R.~L. and {Twicken}, J.~D.},
    title = "{Kepler Presearch Data Conditioning II - A Bayesian Approach to Systematic Error Correction}",
  journal = {\pasp},
     year = 2012,
    month = sep,
   volume = 124,
    pages = {1000},
      doi = {10.1086/667697}
}

@INPROCEEDINGS{Skilling2004,
       author = {{Skilling}, John},
        title = "{Nested Sampling}",
    booktitle = {American Institute of Physics Conference Series},
         year = "2004",
       editor = {{Fischer}, Rainer and {Preuss}, Roland and {Toussaint}, Udo Von},
       series = {American Institute of Physics Conference Series},
       volume = {735},
        month = "Nov",
        pages = {395-405},
          doi = {10.1063/1.1835238},
       adsurl = {https://ui.adsabs.harvard.edu/abs/2004AIPC..735..395S}
}

@article{Skilling2006,
       author = "Skilling, John",
          doi = "10.1214/06-BA127",
     fjournal = "Bayesian Analysis",
      journal = "Bayesian Anal.",
        month = "12",
       number = "4",
        pages = "833--859",
    publisher = "International Society for Bayesian Analysis",
        title = "Nested sampling for general Bayesian computation",
          url = "https://doi.org/10.1214/06-BA127",
       volume = "1",
         year = "2006"
}

@ARTICLE{Socrates2012,
       author = {{Socrates}, Aristotle and {Katz}, Boaz and {Dong}, Subo and {Tremaine}, Scott},
        title = "{Super-eccentric Migrating Jupiters}",
      journal = {\apj},
         year = 2012,
        month = may,
       volume = {750},
       number = {2},
          eid = {106},
        pages = {106},
          doi = {10.1088/0004-637X/750/2/106},
archivePrefix = {arXiv},
       eprint = {1110.1644},
 primaryClass = {astro-ph.EP},
       adsurl = {https://ui.adsabs.harvard.edu/abs/2012ApJ...750..106S}
}

@ARTICLE{Sousa2011,
       author = {{Sousa}, S.~G. and {Santos}, N.~C. and {Israelian}, G. and {Mayor}, M. and {Udry}, S.},
        title = "{Spectroscopic stellar parameters for 582 FGK stars in the HARPS volume-limited sample. Revising the metallicity-planet correlation}",
      journal = {\aap},
         year = 2011,
        month = sep,
       volume = {533},
          eid = {A141},
        pages = {A141},
          doi = {10.1051/0004-6361/201117699},
archivePrefix = {arXiv},
       eprint = {1108.5279},
 primaryClass = {astro-ph.EP},
       adsurl = {https://ui.adsabs.harvard.edu/abs/2011A&A...533A.141S}
}

@ARTICLE{Speagle2019,
       author = {{Speagle}, Joshua S.},
        title = "{DYNESTY: a dynamic nested sampling package for estimating Bayesian posteriors and evidences}",
      journal = {\mnras},
         year = 2020,
        month = apr,
       volume = {493},
       number = {3},
        pages = {3132-3158},
          doi = {10.1093/mnras/staa278},
archivePrefix = {arXiv},
       eprint = {1904.02180},
 primaryClass = {astro-ph.IM},
       adsurl = {https://ui.adsabs.harvard.edu/abs/2020MNRAS.493.3132S}
}

@ARTICLE{spitoni2023,
       author = {{Spitoni}, E. and {Recio-Blanco}, A. and {de Laverny}, P. and {Palicio}, P.~A. and {Kordopatis}, G. and {Schultheis}, M. and {Contursi}, G. and {Poggio}, E. and {Romano}, D. and {Matteucci}, F.},
        title = "{Beyond the two-infall model. I. Indications for a recent gas infall with Gaia DR3 chemical abundances}",
      journal = {\aap},
         year = 2023,
        month = feb,
       volume = {670},
          eid = {A109},
        pages = {A109},
          doi = {10.1051/0004-6361/202244349},
archivePrefix = {arXiv},
       eprint = {2206.12436},
 primaryClass = {astro-ph.GA},
       adsurl = {https://ui.adsabs.harvard.edu/abs/2023A&A...670A.109S}
}

@ARTICLE{Stassun2019,
       author = {{Stassun}, Keivan G. and {Oelkers}, Ryan J. and {Paegert}, Martin and {Torres}, Guillermo and {Pepper}, Joshua and {De Lee}, Nathan and {Collins}, Kevin and {Latham}, David W. and {Muirhead}, Philip S. and {Chittidi}, Jay and {Rojas-Ayala}, B{\'a}rbara and {Fleming}, Scott W. and {Rose}, Mark E. and {Tenenbaum}, Peter and {Ting}, Eric B. and {Kane}, Stephen R. and {Barclay}, Thomas and {Bean}, Jacob L. and {Brassuer}, C.~E. and {Charbonneau}, David and {Ge}, Jian and {Lissauer}, Jack J. and {Mann}, Andrew W. and {McLean}, Brian and {Mullally}, Susan and {Narita}, Norio and {Plavchan}, Peter and {Ricker}, George R. and {Sasselov}, Dimitar and {Seager}, S. and {Sharma}, Sanjib and {Shiao}, Bernie and {Sozzetti}, Alessandro and {Stello}, Dennis and {Vanderspek}, Roland and {Wallace}, Geoff and {Winn}, Joshua N.},
        title = "{The Revised TESS Input Catalog and Candidate Target List}",
      journal = {\aj},
         year = 2019,
        month = oct,
       volume = {158},
       number = {4},
          eid = {138},
        pages = {138},
          doi = {10.3847/1538-3881/ab3467},
archivePrefix = {arXiv},
       eprint = {1905.10694},
 primaryClass = {astro-ph.SR},
       adsurl = {https://ui.adsabs.harvard.edu/abs/2019AJ....158..138S}
}

@ARTICLE{Stevenson2025,
       author = {{Stevenson}, A.~T. and {Haswell}, C.~A. and {Faria}, J.~P. and {Barnes}, J.~R. and {Barstow}, J.~K. and {Dickinson}, H. and {Standing}, M.~R.},
        title = "{RV-exoplanet eccentricities: Good, Beta, and Best}",
      journal = {\mnras},
         year = 2025,
        month = may,
       volume = {539},
       number = {2},
        pages = {727-754},
          doi = {10.1093/mnras/staf502},
archivePrefix = {arXiv},
       eprint = {2503.20385},
 primaryClass = {astro-ph.EP},
       adsurl = {https://ui.adsabs.harvard.edu/abs/2025MNRAS.539..727S}
}

@ARTICLE{Stumpe2012,
   author = {{Stumpe}, M.~C. and {Smith}, J.~C. and {Van Cleve}, J.~E. and 
	{Twicken}, J.~D. and {Barclay}, T.~S. and {Fanelli}, M.~N. and 
	{Girouard}, F.~R. and {Jenkins}, J.~M. and {Kolodziejczak}, J.~J. and 
	{McCauliff}, S.~D. and {Morris}, R.~L.},
    title = "{Kepler Presearch Data Conditioning I{\mdash}Architecture and Algorithms for Error Correction in Kepler Light Curves}",
  journal = {\pasp},
     year = 2012,
    month = sep,
   volume = 124,
    pages = {985},
      doi = {10.1086/667698}
}

@ARTICLE{Stumpe2014,
       author = {{Stumpe}, Martin C. and {Smith}, Jeffrey C. and
         {Catanzarite}, Joseph H. and {Van Cleve}, Jeffrey E. and
         {Jenkins}, Jon M. and {Twicken}, Joseph D. and {Girouard}, Forrest R.},
        title = "{Multiscale Systematic Error Correction via Wavelet-Based Bandsplitting in Kepler Data}",
      journal = {\pasp},
         year = "2014",
        month = "Jan",
       volume = {126},
       number = {935},
        pages = {100},
          doi = {10.1086/674989}
}

@ARTICLE{Strakhov2023,
       author = {{Strakhov}, I.~A. and {Safonov}, B.~S. and {Cheryasov}, D.~V.},
        title = "{Speckle Interferometry with CMOS Detector}",
      journal = {Astrophysical Bulletin},
         year = 2023,
        month = jun,
       volume = {78},
       number = {2},
        pages = {234-258},
          doi = {10.1134/S1990341323020104},
archivePrefix = {arXiv},
       eprint = {2305.00451},
 primaryClass = {astro-ph.IM},
       adsurl = {https://ui.adsabs.harvard.edu/abs/2023AstBu..78..234S}
}

@ARTICLE{Telting2014,
       author = {{Telting}, J.~H. and {Avila}, G. and {Buchhave}, L. and {Frandsen}, S. and {Gandolfi}, D. and {Lindberg}, B. and {Stempels}, H.~C. and {Prins}, S. and {NOT staff}},
        title = "{FIES: The high-resolution Fiber-fed Echelle Spectrograph at the Nordic Optical Telescope}",
      journal = {Astronomische Nachrichten},
         year = 2014,
        month = jan,
       volume = {335},
       number = {1},
        pages = {41},
          doi = {10.1002/asna.201312007},
       adsurl = {https://ui.adsabs.harvard.edu/abs/2014AN....335...41T}
}

@ARTICLE{Tinetti2018,
       author = {{Tinetti}, Giovanna and {Drossart}, Pierre and {Eccleston}, Paul and {Hartogh}, Paul and {Heske}, Astrid and {Leconte}, J{\'e}r{\'e}my and {Micela}, Giusi and {Ollivier}, Marc and {Pilbratt}, G{\"o}ran and {Puig}, Ludovic and {Turrini}, Diego and {Vandenbussche}, Bart and {Wolkenberg}, Paulina and {Beaulieu}, Jean-Philippe and {Buchave}, Lars A. and {Ferus}, Martin and {Griffin}, Matt and {Guedel}, Manuel and {Justtanont}, Kay and {Lagage}, Pierre-Olivier and {Machado}, Pedro and {Malaguti}, Giuseppe and {Min}, Michiel and {N{\o}rgaard-Nielsen}, Hans Ulrik and {Rataj}, Mirek and {Ray}, Tom and {Ribas}, Ignasi and {Swain}, Mark and {Szabo}, Robert and {Werner}, Stephanie and {Barstow}, Joanna and {Burleigh}, Matt and {Cho}, James and {Coud{\'e} du Foresto}, Vincent and {Coustenis}, Athena and {Decin}, Leen and {Encrenaz}, Therese and {Galand}, Marina and {Gillon}, Michael and {Helled}, Ravit and {Morales}, Juan Carlos and {Garc{\'\i}a Mu{\~n}oz}, Antonio and {Moneti}, Andrea and {Pagano}, Isabella and {Pascale}, Enzo and {Piccioni}, Giuseppe and {Pinfield}, David and {Sarkar}, Subhajit and {Selsis}, Franck and {Tennyson}, Jonathan and {Triaud}, Amaury and {Venot}, Olivia and {Waldmann}, Ingo and {Waltham}, David and {Wright}, Gillian and {Amiaux}, Jerome and {Augu{\`e}res}, Jean-Louis and {Berth{\'e}}, Michel and {Bezawada}, Naidu and {Bishop}, Georgia and {Bowles}, Neil and {Coffey}, Deirdre and {Colom{\'e}}, Josep and {Crook}, Martin and {Crouzet}, Pierre-Elie and {Da Peppo}, Vania and {Sanz}, Isabel Escudero and {Focardi}, Mauro and {Frericks}, Martin and {Hunt}, Tom and {Kohley}, Ralf and {Middleton}, Kevin and {Morgante}, Gianluca and {Ottensamer}, Roland and {Pace}, Emanuele and {Pearson}, Chris and {Stamper}, Richard and {Symonds}, Kate and {Rengel}, Miriam and {Renotte}, Etienne and {Ade}, Peter and {Affer}, Laura and {Alard}, Christophe and {Allard}, Nicole and {Altieri}, Francesca and {Andr{\'e}}, Yves and {Arena}, Claudio and {Argyriou}, Ioannis and {Aylward}, Alan and {Baccani}, Cristian and {Bakos}, Gaspar and {Banaszkiewicz}, Marek and {Barlow}, Mike and {Batista}, Virginie and {Bellucci}, Giancarlo and {Benatti}, Serena and {Bernardi}, Pernelle and {B{\'e}zard}, Bruno and {Blecka}, Maria and {Bolmont}, Emeline and {Bonfond}, Bertrand and {Bonito}, Rosaria and {Bonomo}, Aldo S. and {Brucato}, John Robert and {Brun}, Allan Sacha and {Bryson}, Ian and {Bujwan}, Waldemar and {Casewell}, Sarah and {Charnay}, Bejamin and {Pestellini}, Cesare Cecchi and {Chen}, Guo and {Ciaravella}, Angela and {Claudi}, Riccardo and {Cl{\'e}dassou}, Rodolphe and {Damasso}, Mario and {Damiano}, Mario and {Danielski}, Camilla and {Deroo}, Pieter and {Di Giorgio}, Anna Maria and {Dominik}, Carsten and {Doublier}, Vanessa and {Doyle}, Simon and {Doyon}, Ren{\'e} and {Drummond}, Benjamin and {Duong}, Bastien and {Eales}, Stephen and {Edwards}, Billy and {Farina}, Maria and {Flaccomio}, Ettore and {Fletcher}, Leigh and {Forget}, Fran{\c{c}}ois and {Fossey}, Steve and {Fr{\"a}nz}, Markus and {Fujii}, Yuka and {Garc{\'\i}a-Piquer}, {\'A}lvaro and {Gear}, Walter and {Geoffray}, Herv{\'e} and {G{\'e}rard}, Jean Claude and {Gesa}, Lluis and {Gomez}, H. and {Graczyk}, Rafa{\l} and {Griffith}, Caitlin and {Grodent}, Denis and {Guarcello}, Mario Giuseppe and {Gustin}, Jacques and {Hamano}, Keiko and {Hargrave}, Peter and {Hello}, Yann and {Heng}, Kevin and {Herrero}, Enrique and {Hornstrup}, Allan and {Hubert}, Benoit and {Ida}, Shigeru and {Ikoma}, Masahiro and {Iro}, Nicolas and {Irwin}, Patrick and {Jarchow}, Christopher and {Jaubert}, Jean and {Jones}, Hugh and {Julien}, Queyrel and {Kameda}, Shingo and {Kerschbaum}, Franz and {Kervella}, Pierre and {Koskinen}, Tommi and {Krijger}, Matthijs and {Krupp}, Norbert and {Lafarga}, Marina and {Landini}, Federico and {Lellouch}, Emanuel and {Leto}, Giuseppe and {Luntzer}, A. and {Rank-L{\"u}ftinger}, Theresa and {Maggio}, Antonio and {Maldonado}, Jesus and {Maillard}, Jean-Pierre and {Mall}, Urs and {Marquette}, Jean-Baptiste and {Mathis}, Stephane and {Maxted}, Pierre and {Matsuo}, Taro and {Medvedev}, Alexander and {Miguel}, Yamila and {Minier}, Vincent and {Morello}, Giuseppe and {Mura}, Alessandro and {Narita}, Norio and {Nascimbeni}, Valerio and {Nguyen Tong}, N. and {Noce}, Vladimiro and {Oliva}, Fabrizio and {Palle}, Enric and {Palmer}, Paul and {Pancrazzi}, Maurizio and {Papageorgiou}, Andreas and {Parmentier}, Vivien and {Perger}, Manuel and {Petralia}, Antonino and {Pezzuto}, Stefano and {Pierrehumbert}, Ray and {Pillitteri}, Ignazio},
        title = "{A chemical survey of exoplanets with ARIEL}",
      journal = {Experimental Astronomy},
         year = 2018,
        month = nov,
       volume = {46},
       number = {1},
        pages = {135-209},
          doi = {10.1007/s10686-018-9598-x},
       adsurl = {https://ui.adsabs.harvard.edu/abs/2018ExA....46..135T}
}

@ARTICLE{Tokovinin2018,
       author = {{Tokovinin}, Andrei},
        title = "{Ten Years of Speckle Interferometry at SOAR}",
      journal = {\pasp},
         year = 2018,
        month = mar,
       volume = {130},
       number = {985},
        pages = {035002},
          doi = {10.1088/1538-3873/aaa7d9},
archivePrefix = {arXiv},
       eprint = {1801.04772},
 primaryClass = {astro-ph.IM},
       adsurl = {https://ui.adsabs.harvard.edu/abs/2018PASP..130c5002T}
}

@ARTICLE{Tull1995,
       author = {{Tull}, Robert G. and {MacQueen}, Phillip J. and {Sneden}, Christopher and {Lambert}, David L.},
        title = "{The High-Resolution Cross-Dispersed Echelle White Pupil Spectrometer of the McDonald Observatory 2.7-m Telescope}",
      journal = {\pasp},
         year = 1995,
        month = mar,
       volume = {107},
        pages = {251},
          doi = {10.1086/133548},
       adsurl = {https://ui.adsabs.harvard.edu/abs/1995PASP..107..251T}
}

@INPROCEEDINGS{Twicken2010,
       author = {{Twicken}, Joseph D. and {Clarke}, Bruce D. and {Bryson}, Stephen T. and {Tenenbaum}, Peter and {Wu}, Hayley and {Jenkins}, Jon M. and {Girouard}, Forrest and {Klaus}, Todd C.},
        title = "{Photometric analysis in the Kepler Science Operations Center pipeline}",
    booktitle = {Software and Cyberinfrastructure for Astronomy},
         year = 2010,
       editor = {{Radziwill}, Nicole M. and {Bridger}, Alan},
       series = {Society of Photo-Optical Instrumentation Engineers (SPIE) Conference Series},
       volume = {7740},
        month = jul,
          eid = {774023},
        pages = {774023},
          doi = {10.1117/12.856790},
       adsurl = {https://ui.adsabs.harvard.edu/abs/2010SPIE.7740E..23T}

}

@ARTICLE{Udry2003,
       author = {{Udry}, S. and {Mayor}, M. and {Santos}, N.~C.},
        title = "{Statistical properties of exoplanets. I. The period distribution: Constraints for the migration scenario}",
      journal = {\aap},
         year = 2003,
        month = aug,
       volume = {407},
        pages = {369-376},
          doi = {10.1051/0004-6361:20030843},
archivePrefix = {arXiv},
       eprint = {astro-ph/0306049},
 primaryClass = {astro-ph},
       adsurl = {https://ui.adsabs.harvard.edu/abs/2003A&A...407..369U}
}

@ARTICLE{Valenti1996,
       author = {{Valenti}, J.~A. and {Piskunov}, N.},
        title = "{Spectroscopy made easy: A new tool for fitting observations with synthetic spectra.}",
      journal = {\aaps},
         year = 1996,
        month = sep,
       volume = {118},
        pages = {595-603},
       adsurl = {https://ui.adsabs.harvard.edu/abs/1996A&AS..118..595V}
}

@INPROCEEDINGS{VanderPlas2012,
       author = {{VanderPlas}, J. and {Connolly}, A.~J. and {Ivezic}, Z. and {Gray}, A.},
        title = "{Introduction to astroML: Machine learning for astrophysics}",
    booktitle = {Proceedings of Conference on Intelligent Data Understanding (CIDU},
         year = 2012,
        month = oct,
        pages = {47-54},
          doi = {10.1109/CIDU.2012.6382200},
archivePrefix = {arXiv},
       eprint = {1411.5039},
 primaryClass = {astro-ph.IM},
       adsurl = {https://ui.adsabs.harvard.edu/abs/2012cidu.conf...47V}
}

@ARTICLE{VanderPlas2015,
       author = {{VanderPlas}, Jacob T. and {Ivezi{\'c}}, {\v{Z}}eljko},
        title = "{Periodograms for Multiband Astronomical Time Series}",
      journal = {\apj},
         year = 2015,
        month = oct,
       volume = {812},
       number = {1},
          eid = {18},
        pages = {18},
          doi = {10.1088/0004-637X/812/1/18},
archivePrefix = {arXiv},
       eprint = {1502.01344},
 primaryClass = {astro-ph.IM},
       adsurl = {https://ui.adsabs.harvard.edu/abs/2015ApJ...812...18V}
}

@ARTICLE{Vines2022,
       author = {{Vines}, Jose I. and {Jenkins}, James S.},
        title = "{ARIADNE: Measuring accurate and precise stellar parameters through SED fitting}",
      journal = {\mnras},
         year = 2022,
        month = apr,
          doi = {10.1093/mnras/stac956},
archivePrefix = {arXiv},
       eprint = {2204.03769},
 primaryClass = {astro-ph.SR},
       adsurl = {https://ui.adsabs.harvard.edu/abs/2022MNRAS.tmp..920V}
}

@ARTICLE{Wang2024,
       author = {{Wang}, Xian-Yu and {Rice}, Malena and {Wang}, Songhu and {Kanodia}, Shubham and {Dai}, Fei and {Logsdon}, Sarah E. and {Schweiker}, Heidi and {Teske}, Johanna K. and {Butler}, R. Paul and {Crane}, Jeffrey D. and {Shectman}, Stephen and {Quinn}, Samuel N. and {Kostov}, Veselin and {Osborn}, Hugh P. and {Goeke}, Robert F. and {Eastman}, Jason D. and {Shporer}, Avi and {Rapetti}, David and {Collins}, Karen A. and {Watkins}, Cristilyn N. and {Relles}, Howard M. and {Ricker}, George R. and {Seager}, Sara and {Winn}, Joshua N. and {Jenkins}, Jon M.},
        title = "{Single-star Warm-Jupiter Systems Tend to Be Aligned, Even around Hot Stellar Hosts: No T $_{eff}$─{\ensuremath{\lambda}} Dependency}",
      journal = {\apjl},
         year = 2024,
        month = sep,
       volume = {973},
       number = {1},
          eid = {L21},
        pages = {L21},
          doi = {10.3847/2041-8213/ad7469},
archivePrefix = {arXiv},
       eprint = {2408.10038},
 primaryClass = {astro-ph.EP},
       adsurl = {https://ui.adsabs.harvard.edu/abs/2024ApJ...973L..21W}
}

@ARTICLE{Weiss2008,
       author = {{Weiss}, Achim and {Schlattl}, Helmut},
        title = "{GARSTEC{\textemdash}the Garching Stellar Evolution Code. The direct descendant of the legendary Kippenhahn code}",
      journal = {\apss},
         year = 2008,
        month = aug,
       volume = {316},
       number = {1-4},
        pages = {99-106},
          doi = {10.1007/s10509-007-9606-5},
       adsurl = {https://ui.adsabs.harvard.edu/abs/2008Ap&SS.316...99W}
}

@ARTICLE{Winther2023,
       author = {{Winther}, Mark Lykke and {Aguirre B{\o}rsen-Koch}, V{\'\i}ctor and {R{\o}rsted}, Jakob Lysgaard and {Stokholm}, Amalie and {Verma}, Kuldeep},
        title = "{Did Kepler-444 have a long-lived convective core?}",
      journal = {\mnras},
         year = 2023,
        month = oct,
       volume = {525},
       number = {1},
        pages = {1416-1430},
          doi = {10.1093/mnras/stad1802},
archivePrefix = {arXiv},
       eprint = {2306.08430},
 primaryClass = {astro-ph.SR},
       adsurl = {https://ui.adsabs.harvard.edu/abs/2023MNRAS.525.1416W}
}

@ARTICLE{Wu2003,
       author = {{Wu}, Y. and {Murray}, N.},
        title = "{Planet Migration and Binary Companions: The Case of HD 80606b}",
      journal = {\apj},
         year = 2003,
        month = may,
       volume = {589},
       number = {1},
        pages = {605-614},
          doi = {10.1086/374598},
archivePrefix = {arXiv},
       eprint = {astro-ph/0303010},
 primaryClass = {astro-ph},
       adsurl = {https://ui.adsabs.harvard.edu/abs/2003ApJ...589..605W}
}

@ARTICLE{Yee2017,
       author = {{Yee}, Samuel W. and {Petigura}, Erik A. and {von Braun}, Kaspar},
        title = "{Precision Stellar Characterization of FGKM Stars using an Empirical Spectral Library}",
      journal = {\apj},
         year = 2017,
        month = feb,
       volume = {836},
       number = {1},
          eid = {77},
        pages = {77},
          doi = {10.3847/1538-4357/836/1/77},
archivePrefix = {arXiv},
       eprint = {1701.00922},
 primaryClass = {astro-ph.SR},
       adsurl = {https://ui.adsabs.harvard.edu/abs/2017ApJ...836...77Y}
}

@ARTICLE{Zeipel1910,
       author = {{von Zeipel}, H.},
        title = "{Sur l'application des s{\'e}ries de M. Lindstedt {\`a} l'{\'e}tude du mouvement des com{\`e}tes p{\'e}riodiques}",
      journal = {Astronomische Nachrichten},
         year = 1910,
        month = mar,
       volume = {183},
       number = {22},
        pages = {345},
          doi = {10.1002/asna.19091832202},
       adsurl = {https://ui.adsabs.harvard.edu/abs/1910AN....183..345V}
}

@ARTICLE{Ziegler2020,
       author = {{Ziegler}, Carl and {Tokovinin}, Andrei and {Brice{\~n}o}, C{\'e}sar and {Mang}, James and {Law}, Nicholas and {Mann}, Andrew W.},
        title = "{SOAR TESS Survey. I. Sculpting of TESS Planetary Systems by Stellar Companions}",
      journal = {\aj},
         year = 2020,
        month = jan,
       volume = {159},
       number = {1},
          eid = {19},
        pages = {19},
          doi = {10.3847/1538-3881/ab55e9},
archivePrefix = {arXiv},
       eprint = {1908.10871},
 primaryClass = {astro-ph.EP},
       adsurl = {https://ui.adsabs.harvard.edu/abs/2020AJ....159...19Z}
}


\begin{appendix}
    
\FloatBarrier
\section{Additional tables}\label{sec:appB}

\begin{table}[H]
    \centering
    \caption{Space-based photometry.}
    \begin{tabular}{c c c c}
    \toprule
    Facility & Sectors/Date & Cadence & System \\
    & -/(UTC) & (min) & \\
    \midrule
    TESS & 44, 45 & 10 & TOI-5120 \\
    TESS & 71, 72 & 2 & TOI-5120 \\
    CHEOPS & 2022-12-30 & 1 & TOI-5120 \\
    CHEOPS & 2024-02-02 & 1 & TOI-5120 \\
    TESS & 25 & 30 & TOI-5699 \\
    TESS & 51, 52 & 10 & TOI-5699 \\
    TESS & 78, 79 & 2 & TOI-5699 \\
    TESS & 26 & 30 & TOI-2158 \\
    TESS & 40, 53 & 2 & TOI-2158 \\
    \bottomrule
    \end{tabular}
    \label{tab:space_phot}
\end{table}

\begin{table}[H]
    \centering
    \caption{Radial velocities for TOI-5120, TOI-5699, and TOI-2158. This table is available in its entirety online (see Data availability).}
    \begin{tabular}{c c c c c}
    \toprule
    BJD & RV & $\sigma_{\rm RV}$ & Instrument & TOI \\
    (TDB) & (m~s$^{-1}$) & (m~s$^{-1}$) & & \\
    \midrule
    2459620.713441 & 47758.90 & 16.30 & TS23 & 5120 \\ 
    2459621.762558 & 47746.50 & 15.00 & TS23 & 5120 \\ 
    \vdots & \vdots & \vdots & \vdots & \vdots  \\ 
    2459622.444186 & -1.90 & 8.00 & FIES & 5120 \\ 
    2459623.447149 & -22.60 & 7.60 & FIES & 5120 \\ 
    \vdots & \vdots & \vdots & \vdots & \vdots \\     
    2459978.991798 & 65128.50 & 15.50 & TS23 & 5699 \\ 
    2459980.001616 & 65130.10 & 18.90 & TS23 & 5699 \\ 
    \vdots & \vdots & \vdots & \vdots & \vdots \\ 
    2460779.707932 & -30.90 & 9.50 & FIES+ & 5699 \\ 
    2460800.562709 & -25.30 & 9.10 & FIES+ & 5699 \\ 
    \vdots & \vdots & \vdots & \vdots & \vdots \\ 
    2459302.926498 & -64797.20 & 23.70 & TS23 & 2158 \\ 
    2459308.917232 & -64742.00 & 20.70 & TS23 & 2158 \\ 
    \vdots & \vdots & \vdots & \vdots & \vdots \\ 
    2459425.453503 & 2.90 & 5.60 & FIES+ & 2158 \\ 
    2459428.618606 & 144.70 & 14.10 & FIES+ & 2158 \\ 
    \vdots & \vdots & \vdots & \vdots & \vdots \\ 
    2459332.677805 & 0.10 & 10.50 & FIES & 2158 \\ 
    2459339.681805 & -37.90 & 8.70 & FIES & 2158 \\ 
    \vdots & \vdots & \vdots & \vdots & \vdots \\ 
    
    \bottomrule
    \end{tabular}
    \label{tab:rvs_all}
\end{table}

\begin{table}[H]
	\centering
	\caption{Varied parameters of the \texttt{Barbie}-MS grid with \num{5977} tracks. 
    Evolution along the tracks is tracked from the Zero-Age Main-Sequence until $\Delta\nu=\SI{50}{\micro\hertz}$, 
    and includes treatment of microscopic diffusion, but no convective overshoot.}
	\begin{tabular}{p{1.5cm} S[table-format = 2.2, table-column-width=1.5cm] S[table-format = 2.2, table-column-width=1.5cm]}
		\toprule
		Variable	&	{Minimum}	&	{Maximum}\\
		\midrule
		$M\, (\si{\solarmass})$	                &  0.70	&  1.20	\\
		$[\textup{Fe}/\textup{H}]_\textup{ini}$	& -1.0	&  0.6	\\
		$Y_\textup{ini}$		                    &  0.24	&  0.32	\\
		$\alpha_\textup{MLT}$	                    &  1.60	&  2.00	\\
		\bottomrule
	\end{tabular}
	\label{tab:barbiegrid} 
\end{table}

\begin{table*}
	\centering
	\caption{
		Limb-darkening coefficients for the different bandpasses for the observations of TOI-5120.
	}
	\label{tab:ld_all}
	\begin{threeparttable}
		\begin{tabular}{l c c c c}
		\toprule
		
			TOI-5120 & $q_1+q_2$ & $q_1-q_2$ & $q_1$ & $q_2$ \\

		\midrule 
		
			TESS & \LConeqsumfiveone & \LConeqdifffiveone & \LConeqonefiveone & \LConeqtwofiveone \\
            CHEOPS & \LCnineqsumfiveone & \LCnineqdifffiveone & \LCnineqonefiveone & \LCnineqtwofiveone \\
			$B$ & \LCthreeqsumfiveone & \LCthreeqdifffiveone & \LCthreeqonefiveone & \LCthreeqtwofiveone \\
			$z$-short & \LCfiveqsumfiveone & \LCfiveqdifffiveone & \LCfiveqonefiveone & \LCfiveqtwofiveone \\
			$i^\prime$ & \LCsevenqsumfiveone & \LCsevenqdifffiveone & \LCsevenqonefiveone & \LCsevenqtwofiveone \\
			$r^\prime$ & \LCeightqsumfiveone & \LCeightqdifffiveone & \LCeightqonefiveone & \LCeightqtwofiveone \\
		\midrule

			TOI-5699 & $q_1+q_2$ & $q_1-q_2$ & $q_1$ & $q_2$ \\

		\midrule 
			
			TESS & \LConeqsumfivesix & \LConeqdifffivesix & \LConeqonefivesix & \LConeqtwofivesix \\
            $z$-short & \LCfourqsumfivesix & \LCfourqdifffivesix & \LCfourqonefivesix & \LCfourqtwofivesix \\
        \midrule
			TOI-2158 & $q_1+q_2$ & $q_1-q_2$ & $q_1$ & $q_2$ \\

		\midrule 
		
			TESS & \LConeqsumtwoone & \LConeqdifftwoone & \LConeqonetwoone & \LConeqtwotwoone \\
			$B$ & \LCfourqsumtwoone & \LCfourqdifftwoone & \LCfourqonetwoone & \LCfourqtwotwoone \\
			$z$-short & \LCfiveqsumtwoone & \LCfiveqdifftwoone & \LCfiveqonetwoone & \LCfiveqtwotwoone \\
			$i^\prime$ & \LCthreeqsumtwoone & \LCthreeqdifftwoone & \LCthreeqonetwoone & \LCthreeqtwotwoone \\

        \bottomrule
		\end{tabular}
	\end{threeparttable}
	\begin{tablenotes}
		\item[] During the fit, we stepped in the sum of 
		the limb darkening coefficients applying a gaussian prior 
		with a width of 0.1,
		while fixing the difference.
	\end{tablenotes}
\end{table*}

\begin{table*}[]
    \centering
    \caption{Ground-based photometry.}
    \begin{tabular}{l c l c c c l}
        \toprule
        Observatory & Aperture & Location & Date & Filter & Coverage & Planet \\
        & (m) & & (UTC) & & & \\
        \midrule
        Acton Sky Portal & 0.36 & Acton, MA, USA & 2023-04-09 & $i^{\prime}$ & Ingress+80\% & TOI-5120~b \\
        Acton Sky Portal & 0.36 & Acton, MA, USA & 2023-11-19 & $r^{\prime}$ & Ingress+80\% & TOI-5120~b \\
        LCOGT-TFN & 1.0 & Tenerife, Canary Islands & 2024-02-01 & $B$ & Ingress+50\% & TOI-5120~b \\
        LCOGT-TFN & 1.0 & Tenerife, Canary Islands & 2024-02-01 & $z$-short & Ingress+50\% & TOI-5120~b \\
        LCOGT-McD & 1.0 & McDonald Observatory, TX, USA & 2024-02-02 & $B$ & Egress+90\% & TOI-5120~b \\
        LCOGT-McD & 1.0 & McDonald Observatory, TX, USA & 2024-02-02 & $z$-short & Egress+90\% & TOI-5120~b \\
        LCOGT-TFN & 1.0 & Tenerife, Canary Islands & 2023-06-26 & $z$-short & Full & TOI-5699~b \\
        LCOGT-McD & 0.4 & McDonald Observatory, TX, USA & 2020-08-06 & Sloan $i^{\prime}$ & Full & TOI-2158~b \\
        LCOGT-SAAO & 1.0 & Cape Town, South Africa & 2021-06-05 & $B$ & Full & TOI-2158~b \\
        LCOGT-SAAO & 1.0 & Cape Town, South Africa & 2021-06-05 & $z$-short & Full & TOI-2158~b \\
        \bottomrule
    \end{tabular}
    \label{tab:gb_phot}
\end{table*}

\begin{table*}[]
    \centering
    \caption{Spectroscopic parameters for TOI-5120, TOI-5699, and TOI-2158 modelled with SME compared to values from \emph{Gaia} DR3 and EXOFOP.}
    \label{tab:spec_app}
    \begin{threeparttable}
        \begin{tabular}{c c c c c c c}
            \toprule
             Method & $T_{\rm eff}$ & $\log g$ & $\rm [Fe/H]$ & $\rm [Ca/H]$ & $\rm [Na/H]$ & $v \sin i_\star$  \\
             & (K) & (cgs) & (dex) & (dex) & (dex) & (km~s$^{-1}$) \\
             \midrule
             \multicolumn{7}{c}{TOI-5120} \\
             SME & $5763\pm90$ & $4.18\pm0.07$ & $0.20\pm0.07$ & $0.25\pm 0.08$ & $0.40\pm0.09$ & $4.8\pm0.9$ \\
             \emph{Gaia} DR3 & $5823^{+12}_{-4}$ & $4.12$ & $\cdots$ & $\cdots$ & $\cdots$ & $\cdots$ \\
             EXOFOP\tnote{(a)} & $5778\pm103$ & $4.12\pm0.07$ & $0.20\pm0.01$ & $\cdots$ & $\cdots$ & $\cdots$ \\
             \noalign{\smallskip} \noalign{\smallskip}
             \multicolumn{7}{c}{TOI-5699} \\
             SME & $5905\pm100$ & $4.05\pm0.07$ & $0.11\pm0.05$ & $0.20\pm 0.06$ & $0.25\pm0.05$ & $4.3\pm1.2$ \\
             \emph{Gaia} DR3 & $5818\pm3$ & $4.11$ & $\cdots$ & $\cdots$ & $\cdots$ & $\cdots$ \\
             EXOFOP\tnote{(b)} & $5900\pm143$ & $3.99\pm0.08$ & $0.02\pm0.01$ & $\cdots$ & $\cdots$ & $\cdots$ \\
             \noalign{\smallskip} \noalign{\smallskip}
             \multicolumn{7}{c}{TOI-2158} \\
             SME & $5399\pm64$ & $4.00\pm0.05$ & $0.19\pm0.09$ & $0.19\pm 0.11$ & $0.42\pm0.12$ & $3.6\pm1.0$ \\
             \emph{Gaia} DR3 & $5400\pm^{+10}_{-24}$ & $4.23$ & $\cdots$ & $\cdots$ & $\cdots$ & $\cdots$ \\
             EXOFOP\tnote{(c)} & $5389\pm123$ & $4.06\pm0.08$ & $0.39$ & $\cdots$ & $\cdots$ & $\cdots$ \\
             SPC & $5673 \pm 50$ & $4.19 \pm 0.05$ & $0.47 \pm 0.08$ & $\cdots$ & $\cdots$ & $3.7 \pm 0.5$ \\
             \bottomrule
        \end{tabular}
    \end{threeparttable}
	\begin{tablenotes}
		\item[] $\rm^{(a)}$ {\footnotesize \url{https://exofop.ipac.caltech.edu/tess/target.php?id=58723861}}
        \item[] $\rm^{(b)}$ {\footnotesize \url{https://exofop.ipac.caltech.edu/tess/target.php?id=224328450}}
        \item[] $\rm^{(c)}$ {\footnotesize \url{https://exofop.ipac.caltech.edu/tess/target.php?id=342642208}}
	\end{tablenotes}
\end{table*}

\begin{table*}[]
    \centering
    \caption{Comparison of stellar parameters of TOI-2158, TOI-5120, and TOI-5699 modelled using the spectroscopic values from the SME analysis.}
    \label{tab:stars_app}
        \begin{threeparttable}
            \begin{tabular}{c c c c c c}
                \toprule
                 Method & $M_\star$ & $R_\star$ & $\rho_\star$ & $L_\star$ & Age  \\
                 & (M$_\odot$) & (R$_\odot$) & (g~cm$^{-3}$) & (L$_\odot$) & (Gyr) \\
                 \midrule
                 \multicolumn{6}{c}{TOI-5120} \\
                 {\texttt ARIADNE} & $1.154 \pm0.053$ & $1.413\pm 0.035$ &  $0.58 \pm 0.05$ & $2.04\pm 0.10 $ & $5.7\pm 0.8$ \\
                 {\texttt BASTA} & \fiveonemassbasta & \fiveoneradbasta &  \fiveonerhobasta & \fiveonelumbasta  & \fiveoneagebasta  \\
                 \noalign{\smallskip} \noalign{\smallskip}
                 \multicolumn{6}{c}{TOI-5699} \\
                 {\texttt ARIADNE} & $1.158\pm0.065$ & $1.676\pm0.051$ &  $0.35 \pm 0.04$ & $3.10\pm 0.10 $ & $6.0^{+0.5}_{-3.5}$  \\
                {\texttt BASTA} & \fivesixmassbasta & \fivesixradbasta &  \fivesixrhobasta & \fivesixlumbasta  & \fivesixagebasta  \\
                 \noalign{\smallskip} \noalign{\smallskip}
                 \multicolumn{6}{c}{TOI-2158} \\
                 {\texttt ARIADNE} & $0.984\pm 0.032$ & $1.453\pm 0.017$ & $0.35 \pm 0.04$   & $ 1.59 \pm 0.05 $ &  $13.1^{+0.4}_{-1.5}$  \\
                 {\texttt BASTA} & $0.975 \pm 0.019$ & $1.50 \pm 0.05$ & $0.40 \pm 0.04$ & $1.68 \pm 0.08$  & $14.0 \pm 0.9$ \\
                 {\texttt BASTA}\tnote{(a)} & \twoonemassbasta & \twooneradbasta &  \twoonerhobasta & \twoonelumbasta  & \twooneagebasta  \\
                 \bottomrule
            \end{tabular}
        \end{threeparttable}
    	\begin{tablenotes}
    		\item[] $\rm^{(a)}$ {\footnotesize  Spectroscopic values from SPC.}
    	\end{tablenotes}
\end{table*}

\begin{table*}
	\centering
	\caption{
		Posterior values from the joint fit of the RVs and TESS photometry for 
		TOI-5120 and TOI-5699 assuming a circular orbit. Symbols have the same meaning as in \tref{tab:pars}.
    }
	\label{tab:pars_circ}

		\begin{tabular}{c c c}
		\toprule
		
			& TOI-5120 & TOI-5699 \\
		
		\midrule
			
			\multicolumn{3}{c}{Planet b} \\
		$ T_0 $~(BJD$_{\rm TDB}$) & \circmidfiveone & \circmidfivesix \\
		$ P $~(d) & \circperfiveone & \circperfivesix \\
		$ a/R_\star $ & \circarfiveone & \circarfivesix \\
		$ \cos i $ & \circcosifiveone & \circcosifivesix \\
		$ R_{\rm p}/R_\star $ & \circrpfiveone & \circrpfivesix \\
		$ K $~(m~s$^{-1}$) & \circkampfiveone & \circkampfivesix \\

		\midrule

			\multicolumn{3}{c}{System/instrument} \\
		
		$ \gamma_1 $~(m~s$^{-1}$) & \circgammafiveone & \circgammafivesix \\
		$ \gamma_2 $~(m~s$^{-1}$) & \circgammatfiveone & \circgammatfivesix \\
		$ \sigma_{1} $~(m~s$^{-1}$) & \circjitterfiveone & \circjitterfivesix \\
		$ \sigma_{2} $~(m~s$^{-1}$) & \circjittertfiveone & \circjittertfivesix \\

		\bottomrule
		\end{tabular}

\end{table*}


\FloatBarrier

\section{Additional figures}

\begin{figure}[H]
    \centering
    \includegraphics[width=\linewidth]{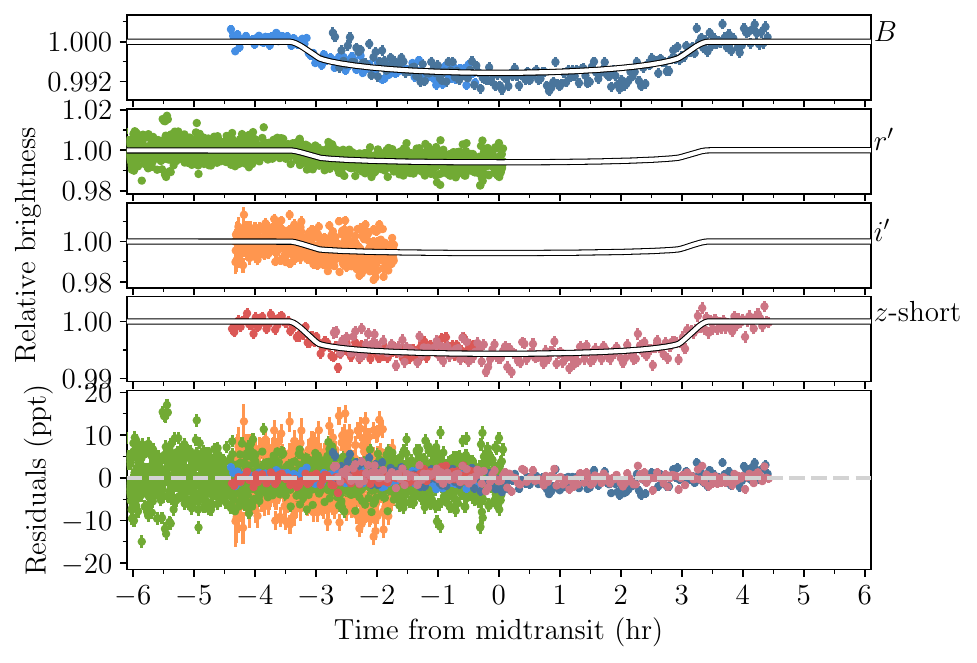}
    \caption{Ground-based follow-up photometry for TOI-5120~b. Observations were carried out in the $B$ (blue), $r^\prime$ (green), $i^\prime$ (orange), and $z$-short (red) bandpasses. The best-fitting transit model for TOI-5120~b for each bandpass is overplotted. For $B$ and $z$-short there are two sets of observations (see \fref{fig:time_5120}).}
    \label{fig:lc_5120_gb}
\end{figure}

\begin{figure}[H]
    \centering
    \includegraphics[width=\linewidth]{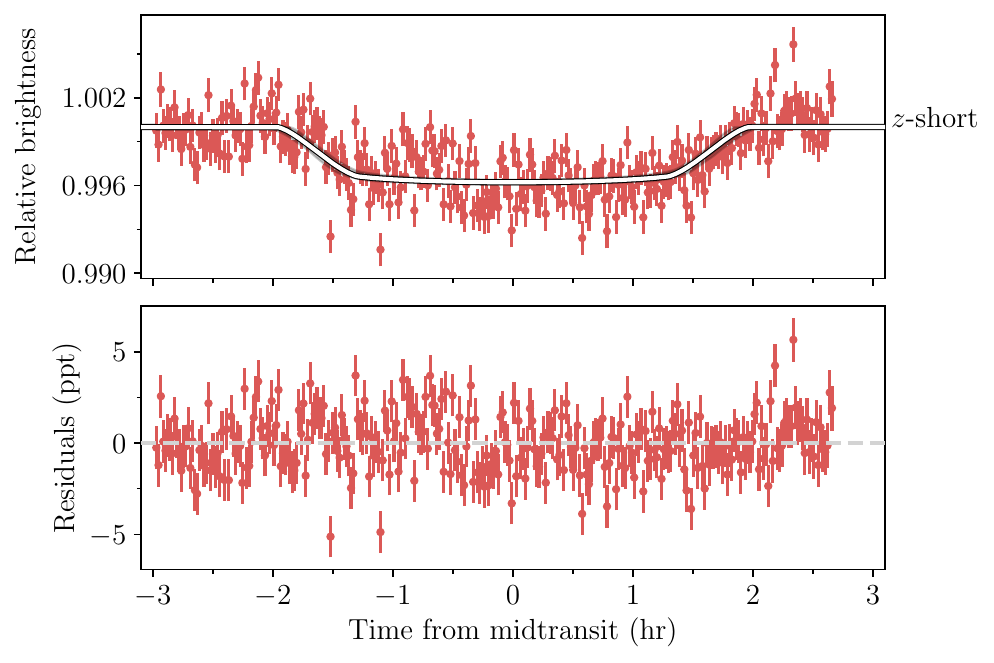}
    \caption{Ground-based follow-up photometry for TOI-5699~b carried out using the $z$-short (red) bandpass. The best-fitting transit model for TOI-5699~b is overplotted.}
    \label{fig:lc_5699_gb}
\end{figure}

\begin{figure}[H]
    \centering
    \includegraphics[width=\linewidth]{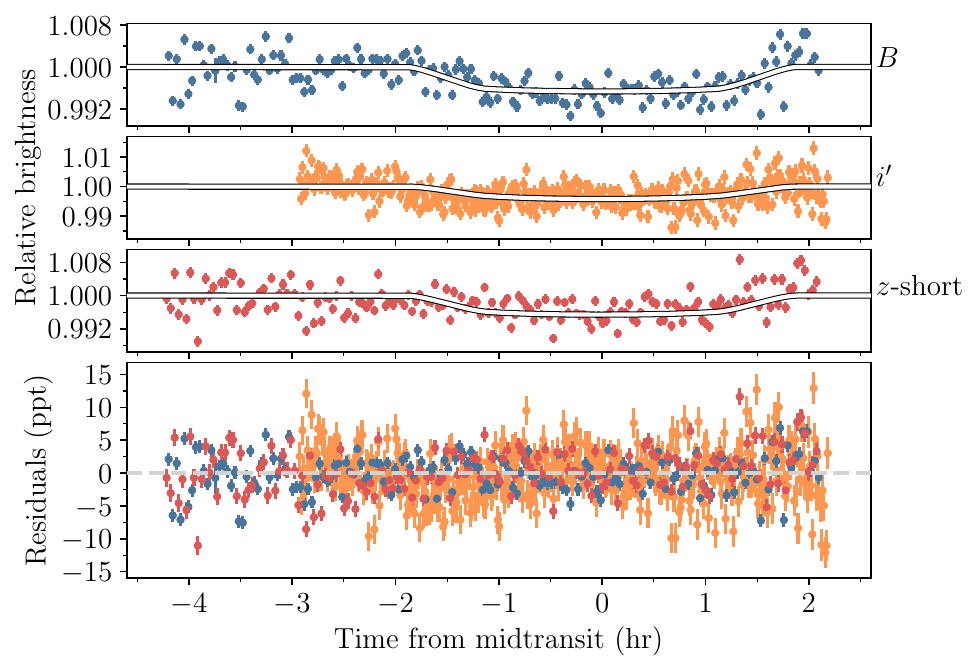}
    \caption{Ground-based follow-up photometry for TOI-2158~b. Observations were carried out in the $B$ (blue), $i^\prime$ (orange), and $z$-short (red) bandpasses. The best-fitting transit model for TOI-2158~b for each bandpass is overplotted.}
    \label{fig:lc_2158_gb}
\end{figure}

\begin{figure}[H]
    \centering
    \includegraphics[width=\linewidth]{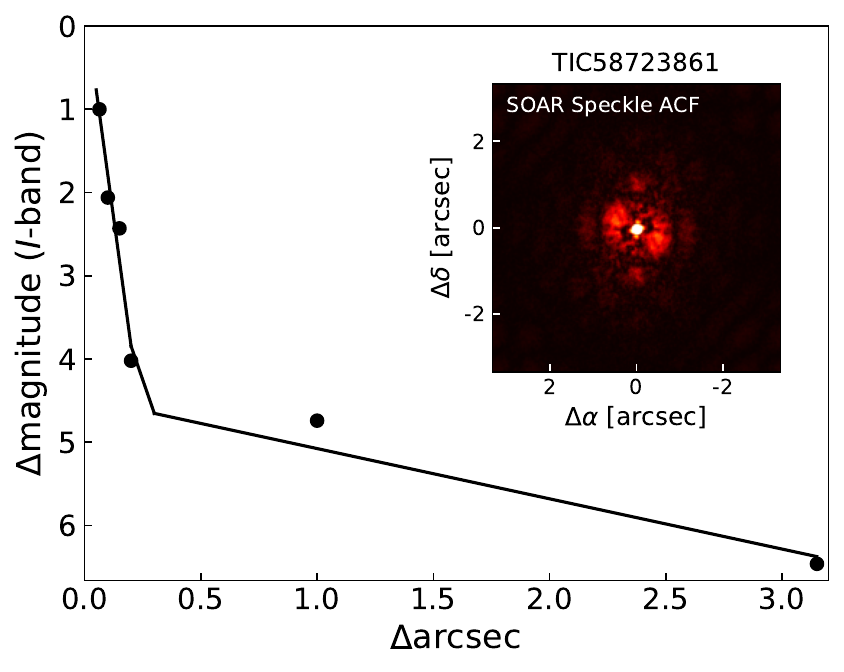}
    \caption{SOAR 5$\sigma$ detection sensitivity and speckle auto-correlation function imaging for the TOI-5120 system.}
    \label{fig:soar_image}
\end{figure}

\begin{figure}[H]
    \centering
    \includegraphics[width=\linewidth]{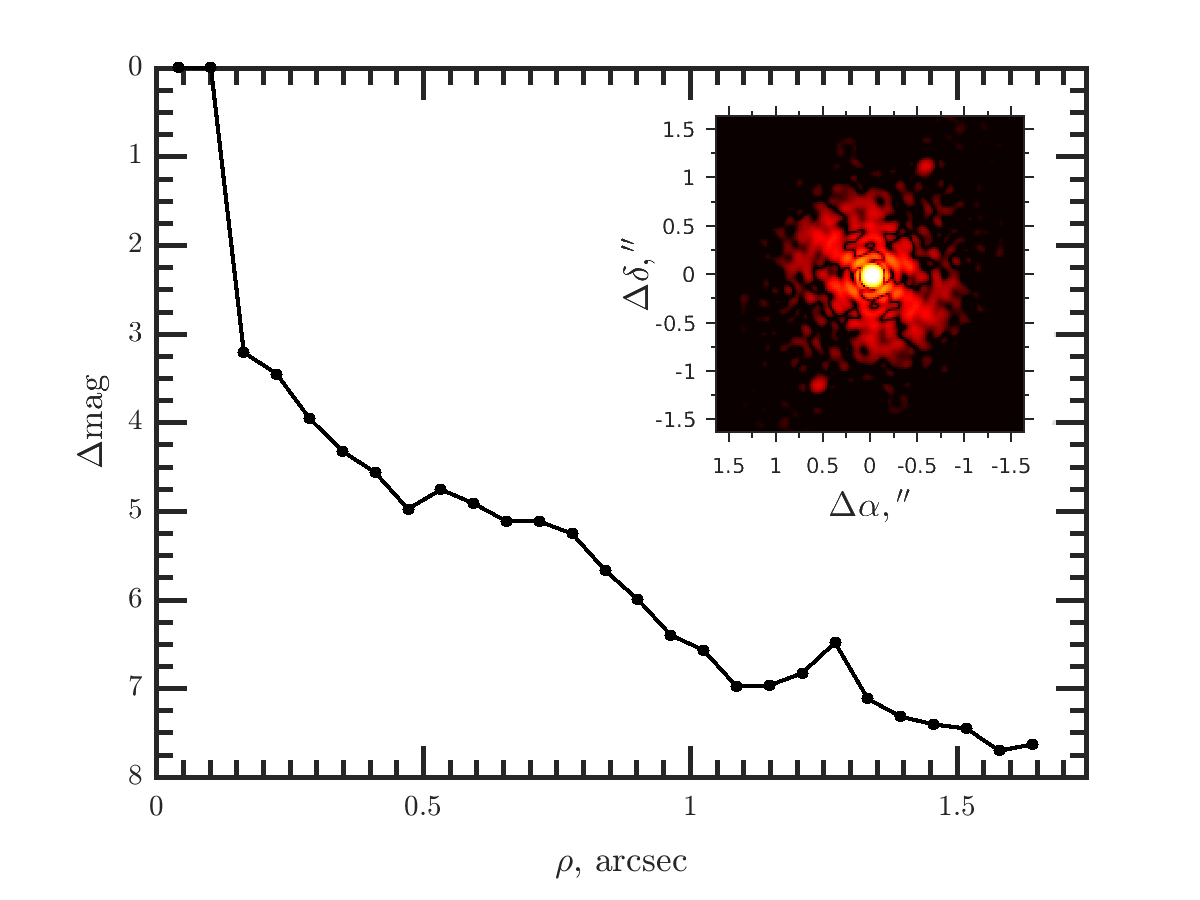}
    \caption{SAI-2.5m speckle sensitivity curve and ACF for TOI-5120 taken in the $I$-band.}
    \label{fig:sai_toi5120}
\end{figure}

\begin{figure}[H]
    \centering
    \includegraphics[width=\linewidth]{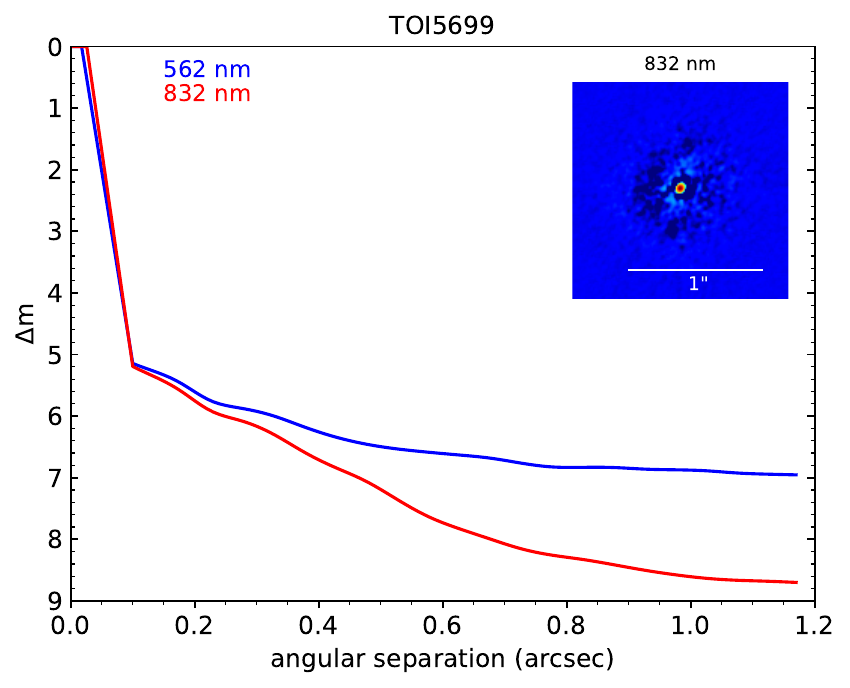}
    \caption{Gemini North speckle optical 5$\sigma$ contrast curves and the 832~nm reconstructed image for the TOI~5699 system.
    }
    \label{fig:gemini_imaging}
\end{figure}

\begin{figure}[H]
    \centering
    \includegraphics[width=\linewidth]{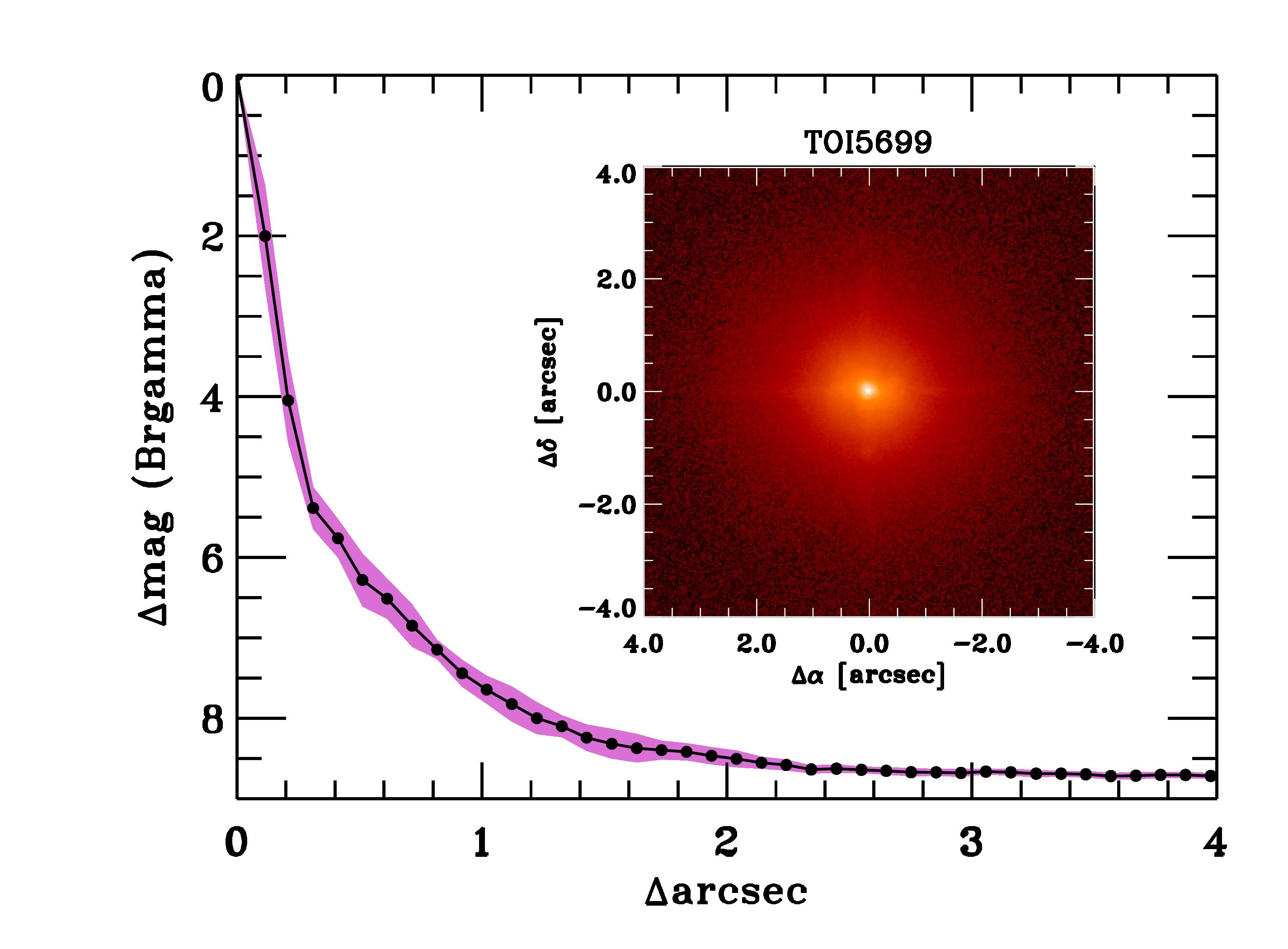}
    \caption{Palomar NIR AO close-in view and sensitivity curve for the TOI-5699 system.
    }
    \label{fig:palomar_aoimaging}
\end{figure}

\begin{figure}[H]
    \centering
    \includegraphics[width=\linewidth]{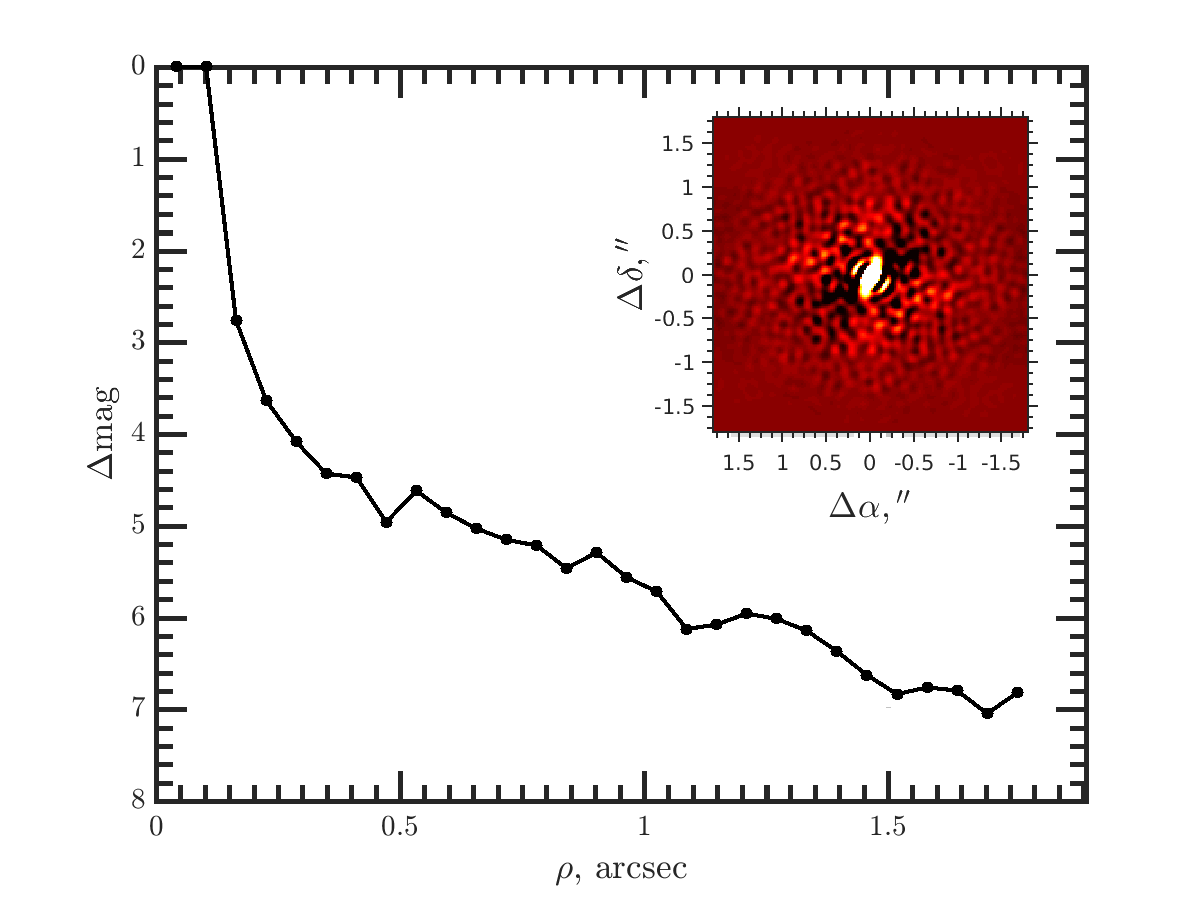}
    \caption{SAI-2.5m speckle sensitivity curve and ACF for TOI-5699 taken in the $I$-band.}
    \label{fig:sai_toi5699}
\end{figure}

\begin{figure}[H]
    \centering
    \includegraphics[width=\linewidth]{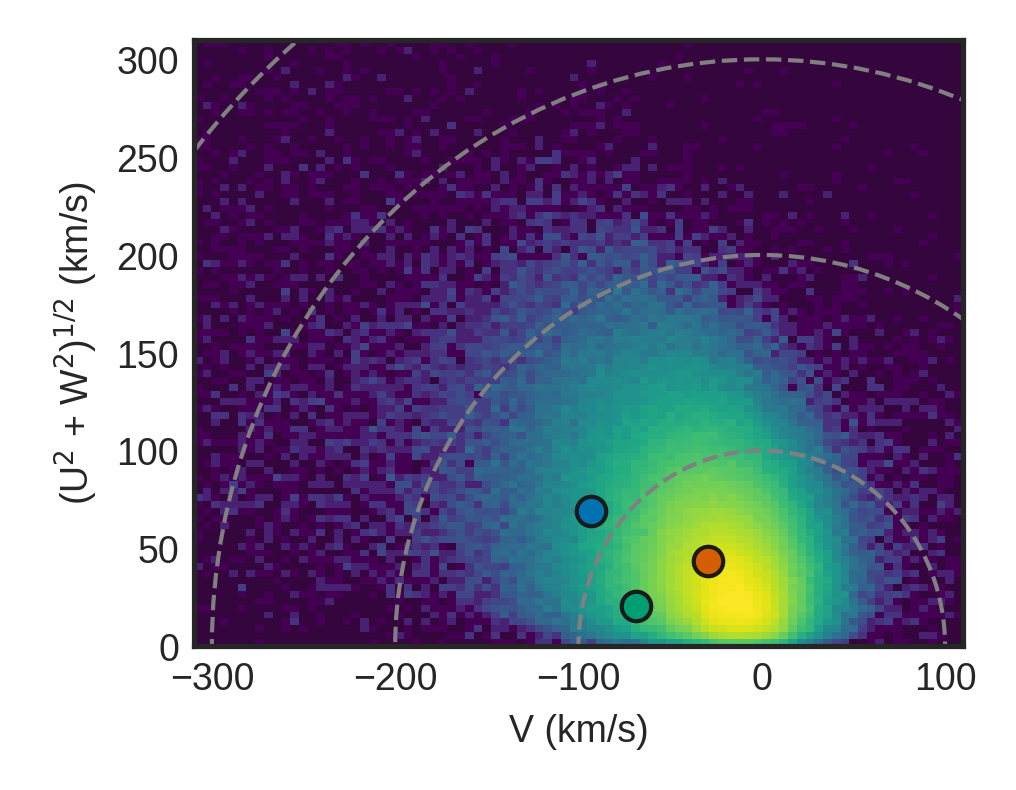}
    \caption{Plot of the Galactic orbital motions of the three stars, compared single stars from \emph{Gaia}~DR3 with reliable astrometric data and available line-of-sight velocities (\texttt{astrometric\_params\_solved = 95}, \texttt{rv\_nb\_transits > 0}, \texttt{ruwe < 1.4}, and \texttt{non\_single\_star = 0}).
The Toomre diagram shows the positions of our host stars in heliocentric Cartesian velocity space, demonstrating that they are members of the Galactic disk. TOI-5120, TOI-5699, and TOI-2158 are shown with green, orange, and blue markers, respectively.}
    \label{fig:toomre}
\end{figure}

\begin{figure*}
    \centering
    \includegraphics[width=\linewidth]{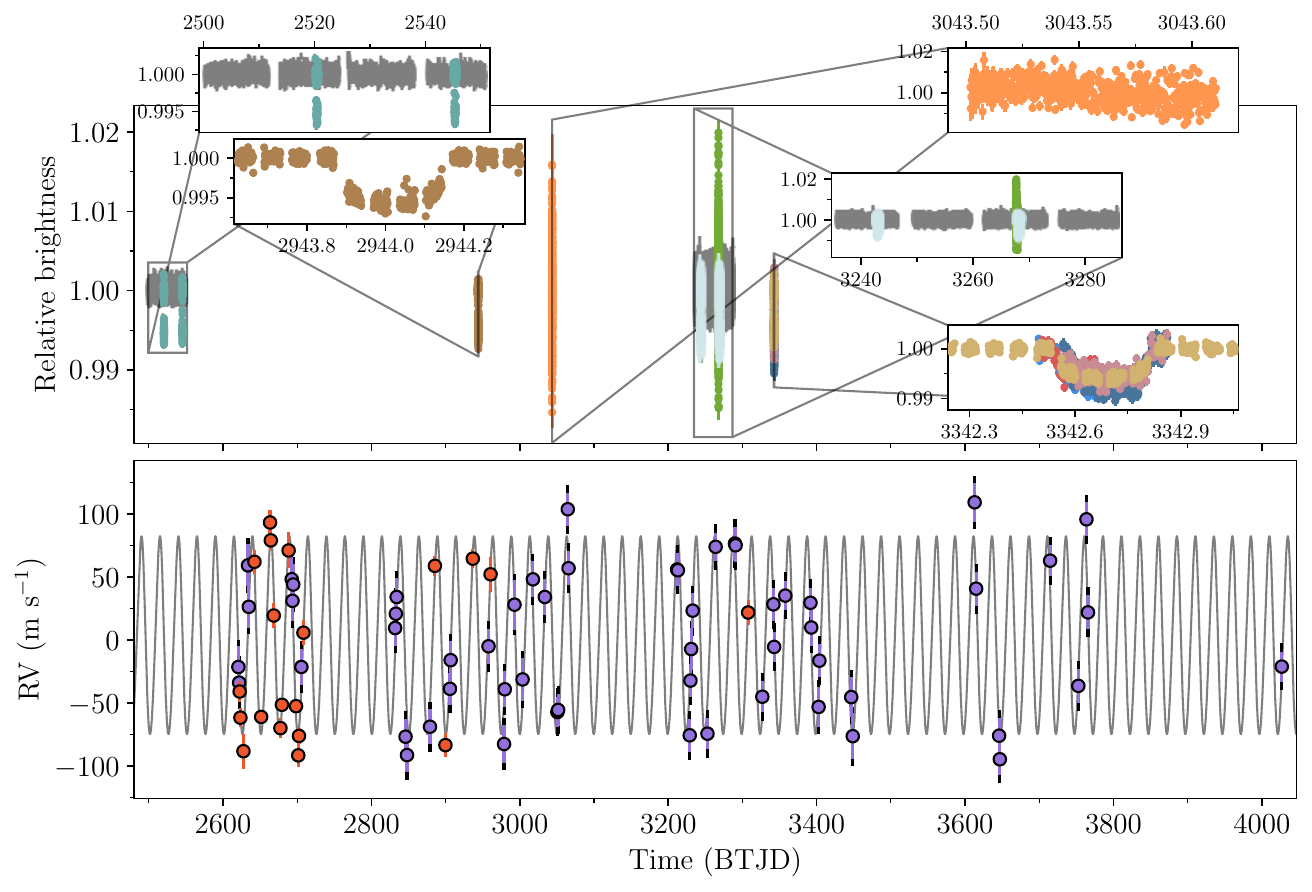}
    \caption{Time series of TOI-5120. {\it Top:} TESS photometry shown in gray with points around transits marked with green colors as shown in \fref{fig:lc_toi5120} (10-min cadence in dark and 2-min cadence in light). The CHEOPS observations are shown as brown and tan markers. Ground-based . {\it Bottom:} RVs obtained with TS23 (purple) and FIES (orange) overplotted on the best-fitting orbital solution.}
    \label{fig:time_5120}
\end{figure*}

\begin{figure*}
    \centering
    \includegraphics[width=\linewidth]{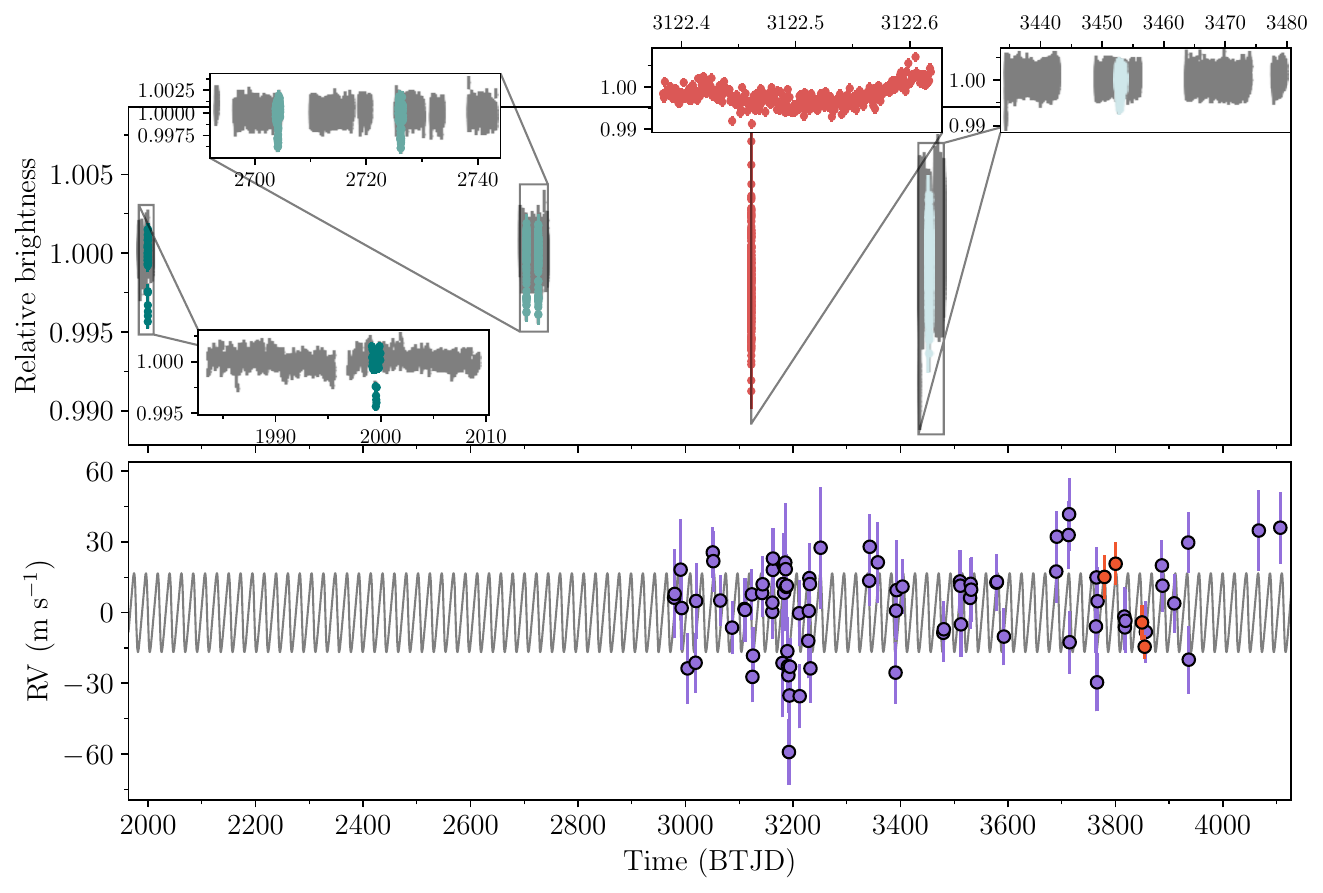}
    \caption{Time series of TOI-5699. {\it Top:} TESS photometry shown in gray with markers around transits colored as in \fref{fig:lc_toi5699} (30-min cadence in dark green, 10-min cadence in lighter green, and 2-min in pale blue). The single ground-based transit taken with the $z$-short filter from LCOGT is shown in red. {\it Bottom:} RVs obtained with TS23 overplotted on the best-fitting orbital solution.}
    \label{fig:time_5699}
\end{figure*}

\begin{figure*}
    \centering
    \includegraphics[width=\linewidth]{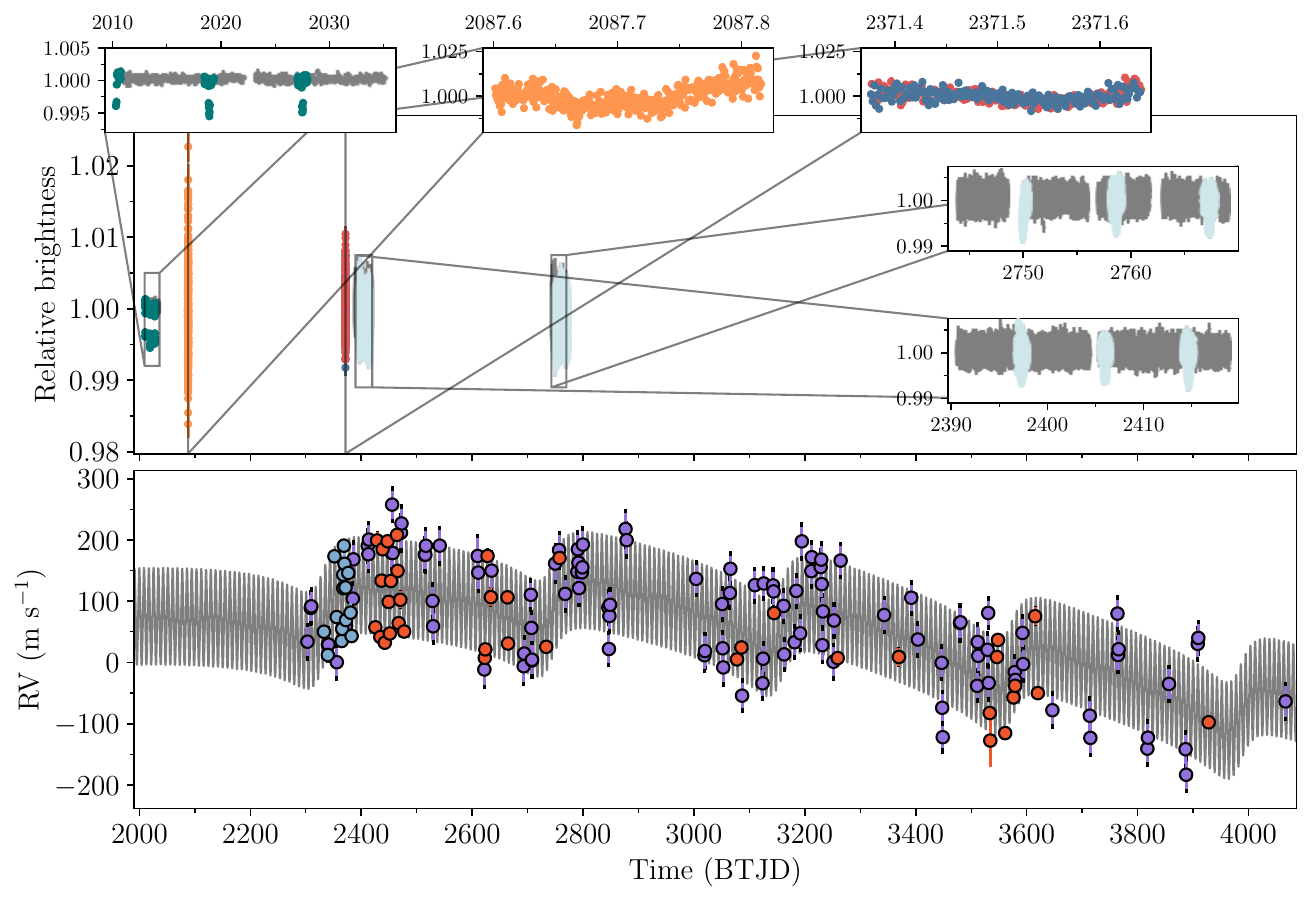}
    \caption{Time series of TOI-2158. {\it Top:} TESS photometry shown in gray with points around transits marked with green colors as shown in \fref{fig:lc_toi2158} (30-min cadence in dark and 2-min cadence in light). Ground-based photometry from LCOGT are shown as red ($i^{\prime}$), blue ($B$), and orange ($z$-short) markers. The insets show the raw, continuous photometric time series. {\it Bottom:} RVs obtained with TS23 (purple), FIES (light blue), and FIES+ (orange) overplotted on the best-fitting orbital solution as \fref{fig:rv_toi2158_b} and \fref{fig:rv_toi2158_c}. The white line denotes the long-term trend.}
    \label{fig:time_2158}
\end{figure*}

\begin{figure}[H]
    \centering
    \includegraphics[width=\columnwidth]{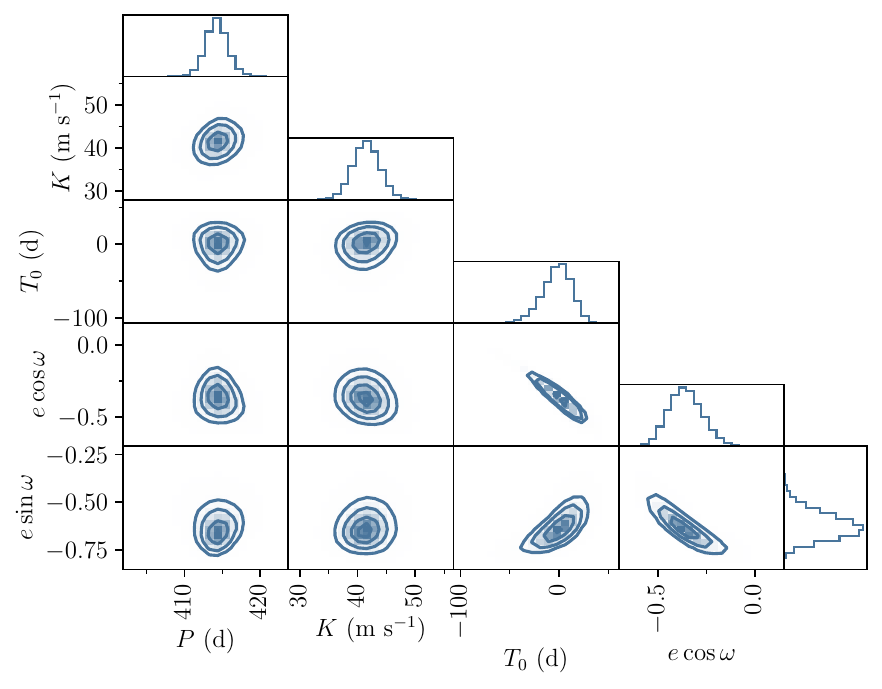}
    \caption{Correlation plot of the key stepping parameters of the MCMC carried out for TOI-2158~c. $T_0$ is offset by $2459620$.
    }
    \label{fig:corner_toi2158c}
\end{figure}

\begin{figure}[H]
    \centering
    \includegraphics[width=\columnwidth]{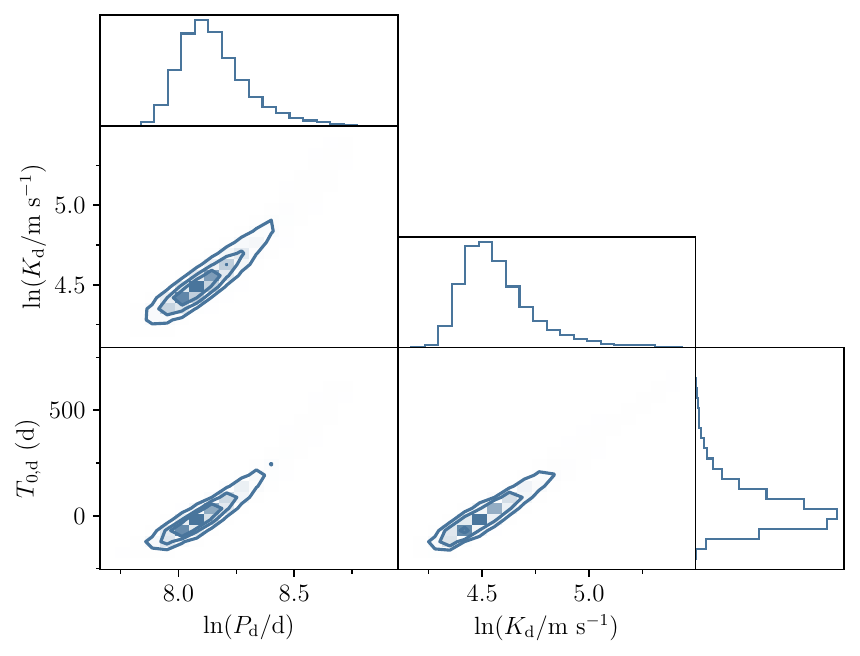}
    \caption{Correlation plot of the key stepping parameters of the MCMC carried out for TOI-2158~d. $T_0$ is offset by $2460506$.
    }
    \label{fig:corner_toi2158d}
\end{figure}

\begin{figure*}
    \centering
    \includegraphics[width=\textwidth]{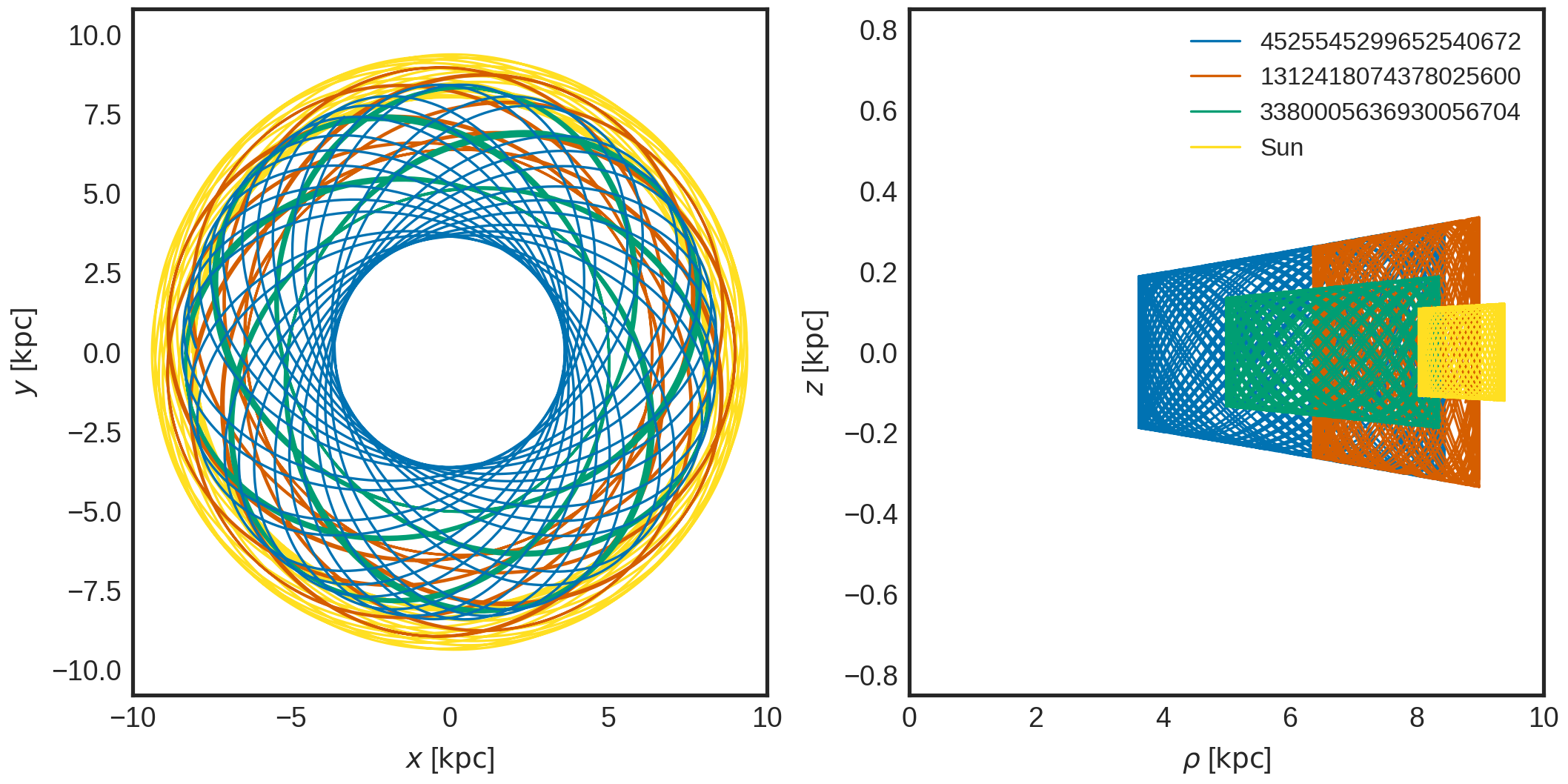}
    \caption{The Galactic orbits of the three host stars around the Milky Way's center. The orbit of the Sun is shown for comparison. Colour-coding is the same as in \fref{fig:toomre}.}
    \label{fig:gal_orbits}
\end{figure*}

\begin{figure*}
    \centering
    \includegraphics[width=\textwidth]{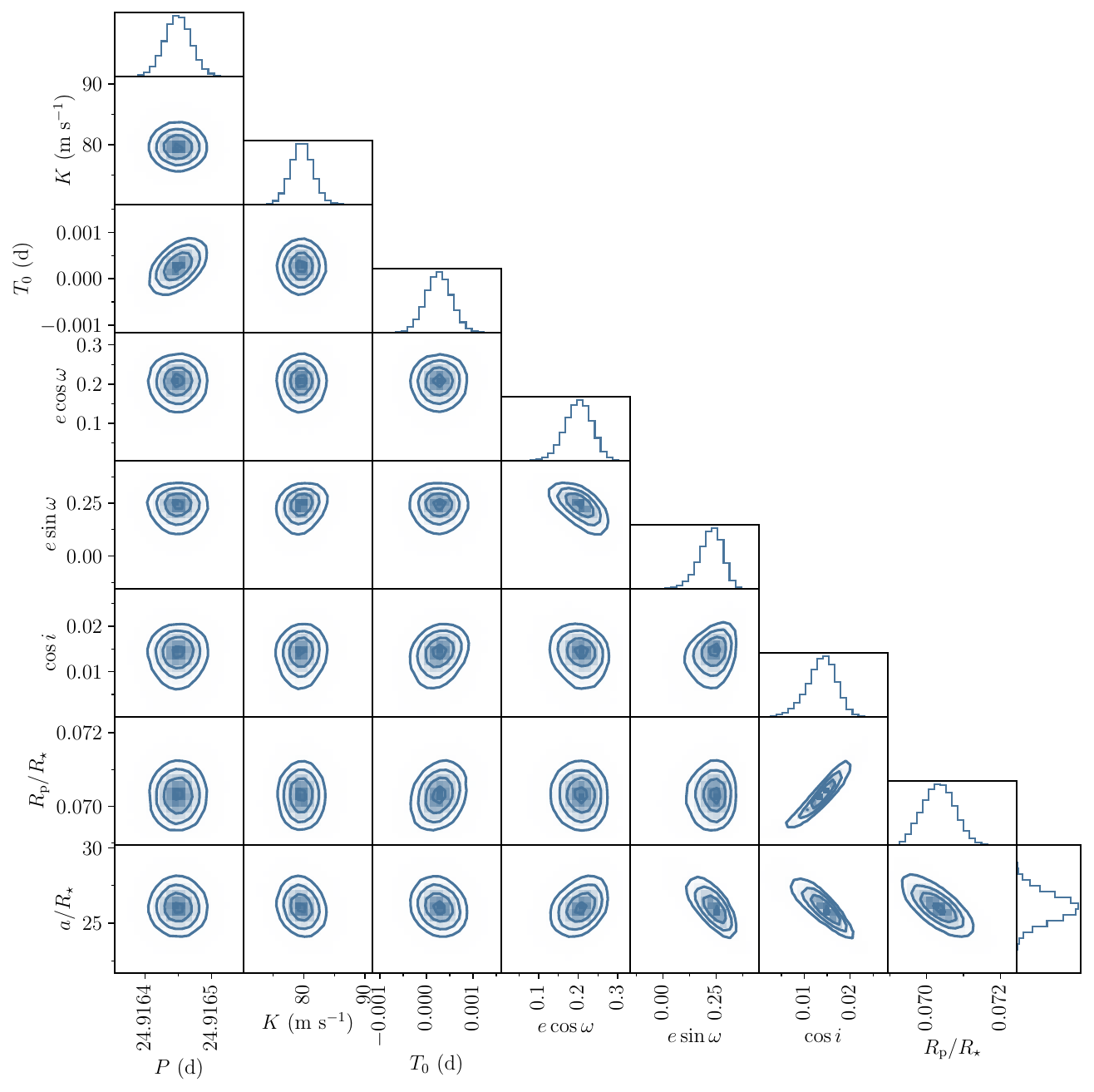}
    \caption{Correlation plot of the key stepping parameters of the MCMC carried out for TOI-5120. $T_0$ is offset by $2460267.93$.}
    \label{fig:corner_toi5120}
\end{figure*}

\begin{figure*}
    \centering
    \includegraphics[width=\textwidth]{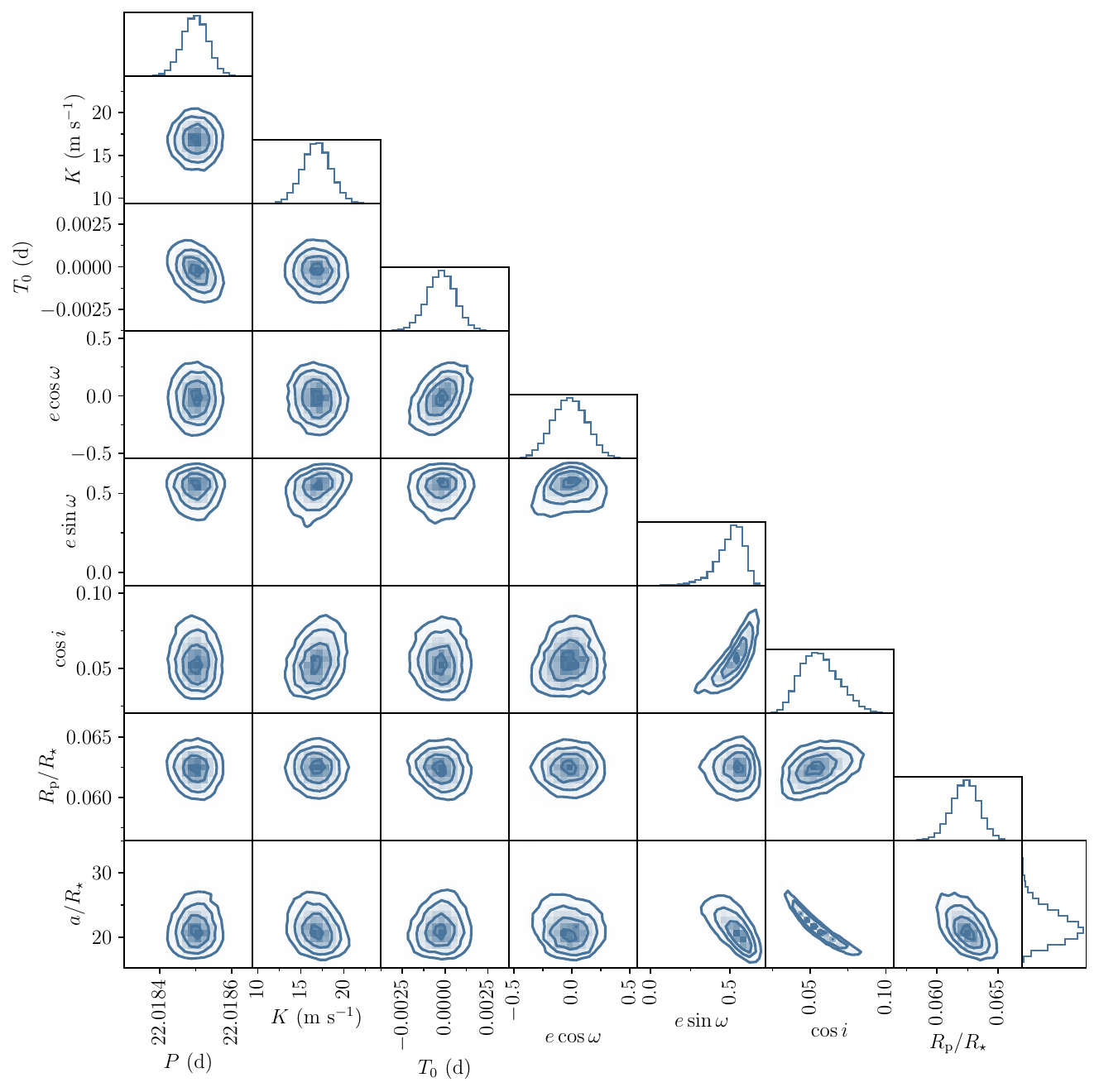}
    \caption{Correlation plot of the key stepping parameters of the MCMC carried out for TOI-5699. $T_0$ is offset by $2459704.156$.}
    \label{fig:corner_toi5699}
\end{figure*}

\begin{figure*}
    \centering
    \includegraphics[width=\textwidth]{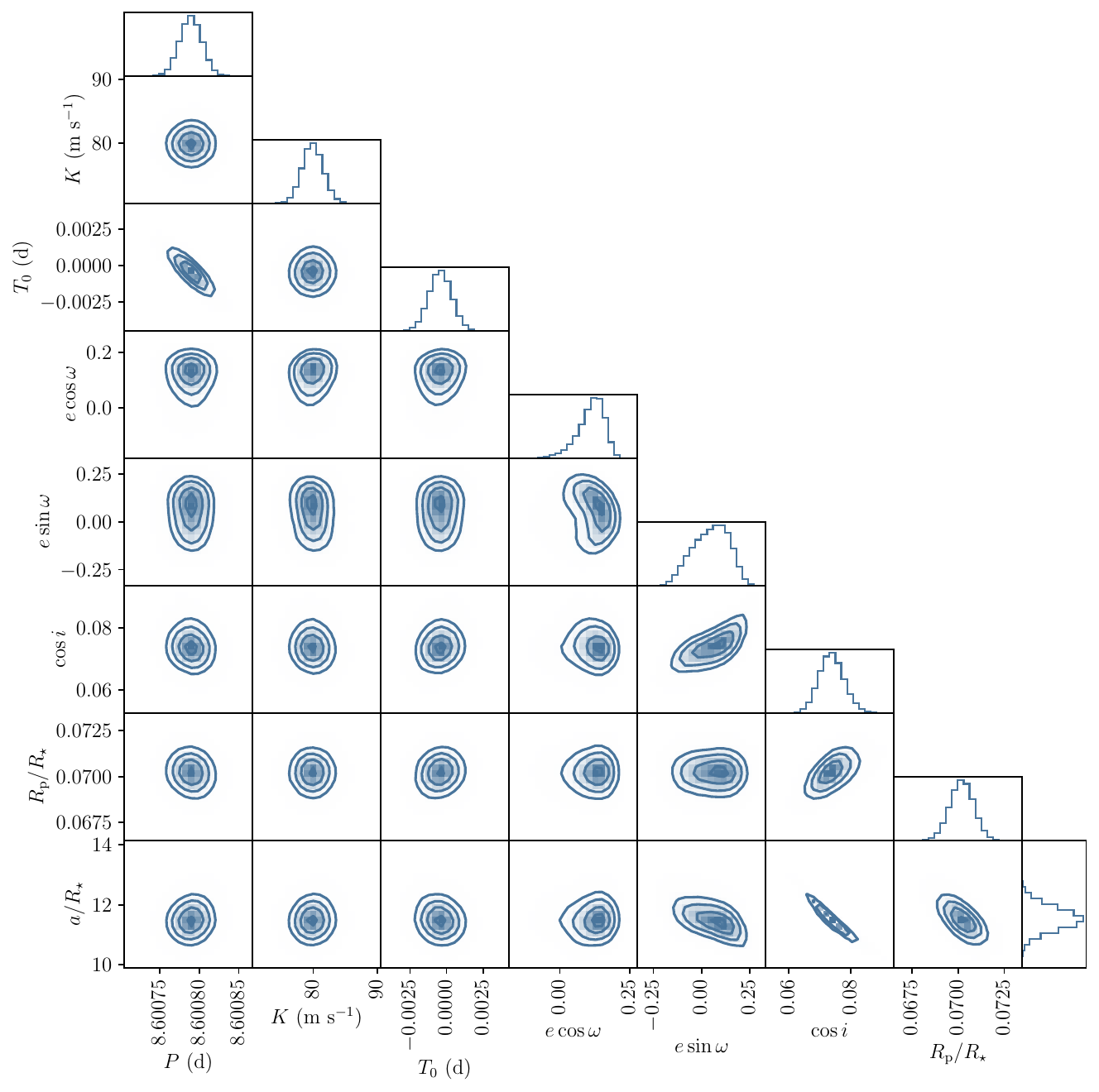}
    \caption{Correlation plot of the key stepping parameters of the MCMC carried out for TOI-2158~b. $T_0$ is offset by $2459018.92$.
    }
    \label{fig:corner_toi2158b}
\end{figure*}

\end{appendix}


\end{document}